\documentclass[12pt]{article}
\usepackage{axodraw}
\usepackage{bbm}
\usepackage{epsfig}
\usepackage{array}
\usepackage{float}
\usepackage{dsfont}
\usepackage{amstext}
\usepackage{rotating}
\usepackage{a4}
\usepackage{a4wide}
\usepackage{cite}

\newcommand{\be}{\begin{eqnarray}}
\newcommand{\ee}{\end{eqnarray}}

\def\nue{{\nu_e}}
\def\anue{{\bar\nu_e}}
\def\numu{{\nu_{\mu}}}
\def\anumu{{\bar\nu_{\mu}}}
\def\nutau{{\nu_{\tau}}}
\def\anutau{{\bar\nu_{\tau}}}

\newcommand{\ms}{\Delta m^2_{21}}
\newcommand{\ma}{\Delta m^2_{31}}

\newcommand{\sss}{\sin^2 \theta_{12}}
\newcommand{\sch}{\sin^2 \theta_{13}}

\newcommand{\sa}{\sin^2 \theta_{23}}

\newcommand{\mt}{$\mu$-$\tau$~}

\newcommand{\s}{\Sigma}
\def\nn{\nonumber}

\def\gtap{\ \raisebox{-.4ex}{\rlap{$\sim$}} \raisebox{.4ex}{$>$}\ }

\begin{document}

\title{\vspace{-2cm}
\hfill {\small \texttt{HRI-P-09-06-001}} 
\vskip 0.4cm
\Large Two Higgs Doublet Type III Seesaw with $\mu$-$\tau$ symmetry 
at LHC}
\author{
Priyotosh Bandyopadhyay\thanks{email: \tt priyotosh@hri.res.in},~
Sandhya Choubey\thanks{email: \tt sandhya@hri.res.in},~
Manimala Mitra\thanks{email: \tt mmitra@hri.res.in}
\\\\
{\normalsize \it Harish--Chandra Research Institute,}\\
{\normalsize \it Chhatnag Road, Jhunsi, 211019 Allahabad, India }\\ \\ 
}
\date{ \today}
\maketitle
\thispagestyle{empty}
\vspace{-0.8cm}
\begin{abstract}
\noindent  
We propose a two Higgs doublet Type III seesaw model with 
$\mu$-$\tau$ flavor symmetry. We add an additional SU(2) Higgs 
doublet and three SU(2) fermion triplets in our model. 
The presence of two Higgs doublets 
allows for natural explanation of small neutrino masses with 
triplet fermions in the 100 GeV mass range, without 
fine tuning of the Yukawa couplings to extremely small 
values. The triplet fermions couple to the gauge bosons and 
can be thus produced at the LHC. We study in detail the 
effective cross-sections for the production and 
subsequent decays of these heavy exotic fermions. 
We show for the first time that the $\mu$-$\tau$ 
flavor symmetry in the low energy neutrino mass matrix 
results in mixing matrices for the 
neutral and charged heavy fermions that are not unity and 
which carry the flavor symmetry pattern. This flavor 
structure can be observed in the decays of the heavy fermions 
at LHC.  
The large Yukawa couplings in our model result in the 
decay of the heavy fermions into lighter leptons and Higgs
with a decay rate which is about $10^{11}$ 
times larger than what is expected for the one Higgs 
Type III seesaw model with 100 GeV triplet fermions.
The smallness of neutrino masses 
constrains the neutral Higgs mixing angle $\sin\alpha$ in our 
model in such a way that the heavy fermions 
decay into the lighter neutral CP even Higgs $h^0$, 
CP odd Higgs $A^0$ and the charged Higgs $H^\pm$, 
but almost never to the heavier neutral CP even Higgs $H^0$. 
The small value for $\sin\alpha$ also results in a 
very long lifetime for $h^0$. This displaced 
decay vertex should be visible at LHC. We provide an 
exhaustive list of collider signature 
channels for our model and 
identify those that 
have very large effective cross-sections at LHC and 
almost no standard model background.

\end{abstract}

\newpage

\section{Introduction}


Understanding of the flavor structure of the fermions has 
emerged as one of the most formidable problems in 
particle physics. While all fermions are expected 
to attain masses in the standard model through their Yukawa 
couplings with the standard Higgs doublet, it is not clear 
why the mass of the electron should be six order of magnitude 
smaller than that of the top quark. The extremely tiny 
neutrino masses pose a further challenge and demand an 
explanation.  
The ``seesaw'' mechanism \cite{seesaw} 
has been the most widely accepted method of 
explaining the smallness of the neutrino mass compared 
to that of the charge leptons. 
The seesaw 
mechanism was so named because the lightness of the 
standard neutrino 
is explained due to the 
heaviness of an additional 
particle beyond the standard model of particle physics. 
The new mass scale could be associated with a GUT scale, 
or in general with any intermediate mass scale. Being 
much heavier than the rest of the standard model particles, 
this additional field can be integrated out, giving a 
dimension five Majorana mass term for the neutrinos \cite{dim5}.
This mass term is inversely proportional to the mass of the 
heavy particle, and hence neutrinos become naturally light.  
There are three variants of the seesaw mechanism. 
These come from the fact that one can obtain the dimension 
five effective operator by integrating out either 
SU(2)$_L\times$ U(1)$_Y$  
(i) singlet fermions \cite{seesaw}, or (ii) triplet scalars 
\cite{type2}, 
or triplet fermions \cite{type3} (also \cite{type3ma}). 
The three variants are 
commonly known as Type I, Type II and Type III seesaw mechanism, 
respectively. 
While Type I and Type II scenarios have been 
extensively explored in the literature for a long time, 
focus has only recently 
shifted to the Type III seesaw mechanism, and a plethora of 
papers have appeared of late. 
While the possibility of gauged U(1) symmetry with 
fermion triplets was studied in \cite{type3maroy}, 
authors of \cite{type3lepto} studied for the first time 
predictions for leptogenesis within the framework of 
Type III seesaw. 
A hybrid Type I+III seesaw framework is shown to result 
within a SU(5) GUT model in \cite{type3su5,type3gmsb}, and within a 
left-right symmetric model with 
spontaneously broken parity in \cite{type3LR}. The effect of 
the additional fermions on the Higgs mass bounds was studied through 
renormalization group equations in \cite{type3rghiggsmass}, while  
the renormalization group evolution of the neutrino 
mass matrix within the Type III seesaw framework was 
performed in \cite{type3rg}. In \cite{type3radiative} the authors 
work with just one extra heavy fermion triplet and 
generate the addition light neutrino masses 
at the loop level. The phenomenology of the Type III seesaw in lepton 
flavor violating processes was studied in 
great depths in \cite{type3lowE,type3muegamma} and also recently in 
\cite{type3taulfv}.

The most crucial feature concerning the 
Type III seesaw is the following. 
Since the additional heavy fermions belong to the 
adjoint representation of SU(2), 
they have gauge interactions. This makes it easier to produce them 
in collider experiments. With the LHC all set to take data, it is 
pertinent to check the viability of testing the seesaw models 
at colliders. 
The implications of the Type III seesaw 
at LHC was first studied in \cite{type3lhc1} and \cite{type3lhc2}
in the context of a SU(5) GUT model. In the SU(5) model it is 
possible to naturally have the adjoint fermions in the 100 GeV to 1 TeV 
mass range, throwing up the possibility of observing them at LHC. 
The authors of these papers 
identified the dilepton channel with 4 jets as the signature 
of the triplet fermions. Subsequently, a lot of work has followed 
on testing Type III seesaw at LHC 
\cite{type3lhchambyestrumia,type3lhcaguila,type3lhc3}. 

In the usual Type III (and also Type I) version of the 
seesaw models with one Higgs doublet, 
the neutrino mass is given by 
\be
m_\nu = - v^2 Y_\Sigma^T\,\frac{1}{M}\,Y_\Sigma
\label{eq:one}
\ee
where, $v$ is the Higgs Vacuum Expectation Value (VEV), 
$M$ is the mass (matrix) of the adjoint fermions and 
$Y_\Sigma$ is the Yukawa coupling (matrix) of these fermions with the 
standard model lepton doublets and Higgs. To predict neutrino 
masses $\sim 0.1$ eV without fine tuning the Yukawas, one 
requires that $M\sim 10^{14}$ GeV. On the other hand, 
an essential requisite of producing 
the heavy fermion triplet signatures at the LHC, is that they 
should not be heavier 
than a few hundred GeV. One can immediately see that 
if $M\sim 300$ GeV, then $m_\nu \sim 0.1$ eV demands that the 
Yukawa coupling $Y_\Sigma\sim 10^{-6}$.  This in a way tentamounts 
to fine tuning of the Yukawas, and smothers out 
the very motivation for the seesaw mechanism -- which was to explain the 
smallness of the neutrino mass without unnaturally reducing the 
Yukawa couplings. 

In this paper, we propose a seesaw model 
with 300-800 GeV mass range triplet fermions, 
without any drastic reduction of the Yukawa 
couplings. We do that by introducing an additional Higgs doublet in 
our model. We impose a $Z_2$ symmetry which ensures that 
this extra Higgs doublet couples to only the exotic 
triplet fermions, while the standard Higgs couples to all 
other standard model particles \cite{gabrielnandi}. 
As a result the smallness of 
the neutrino masses can be explained from the the smallness of 
the VEV of the second Higgs doublet, while all standard model 
fermions get their masses from the VEV of the standard Higgs. 
Therefore, we use the presence of two 
different VEVs in our model 
to explain the smallness of the neutrino 
masses compared to all others, 
without resorting to unnatural 
suppression of the neutrino Yukawa couplings. We show that 
these large Yukawas result in extremely fast decay rates for 
the heavy fermions in our model and hence 
have observational consequences for the 
heavy fermion phenomenology at LHC. We show how this can be used 
to distinguish our two Higgs doublet Type III seesaw model from 
the usual one Higgs doublet models.\footnote{The largeness 
of the Yukawa couplings 
(along with the smallness of the heavy fermion 
masses), also brings in larger 
non-unitarity and larger lepton flavor violation in our 
model, compared to the earlier Type III seesaw models. However, these 
are still well below the sensitivity of the current and 
upcoming future experiments.}

The presence of two Higgs doublets in our model also enhances the 
richness of the phenomenology at LHC. 
We have in our model 
two neutral physical scalar and one neutral physical 
pseudoscalar and a pair of charged scalars. 
We will work out in detail our 
Higgs mass spectrum by imposing constraints coming from the 
neutrino masses. We will show that due to these constraints, 
our Higgs mixing angle is very small and the Higgs behave in a 
very peculiar way and have collider signatures which are 
very different from the usual two Higgs doublet models 
in the market
\cite{hunter,Atwood:2005bf,Antusch:2001vn,Matsuda:2001bg,Aoki:2009vf}.
We will study this crucial link between neutrino 
and Higgs physics in our model and its implications for LHC 
in detail. 

Another feature associated with neutrinos which has puzzled 
model builders is it unique mixing pattern. While all mixing 
angles are tiny in the quark sector, for the leptons we 
have observed two large and one small mixing angle. 
In its standard parametrization 
with mixing angles $\theta_{12}$, 
 $\theta_{23}$ and $\theta_{13}$ and phases $\delta$ (Dirac), 
$\alpha$ and $\beta$ (Majorana), the neutrino mixing matrix is given as
\be
 U_{PMNS} \!= \! \left(
 \begin{array}{ccc}
 c_{12} \, c_{13} & s_{12}\, c_{13} & s_{13}\, e^{-i \delta}\\
 -c_{23}\, s_{12}-s_{23}\, s_{13}\, c_{12}\, e^{i \delta} &
 c_{23}\, c_{12}-s_{23}\, s_{13}\, s_{12}\, 
e^{i \delta} & s_{23}\, c_{13}\\
 s_{23}\, s_{12}-\, c_{23}\, s_{13}\, c_{12}\, e^{i \delta} &
 -s_{23}\, c_{12}-c_{23}\, s_{13}\, s_{12}\, 
e^{i \delta} & c_{23}\, c_{13}
 \end{array}
 \right) 
\!\! \left(
\begin{array}{ccc}
1 & 0 & \cr
0 & e^{i \alpha} & 0 \cr
0 & 0 &  e^{i (\beta + \delta)}
 \end{array}
\!\!\! \right) 
.
\label{eq:upmns}
\ee
In this parametrization, 
the mixing angle $\theta_{23}$ is observed to be very close to 
$\pi/4$, while $\theta_{13}$ has so far been seen to be 
consistent with zero. 
This indicates that there should be some 
underlying symmetry which drives one mixing angle to be maximal and 
another to be zero. The most simple way of generating this 
is by imposing a \mt exchange symmetry on the low energy 
neutrino mass matrix \cite{mutau}. 

In this paper we will impose the 
$\mu$-$\tau$ symmetry on the Yukawa couplings and the 
heavy fermion mass matrices. This leads to $\mu$-$\tau$ symmetry 
in the light neutrino mass matrix and hence the correct 
predictions for the neutrino oscillation data. 
We discuss in detail the 
light as well as heavy neutrino mixing. We first provide 
general expressions for all mass eigenvalues and mixing 
matrices and then study the 
experimental consequences for our model. 
We show that due to the $\mu$-$\tau$ symmetry, the 
mixing matrices of the heavy fermions turn out to be 
non-trivial. In particular, they are also \mt symmetric 
and hence much deviated from unity, even though 
we start with a real and diagonal Majorana 
mass matrix for the heavy triplets. This is a new 
result and 
we will show that this affects the flavor structure of the 
heavy fermion decays at colliders, which 
can be used to test \mt symmetry in neutrinos at LHC. 
We study in detail the collider 
phenomenology of this $\mu$-$\tau$ symmetric 
model with three heavy fermion 
SU(2) triplets and two Higgs SU(2) doublets and 
give predictions for LHC. 


The paper is organized as follows. In section 2, we present the 
lepton Yukawa part of the model within a general framework 
and give expressions for the 
masses and mixings of the 
charged and neutral components of both light as well as heavy 
leptons. In section 3, we present our \mt symmetric model and 
give specific forms for the mass and mixing parameters. We show 
that the mixing for heavy fermions is non-trivial and 
\mt symmetric. In section 4, we study the cross-section for 
the heavy fermion production and LHC, as a function of the 
fermion mass. In section 5, we study the decay rates of these 
heavy fermions into Higgs and gauge bosons. We compare and 
contrast our model against the usual Type III seesaw models with 
only one Higgs. We also show the consequences of non-trivial 
mixing of the heavy fermions on the flavor structure of their 
decays. In section 6, we discuss the decay rates and branching 
ratios of the Higgs decays. We probe issues on Higgs decays, 
which are specific and unique to our model. 
Section 7 is devoted to the discussion of displaced 
decay vertices as a result of the very long living $h^0$ in 
our model. In section 8, we list all possible final state 
particles and their corresponding collider signature channels 
which could be used to test our model. We calculate the effective 
cross-sections for all channels at LHC. 
We highlight some of the 
channels with very large effective 
cross-sections and discuss 
the standard model backgrounds. 
Finally, in section 9 we present our conclusions. 
Discussion of the scalar potential, the Higgs mass spectrum 
and the constraints from neutrino data on the Higgs sector is 
discussed in detail in Appendix A. The lepton-Higgs coupling 
vertices are listed in Appendix B.1, the lepton-gauge 
coupling vertices are listed in Appendix B.2, and the 
quark-Higgs coupling vertices are listed in Appendix B.3.

\section{Yukawa Couplings and Lepton Masses and Mixing}

We add three extra SU(2) triplet fermions 
to our standard model particle content. 
These fermions belong to the adjoint 
representation of SU(2) and are assigned hypercharge $Y=0$. 
This makes each of them self conjugate. 
We will denote their 
Cartesian components  
as\footnote{Throughout this 
paper we denote particles in their weak eigenbasis by primed  
and mass eigenbasis by unprimed notation.}
\be
\Psi'_i = 
\pmatrix{
{\s'}_i^1 \cr
{\s'}_i^2 \cr
{\s'}_i^3 
},
\ee
where $i=1,2,3$ and $\Psi'_i = {\Psi'_i}^C$. In the compact 
$2\times 2$ notation they will be represented in our 
convention as
\be
\Sigma'_i =\frac{1}{\sqrt{2}} \sum_j {\s'}^j_i \cdot \sigma_j 
,
\ee
where $\sigma_j$ are the Pauli matrices.  
The right-handed component of this multiplet in the $2\times 2$ notation 
is then given by
\be
{\Sigma'_R}_i = 
\pmatrix{
 {{\Sigma'}^0_R}_i/\sqrt{2} & {{\s'}^+_R}_i\cr
{{\Sigma'}^-_R}_i & -{{\s'}^0_R}_i/\sqrt{2}\cr
},
\label{eq:sigmadef}
\ee
where 
\be
{\s'^{\pm}_R}_i = \frac{{{\Sigma'}^1_R}_i 
\mp i{{\Sigma'}^2_R}_i}{\sqrt{2}}
~~{\rm and}~~
{{\Sigma'}^0_R}_i = {{\Sigma'}^3_R}_i
\ee
are the components 
of the triplet in the charge eigenbasis. 
The corresponding charge-conjugated multiplet will then be
\be
{{\Sigma'}_R}_i^C = C \overline{{\Sigma'}_R}^T=
\pmatrix{
{{{\Sigma'}^0_R}_i}^C/\sqrt{2} & {{{\s'}^+_R}_i}^C\cr
{{{\Sigma'}^-_R}_i}^C & -{{{\s'}^0_R}_i}^C/\sqrt{2}\cr
}.
\ee
The object which transforms as the left-handed component of the 
$\Sigma$ multiplet can then be written as
\be
{\tilde{\Sigma'}_{R_i}}^C = i\sigma_2 \,{{\Sigma'}_R}_i^C \,i\sigma_2 = 
\pmatrix{
{{{\Sigma'}^0_R}_i}^C/\sqrt{2} & {{{\s'}^-_R}_i}^C\cr
{{{\Sigma'}^+_R}_i}^C & -{{{\s'}^0_R}_i}^C/\sqrt{2}\cr
},
\ee
such that ${\Sigma'}_i = {{\Sigma'}_{R_i}} + {{\tilde{\Sigma'}_{R_i}}^C}$. 

\vglue 0.5cm
As discussed in the introduction, we include 
in our model a new SU(2) scalar doublet,  
$\Phi_2$, in addition to the usual standard model doublet $\Phi_1$. 
This new doublet couples only to the triplet fermions 
introduced above. The triplet fermions on the other hand are 
restricted to couple with only the new $\Phi_2$ 
doublet and not with $\Phi_1$. 
This can be 
ensured very easily by giving $Z_2$ charge of $-1$ 
to the triplet fermions $\Sigma'_i$ and the scalar doublet 
$\Phi_2$, and $Z_2$ charge $+1$ to all standard model 
particles\footnote{We will break this $Z_2$ symmetry 
mildly in the scalar potential. We 
discuss the phenomenological consequences 
of this $Z_2$ symmetry and its breaking when we introduce the 
scalar potential and present the Higgs mass spectrum 
in Appendix A.}. 
The part of the Lagrangian responsible for the 
lepton masses can then be written as
\be
-{\cal L}_Y = \left[Y_{l_{ij}}\overline l'_{R_i} \Phi_1^\dagger L'_j 
+ Y_{\Sigma_{ij}}
{\tilde{\Phi}_2}^\dagger \overline \Sigma'_{R_i} L'_j +h.c. \right]
+ \frac{1}{2} M_{ij} \,{\rm Tr}\left[\overline {\Sigma'}_{R_i}  
{\tilde{\Sigma'}_{R_j}^C} + h.c. \right]
,
\label{eq:yukawa}
\ee
where $L'$ and $l'_R$ are the usual left-handed lepton doublet and 
right-handed charged leptons respectively, $Y_l$ and $Y_\Sigma$ 
are the $3\times 3$ Yukawa coupling matrices, and 
$\tilde\Phi_2 = i\sigma_2\Phi_2^*$.
Once the Higgs doublets $\Phi_1$ and $\Phi_2$ take Vacuum Expectation 
Value (VEV) 
\be
\langle \Phi_1 \rangle =
\pmatrix
{0\cr v},~~~~
 \langle \Phi_2 \rangle =
\pmatrix
{0\cr v'},
\ee
we generate the following neutrino mass matrix
\be
{\cal L}_{\nu} = \frac{1}{2}
\pmatrix {\overline{{\nu'}_{L_i}^C} & \overline{{{\Sigma'}_{R_i}^0}}}
\pmatrix {0 & \frac{v'}{\sqrt{2}}Y^T_{\Sigma_{ij}}\cr
\frac{v'}{\sqrt{2}}Y_{\Sigma_{ij}} & M_{ij}}
\pmatrix{\nu'_{L_j} \cr {{\Sigma'}_{R_j}^0}^{\!\!\!\!C}} + h.c. ,
\label{eq:numass}
\ee
and the following charged lepton mass matrix
\be
{\cal L}_{l} &=& 
\pmatrix {\overline{l'_{R_i}} & \overline{{\Sigma'}_{R_i}^-}}
\pmatrix {v Y_{l_{ij}} & 0\cr
v' Y_{\Sigma_{ij}} & M_{{ij}}}
\pmatrix{l'_{L_j} \cr {{\Sigma'}_{R_j}^+}^{\!\!\!C}} + h.c.,
\\
&=& 
\pmatrix {\overline{l'_{R_i}} & \overline{{\Sigma'}_{R_i}^-}}
M_l
\pmatrix{l'_{L_j} \cr {{\Sigma'}_{R_j}^+}^{\!\!\!C}} + h.c.,
\label{eq:chargedleptonmass}
\ee
Note that due to the 
imposed $Z_2$ symmetry 
neutrino masses depend {\it only} on  the 
new Higgs VEV $v'$ while in the charged lepton mass matrix  
both the VEV's enter. The value of $v'$ is determined by 
the scale of the neutrino masses and is independent of the 
mass scale of all other fermions. Therefore, the neutrino 
masses can be naturally light, without having to fine 
tune the Yukawas $Y_\Sigma$ to unnaturally small values. 

The $6\times 6$ neutrino matrix (\ref{eq:numass}) 
can be diagonalized to yield 3 light and 3 heavy Majorana 
neutrinos. The $6\times 6$ { unitary} matrix $U$ which 
accomplishes this is defined as
\be
U^T \pmatrix {0 & m_D^T \cr
m_D & M} U = \pmatrix{D_m & 0 \cr 0 & D_M},
~~{\rm and}~~
\pmatrix{\nu'_{L_j} \cr {{\Sigma'}_{R_j}^0}^{\!\!\!C}} 
= U\, \pmatrix{\nu_{L_j} \cr {{\Sigma}_{R_j}^0}^{\!\!\!C}} 
,
\ee
where $m_D = v'Y_\Sigma/\sqrt{2}$, and 
\be
D_m = \pmatrix{m_1 & 0 & 0 \cr
                 0 & m_2 & 0 \cr
                 0 & 0 & m_3}
,~~~~
D_M = \pmatrix{M_{\Sigma_1} & 0 & 0 \cr
                 0 & M_{\Sigma_2} & 0 \cr
                 0 & 0 & M_{\Sigma_3}}
.
\ee
Here $m_i$ and $M_{\Sigma_i}$ ($i=1,2,3$) are the low and high 
energy mass eigenvalues of the Majorana neutrinos respectively.
We reiterate that the primed and unprimed notations represent the 
weak and mass eigenbases respectively.  
The mixing matrix $U$ can be parameterized as a product of 
two matrices
\be
U = W_\nu \, U_\nu
\ee
where $W_\nu$ is the matrix which brings the  $6\times 6$ 
neutrino matrix given by  Eq. (\ref{eq:numass}) in its block 
diagonal form as
\be
W_\nu^T \pmatrix {0 & m_D^T \cr
m_D & M} W_\nu &=& \pmatrix{\tilde{m} & 0 \cr 0 & 
\tilde{M}}
,
\ee
while $U_\nu$ diagonalizes $\tilde{m}_\nu$ and $\tilde{M}_\Sigma$ as
\be
U_\nu^T \pmatrix {\tilde{m} & 0 \cr 0 & 
\tilde{M}} U_\nu &=& \pmatrix{D_m & 0 \cr 0 & 
D_M}
.
\ee
The above parameterization therefore enables us to analytically 
estimate the mass eigenvalues and the mixing 
matrix $U$ in terms of $W_\nu$ and $U_\nu$ by  
a two step process, by first calculating $W_\nu$ and then 
$U_\nu$. 
Since the unitary matrix $U$ has $6^2=36$ free parameters and 
the matrix $U_\nu$ has $2\times 3^2=18$ parameter, the matrix 
$W_\nu$ should have $36-18=18$ free parameters. This matrix 
therefore can be parameterized as \cite{grimuslavoura}
\be
W_{\nu} = 
\pmatrix{\sqrt{1-BB^\dagger} & B \cr
-B^\dagger & \sqrt{1-B^\dagger B}}
,
\ee
where $B= B_1 +B_2 +B_3 +...$ 
and $B_j \sim (1/m_{\Sigma})^j$, 
where $m_{\Sigma}$ is the scale of the heavy 
Majorana fermion mass. Using an expansion in $1/m_{\Sigma}$ and 
keeping only terms second order or lower in $1/m_{\Sigma}$, we 
get
\be
W_\nu \simeq
\pmatrix{1-\frac{1}{2}m_D^{\dagger}({M^{-1}})^*{M^{-1}}m_D 
& m_D^\dagger ({M^{-1}})^* \cr
-{M^{-1}}m_D  & 1-\frac{1}{2}{M^{-1}}m_D m_D^\dagger 
({M^{-1}})^*}
.
\label{eq:wnu}
\ee
The light and heavy neutrino mass matrices obtained at 
this block diagonal stage are given by
(only second 
order terms in  $1/m_{\Sigma}$ are kept)
\be
\tilde{m}_\nu &=& -m_D^T {M}^{-1} m_D ,
\label{eq:lownu}
\\
\tilde{M}  &=& M + \frac{1}{2}\left(
m_Dm_D^\dagger({M^{-1}})^* + ({M^{-1}})^*m_D^*m_D^T\right)
.
\label{eq:msigma0}
\ee
Note that Eq. (\ref{eq:lownu}) is the standard seesaw formula 
for the light neutrino mass matrix, while 
Eq. (\ref{eq:msigma0}) gives the heavy neutrino mass matrix. 
These can be diagonalized by two $3\times 3$ unitary matrices
$U_0$ and $U_\Sigma$, respectively. This yields 
\be
U_\nu = \pmatrix {U_{0} & 0 \cr
0 & U_\Sigma}
,
\ee

For the charged leptons we follow an identical method for 
determining the mass eigenvalues and the mixing matrices. 
However, since the charged lepton mass matrix $M_l$ 
given by Eq. (\ref{eq:chargedleptonmass}) is a Dirac mass matrix, 
one has to diagonalize it using a bi-unitary transformation
\be
T^\dagger \pmatrix {m_l & 0 \cr
\sqrt{2} m_D & M} S =\pmatrix{D_l & 0 \cr 0 & D_H}
=M_{l_d}
,
\label{eq:biunitary}
\ee
where $m_l=v Y_{l}$, while $D_l$ 
and $D_H$ are diagonal matrices containing the light and 
heavy charged lepton mass eigenvalues. 
With the above definition for the diagonalization, the 
right-handed and left-handed weak and mass 
eigenbases for the charged leptons are related respectively as,  
\be
\pmatrix{l'_{L} \cr {{\Sigma'}_{R}^+}^{\!C}} = 
S\pmatrix{l_{L} \cr {{\Sigma}_{R}^+}^{\!C}}
,~~{\rm and}~~
\pmatrix{l'_{R} \cr {{\Sigma'}_{R}^+}^{\!C}} = 
T\pmatrix{l_{R} \cr {{\Sigma}_{R}^+}}
.
\ee
Instead of using Eq. (\ref{eq:biunitary}) for the diagonalization, 
we will work with the matrices
\be
M_l^{\dagger}M_l=S\,M_{l_d}^{\dagger} M_{l_d} \, S^{\dagger}
,~~{\rm and}~~
M_l M_l^\dagger = T\,M_{l_d} M_{l_d}^{\dagger}\,T^{\dagger}
,
\ee
to obtain $S$ and $T$ respectively. As for the neutrinos, we 
parameterize
\be
S=W_LU_L,~~{\rm and}~~T=W_RU_R
,
\ee
where $W_L$ and $W_R$ are the unitary matrices which bring 
$M_l^{\dagger}M_l$ and $M_l M_l^\dagger$ to their block 
diagonal forms, respectively,
\be
W_L^\dagger\, M_l^{\dagger}M_l\, W_L = 
\pmatrix{\tilde{m_l}^\dagger \tilde{m_l} & 0 \cr 0 & 
\tilde{M_H}^\dagger \tilde{M_H}}
,~~{\rm and}~~
W_R^\dagger\, M_l M_l^{\dagger} \,W_R = 
\pmatrix{\tilde{m_l} \tilde{m_l}^\dagger & 0 \cr 0 & 
\tilde{M_H} \tilde{M_H}^\dagger }
.
\ee
Using arguments similar to that used for the neutrino 
sector, and keeping terms up to second order in $1/m_{\Sigma}$, 
we obtain
\be
W_L &=& \pmatrix{
1-m_D^\dagger({M^{-1}})^*M^{-1} m_D &
\sqrt{2} m_D^\dagger({M^{-1}})^* \cr
-\sqrt{2} M^{-1} m_D & 
1-M^{-1} m_Dm_D^\dagger({M^{-1}})^* 
}
,
\label{eq:wl}
\ee
\be
W_R &=& \pmatrix{
1 & \sqrt{2} m_l m_D^\dagger ({M^{-1}})^* M^{-1} \cr
-\sqrt{2}({M^{-1}})^* M^{-1} m_D m_l^\dagger & 1}
,
\label{eq:wr}
\ee
The square of the mass matrices for the 
light and heavy charged leptons in the flavor basis obtained 
after block diagonalization by $W_R$ and $W_L$ are 
given by
\be
\tilde{m}_l \tilde{m}^\dagger_l 
&=& m_l m_l^\dagger - 2 m_l m_D^\dagger ({M^{-1}})^*
{M^{-1}} m_D m_l^\dagger ,
\label{eq:leptonmassesr}
\\ 
\tilde{M}_H \tilde{M}^\dagger_H &=& M M^\dagger + 2 m_Dm_D^\dagger + 
 ({M^{-1}})^* M^{-1} m_D m_l^\dagger m_l m_D^\dagger
\nonumber\\
&+&  m_D m_l^\dagger m_l m_D^\dagger ({M^{-1}})^* M^{-1}
,
\label{eq:msigmapm}
\ee
and
\be
\tilde{m}^\dagger_l \tilde{m}_l 
&=& m^\dagger_l m_l -[m_D^{\dagger}{M^*}^{-1}M^{-1}m_Dm_l^{\dagger}m_l
+ \rm{h.c}]    
\label{eq:leptonmassesl}
\\ 
\tilde{M}^\dagger_H \tilde{M}_H &=& 
M^\dagger M +  M^{-1} m_D m_D^\dagger 
M
+ M^\dagger m_D m_D^\dagger ({M^{-1}})^*+  M^{-1} (m_D m_D^\dagger)^2 
({M^{-1}})^*\nonumber\\
+&&\!\!\!\! \!\! \!\!\!\! \!\!\! [ M^{-1} ({M^{-1}})^* M^{-1} 
m_D (m^\dagger_l m_l) m_D^\dagger M 
\!\!-\!\!\frac{1}{2} M^{-1}m_Dm_D^{\dagger}{M^*}^{-1}M^{-1}
m_Dm_D^{\dagger}M+\rm{h.c}]
\label{eq:msigmapm2}
\ee
One can explicitly check that the masses of the 
heavy charged leptons obtained from 
Eqs. (\ref{eq:msigmapm}) and (\ref{eq:msigmapm2}) 
are approximately the same as that obtained for the 
neutral heavy fermion using Eq. (\ref{eq:msigma0}). 
Indeed a comparison of these equations show that 
at tree level, the difference between 
the neutral and charged heavy fermions are of the order of the 
neutrino mass and can be hence neglected.   
One-loop effects bring a small 
splitting between the masses of the heavy and neutral 
fermions of $\approx 166 $ MeV. 
This allows the decay channel $\Sigma^{\pm} = \Sigma^0 + \pi^\pm$
at colliders, 
as discussed in detail in  \cite{type3lhc2,type3lhchambyestrumia}. 
In this paper we will 
neglect this tiny difference and assume that the masses of all 
heavy fermions are the same. 

\vglue 0.3cm
\noindent
The matrices $\tilde{m}_l^\dagger \tilde{m}_l$ and 
$\tilde{M}_H^\dagger \tilde{M}_H$ are diagonalized by $U_l$ and 
$U_h^L$ giving, 
\be
U_L = \pmatrix{U_l & 0 \cr 0 & U_h^L}
.
\ee
Similarly the $\tilde{m}_l \tilde{m}_l^\dagger $ and 
$\tilde{M}_H \tilde{M}_H^\dagger$ matrices are diagonalized by 
$U_r$ and 
$U_h^R$ and hence give, 
\be
U_R = \pmatrix{U_r & 0 \cr 0 & U_h^R}
.
\ee
Finally, the low energy observed neutrino mass matrix is given by
\be
U_{PMNS} = U_l^\dagger U_0
.
\ee
Note that both $U_l$ and $U_0$ are unitary matrices and hence 
$U_{PMNS}$ is unitary.


\section{A $\mu$-$\tau$ Symmetric Model}
\begin{figure}
\includegraphics[width=0.24\textwidth,height=5.1cm,angle=270]{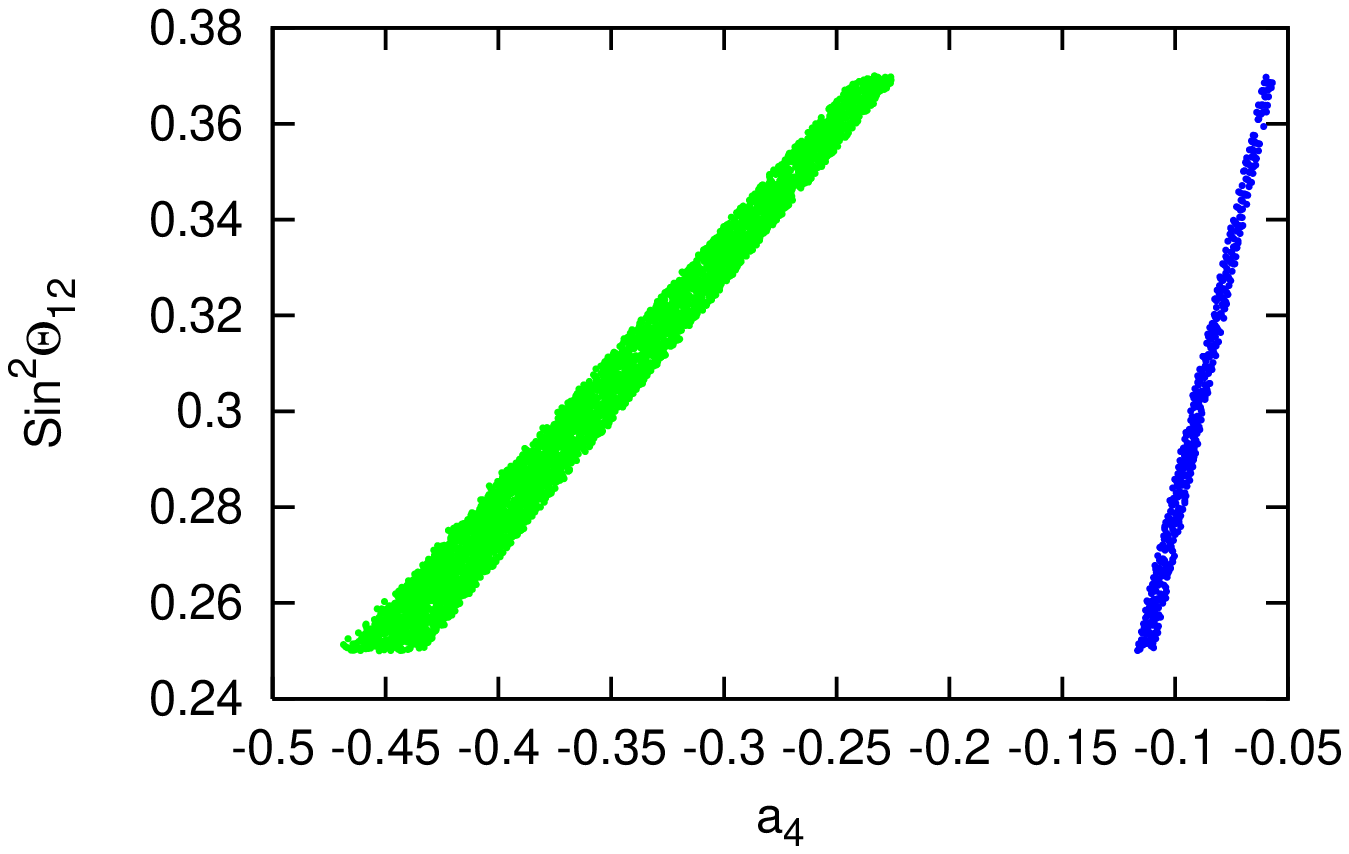}
\includegraphics[width=0.24\textwidth,height=5.1cm,angle=270]{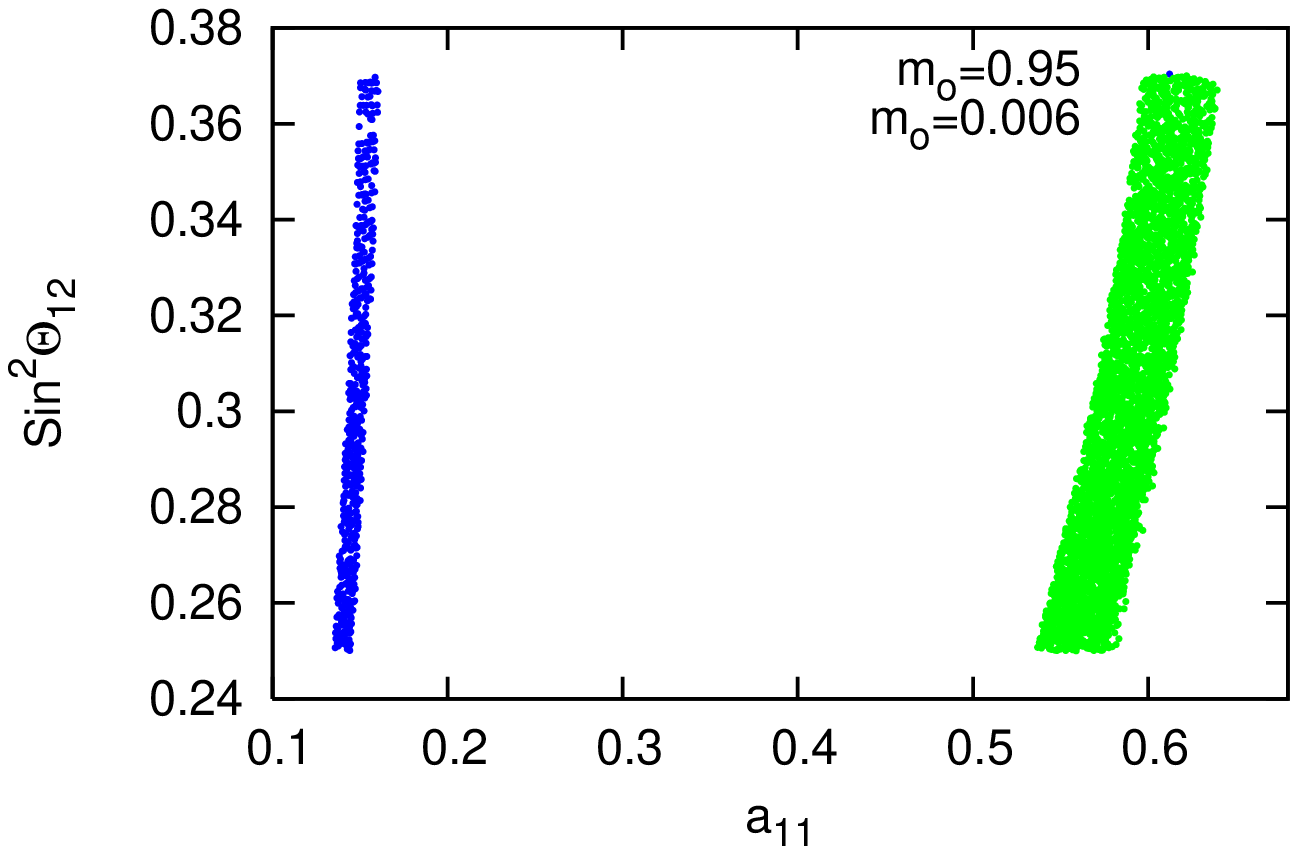}
\includegraphics[width=0.24\textwidth,height=5.1cm,angle=270]{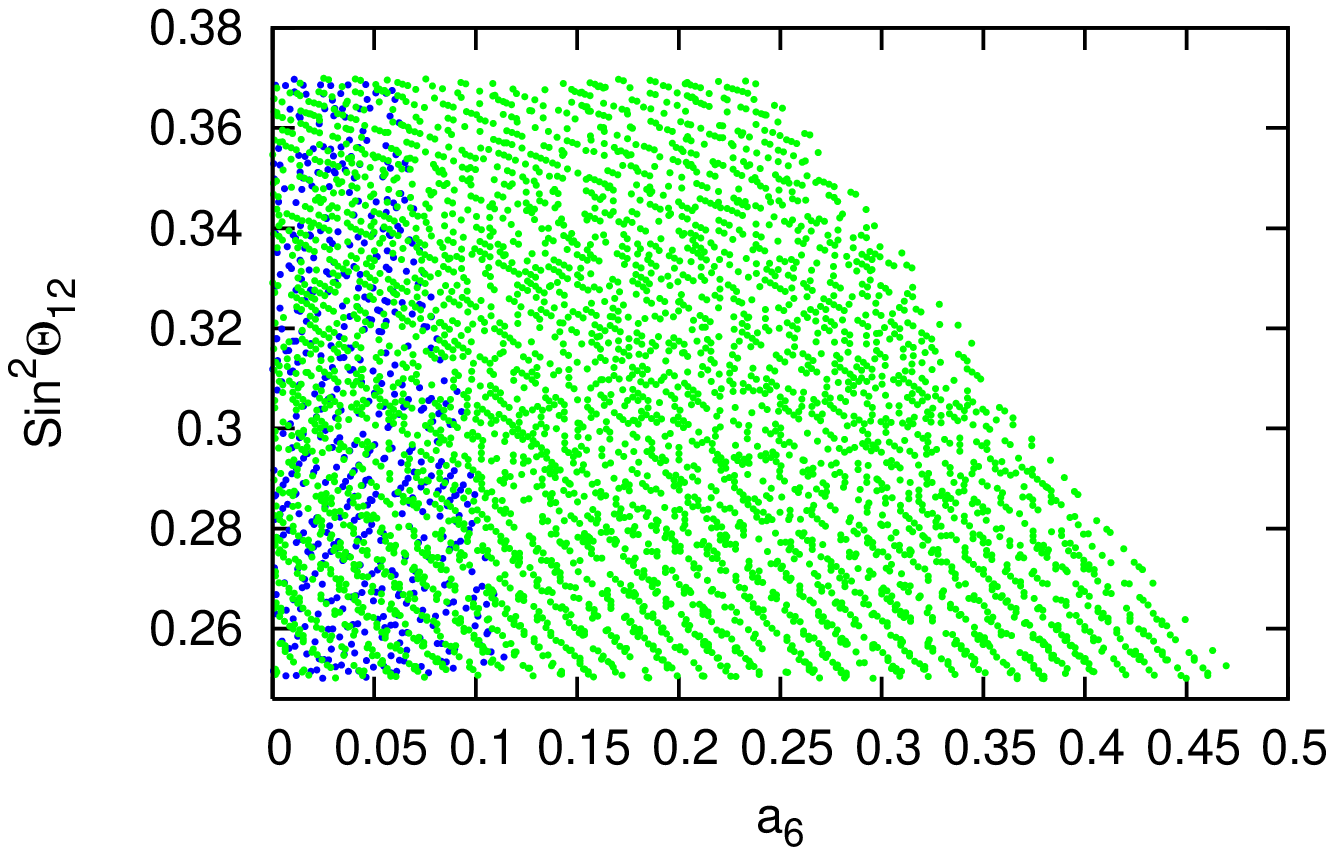}
\vglue 0.1cm
\includegraphics[width=0.24\textwidth,angle=270]{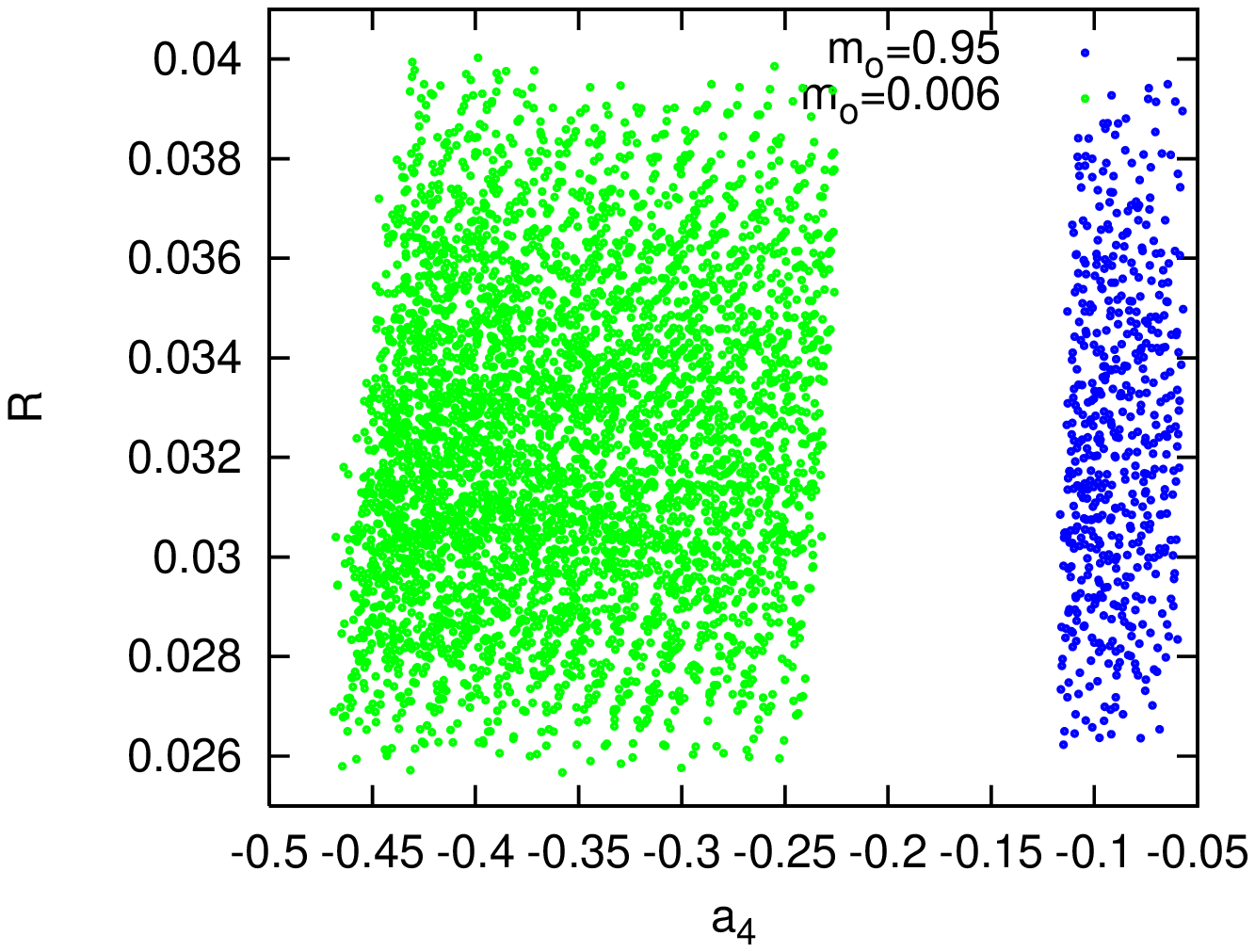}
\includegraphics[width=0.24\textwidth,angle=270]{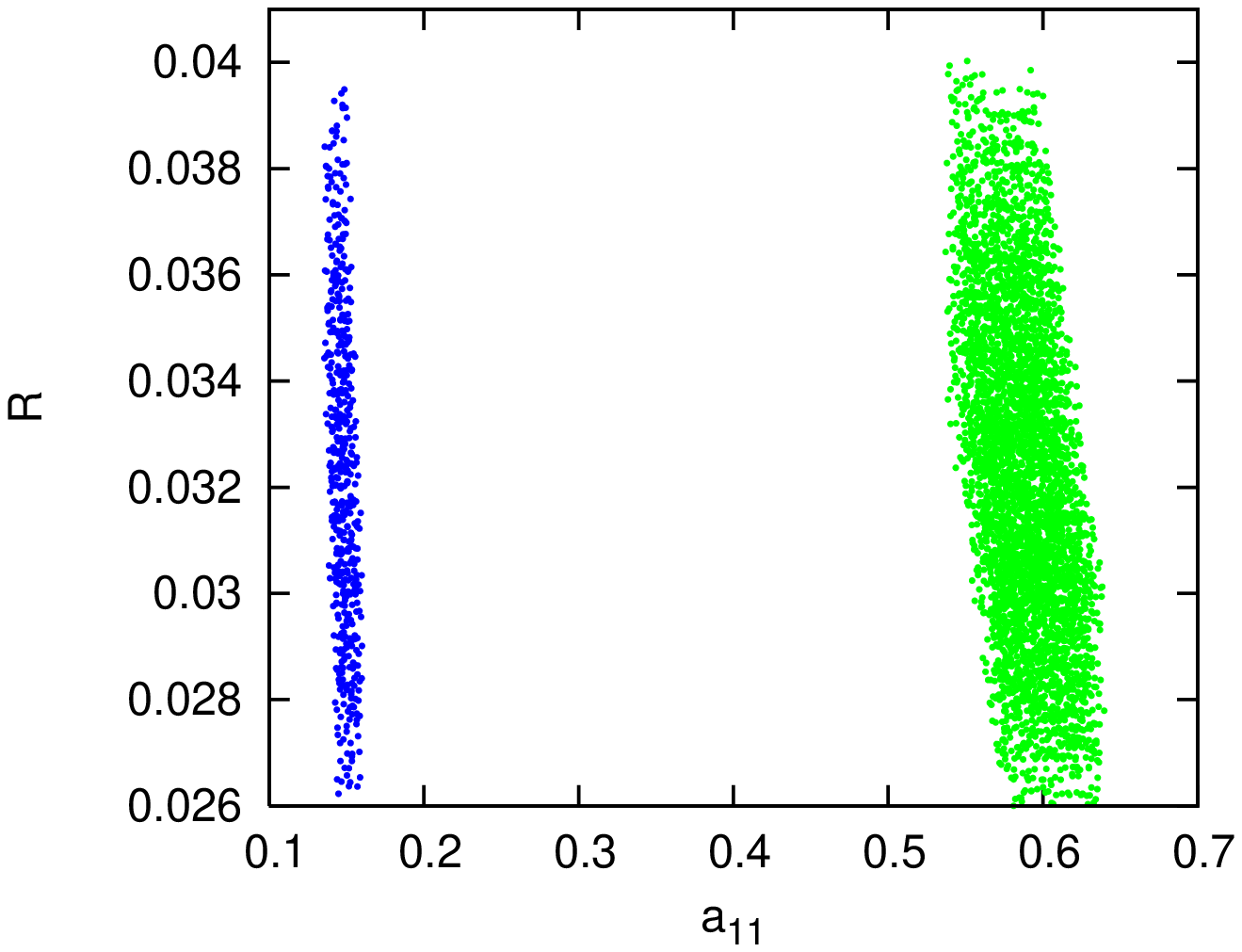}
\includegraphics[width=0.24\textwidth,angle=270]{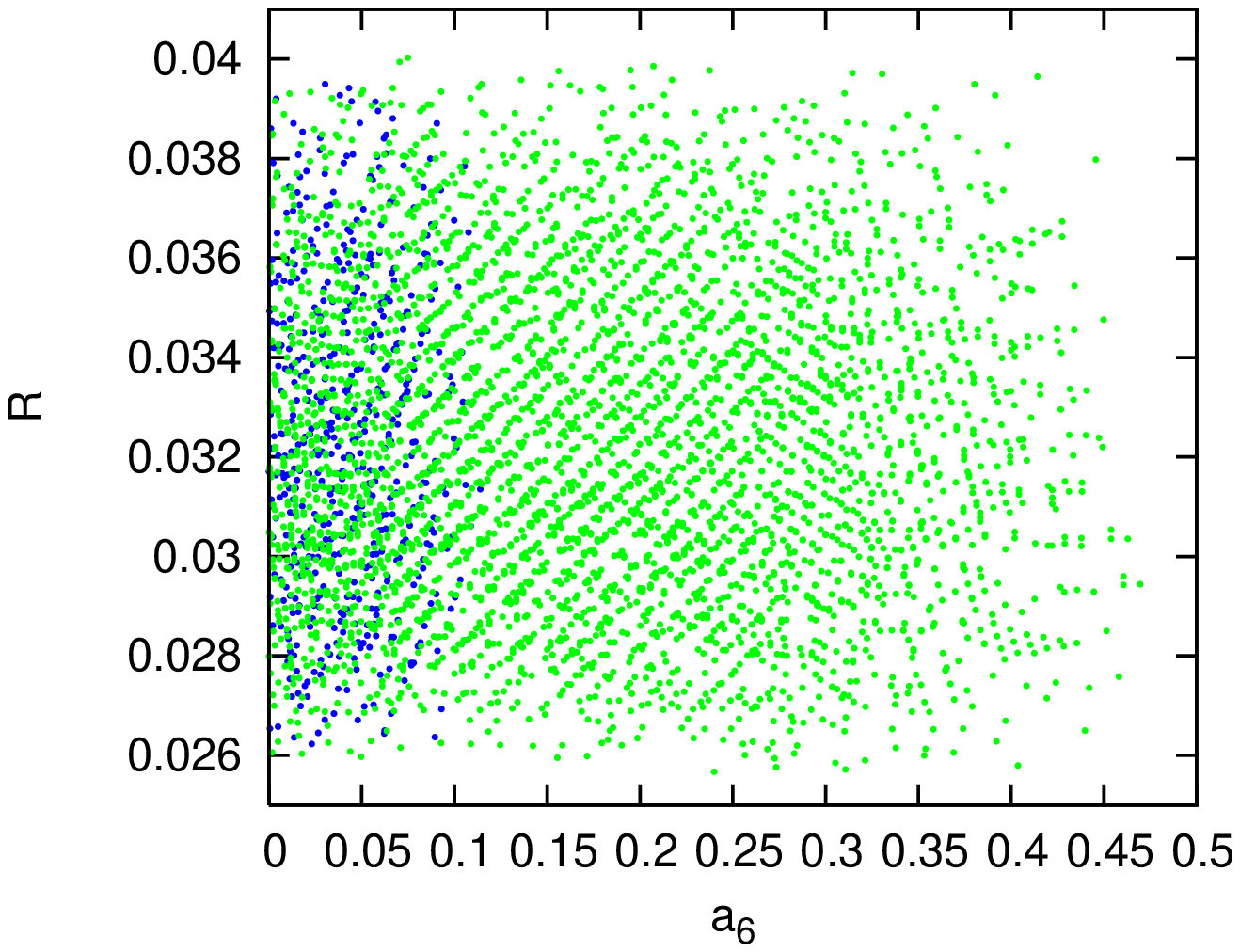}
\caption{\label{fig:ssqth12}
Scatter plots showing variation of $\sss$ (upper panels) 
and $R=\ms/|\ma|$ (lower panels) 
as a function of $a_4$, 
$a_{11}$ and $a_6$. 
All Yukawa couplings 
apart from the one plotted on the x-axis, are allowed to vary 
freely. 
Only points which predict oscillation parameters 
within their current 
$3\sigma$ values are shown. Blue points are 
for $m_0=v'^2/(2M_1) = 0.95$ eV 
while the green points are for $m_0=0.006$ eV. 
}
\end{figure}

\begin{figure}
\includegraphics[width=0.23\textwidth,angle=270]{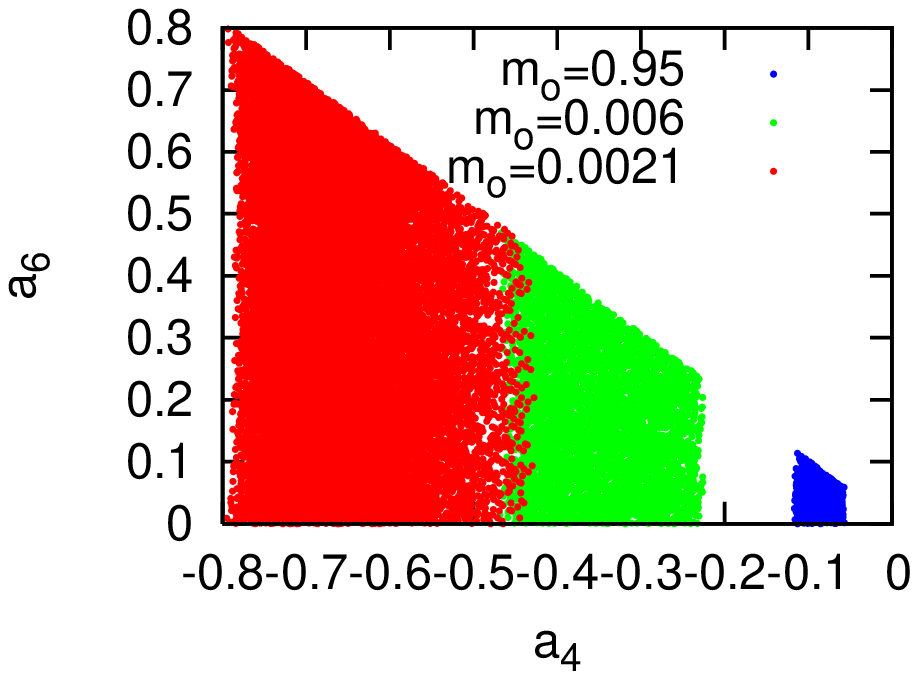}
\includegraphics[width=0.23\textwidth,angle=270]{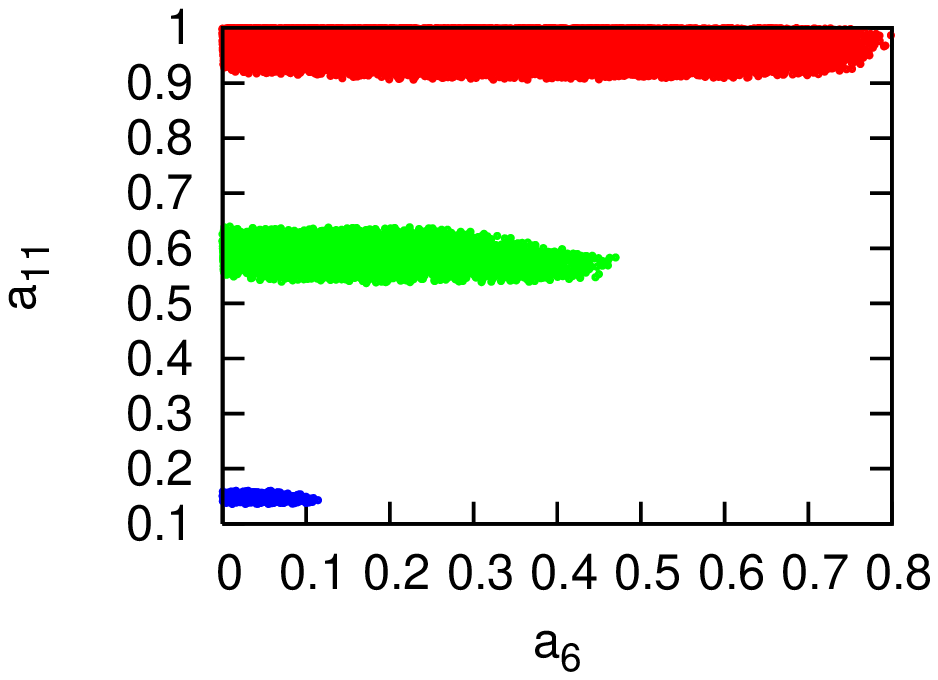}
\includegraphics[width=0.23\textwidth,angle=270]{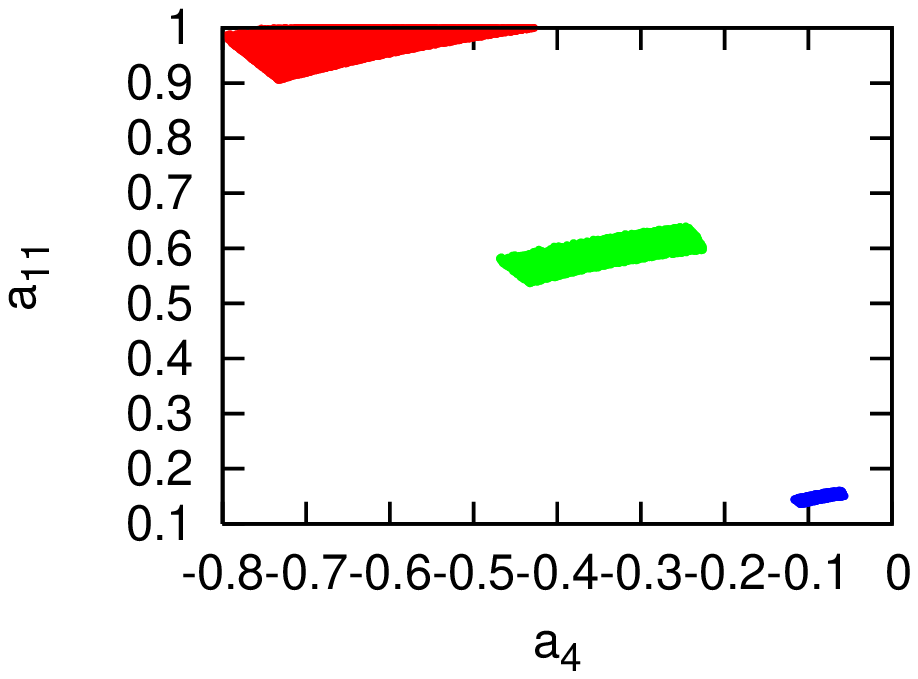}
\includegraphics[width=0.23\textwidth,angle=270]{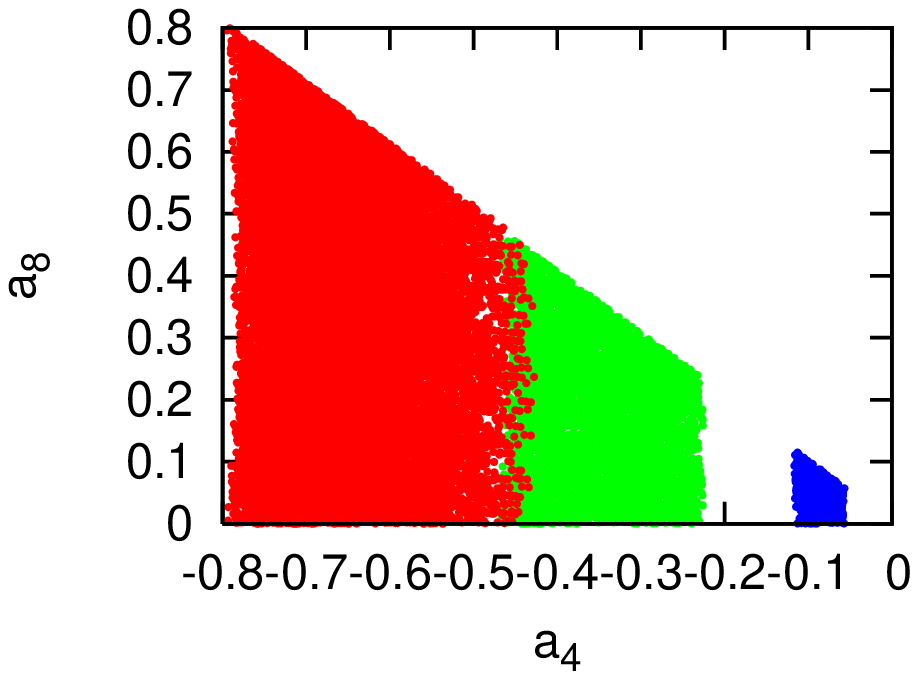}
\includegraphics[width=0.23\textwidth,angle=270]{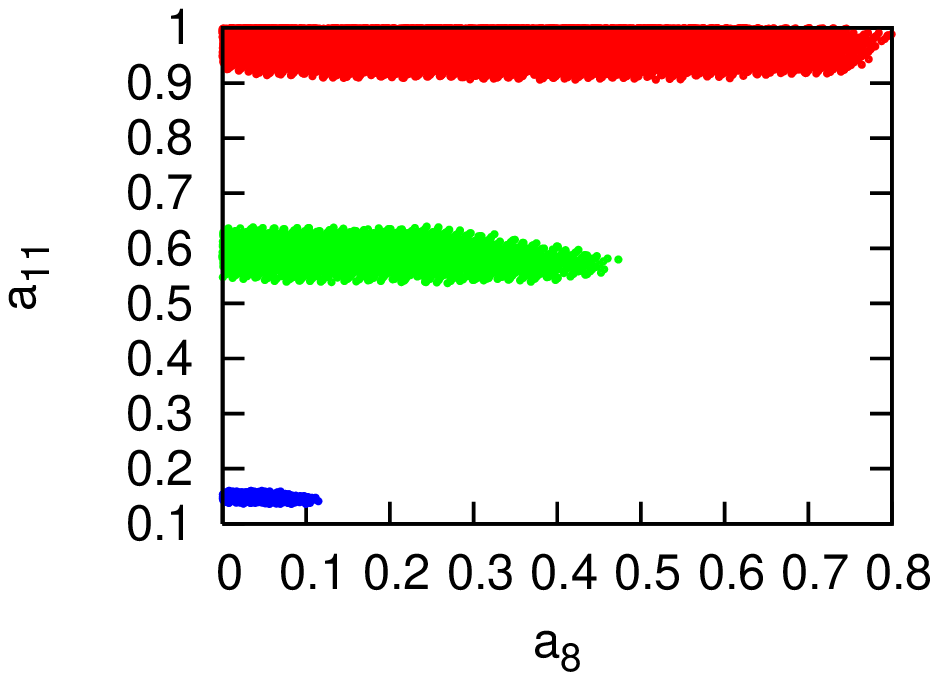}
\includegraphics[width=0.23\textwidth,angle=270]{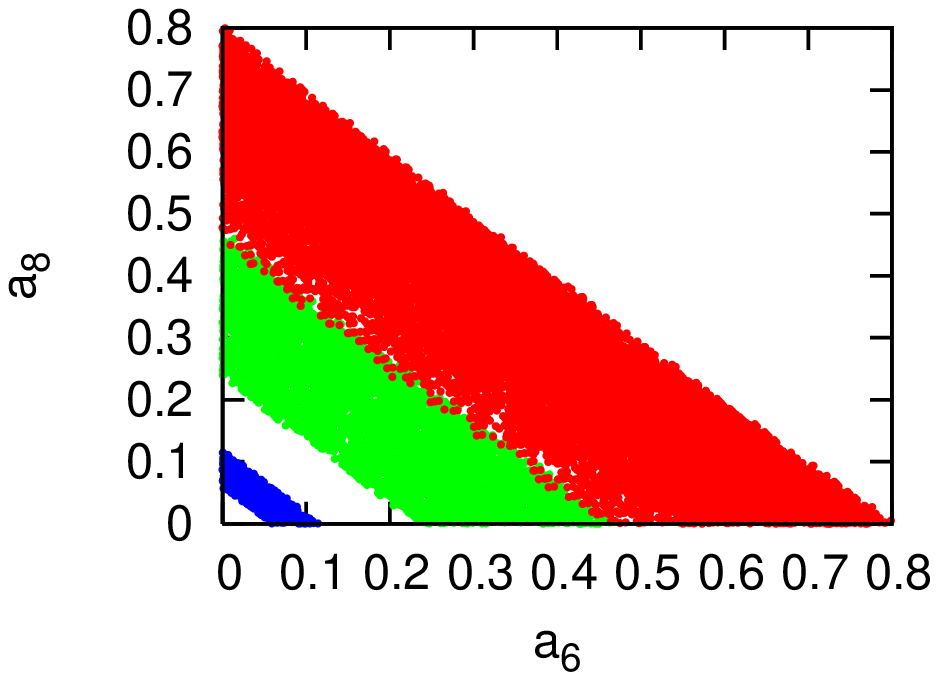}
\caption{\label{fig:param}
Scatter plot showing the 
values of the Yukawa couplings which give all oscillation 
parameters within their current $3\sigma$ allowed ranges. 
Allowed points are shown 
for $m_0=0.96$ eV (blue), 0.006 eV (green) and 0.0021 eV (red). 
All Yukawa couplings apart from the ones plotted in the x-axis 
and y-axis are allowed to vary freely, in each panel. 
}
\end{figure}

\begin{figure}
\includegraphics[width=0.35\textwidth,angle=270]{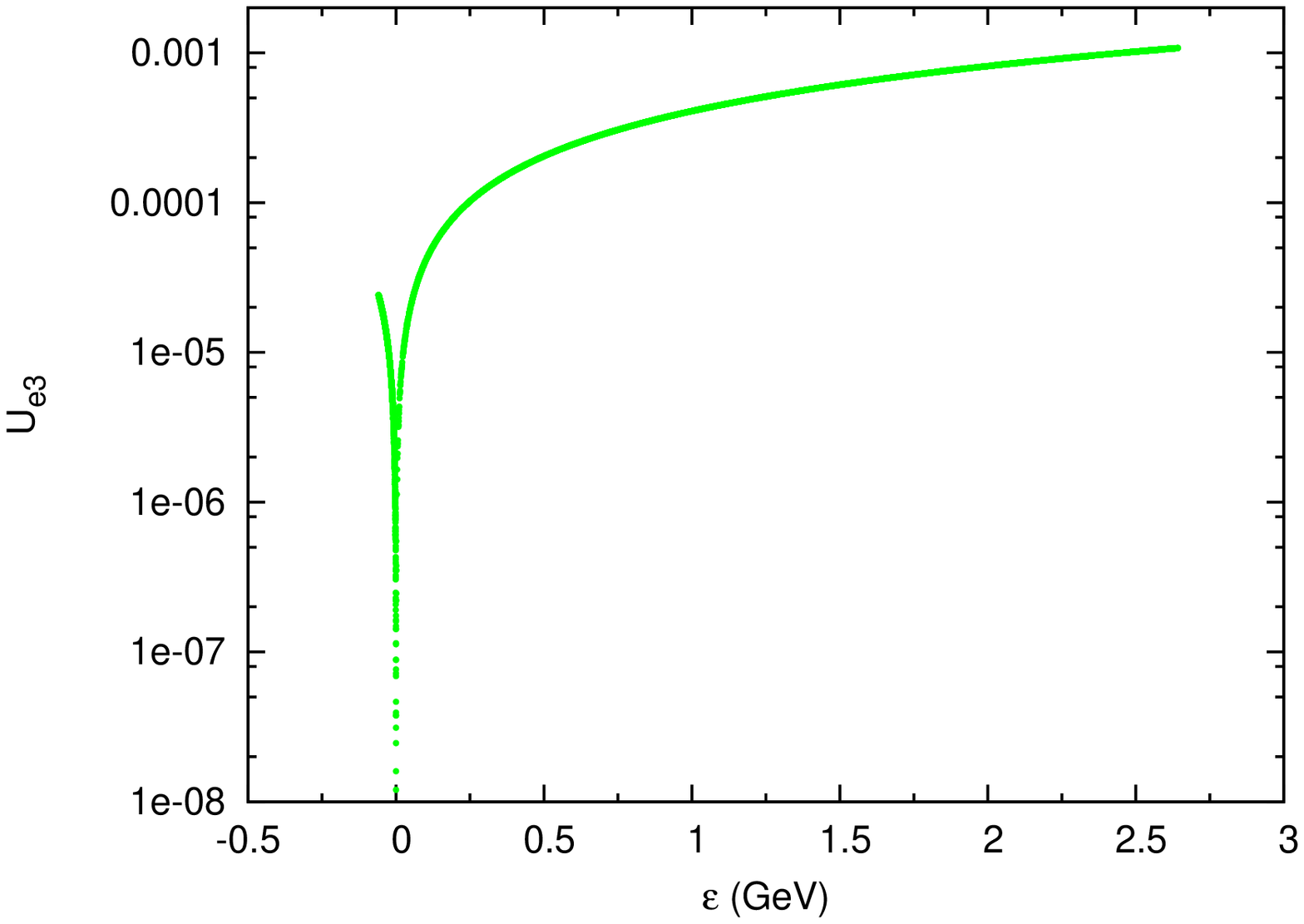}
\includegraphics[width=0.35\textwidth,angle=270]{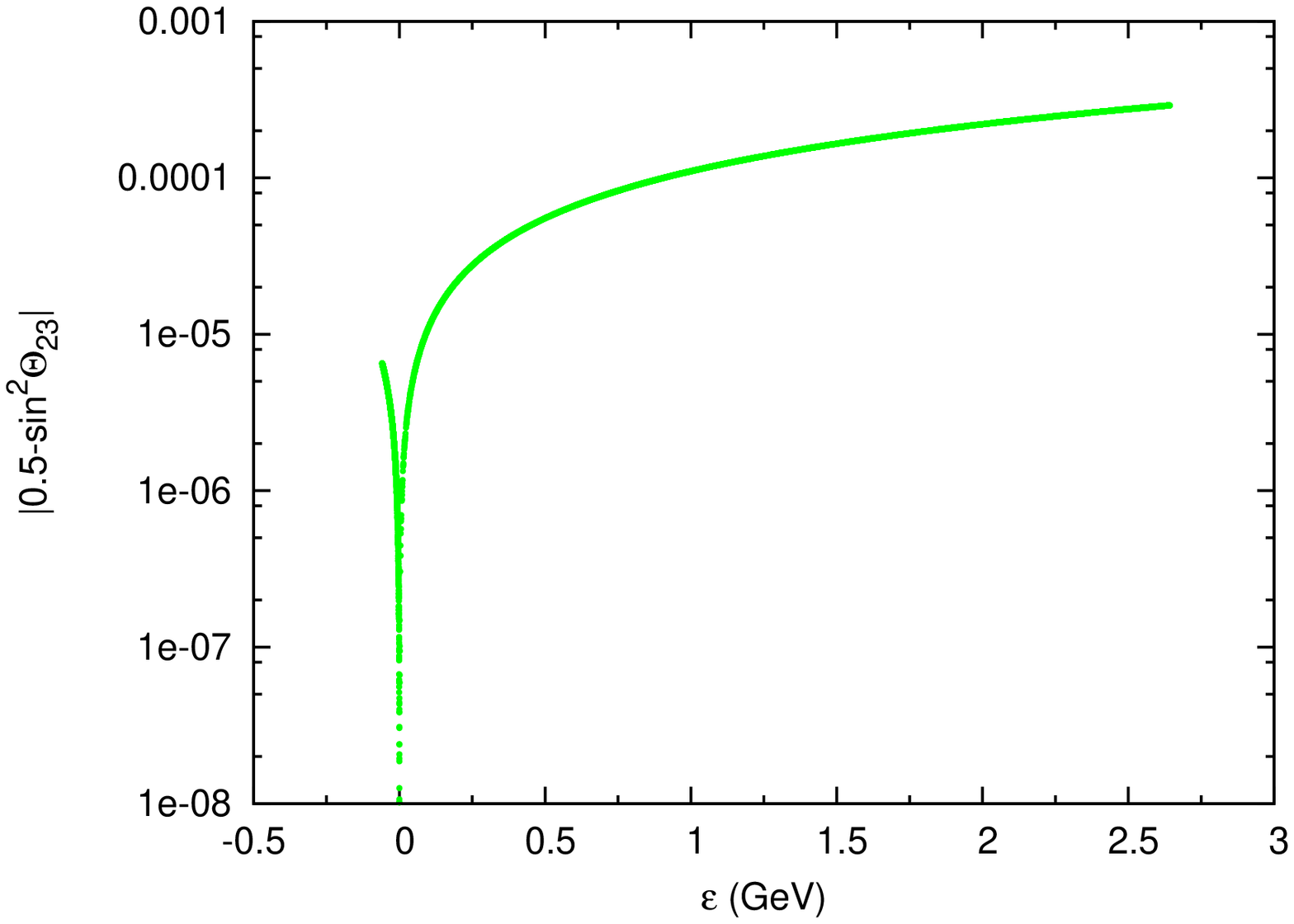}
\caption{\label{fig:mutaubreak}
Non-zero values of $U_{e3}$ and $|0.5 - \sa|$ 
predicted when $\mu$-$\tau$ symmetry is broken. Shown 
are the oscillation parameters against the $\mu$-$\tau$ symmetry
breaking parameter $\epsilon = M_3 - M_2$.  
Only points which reproduce the current neutrino observations 
within their $3\sigma$ C.L. are shown. The plot is generated 
at a fixed set of Yukawa couplings and heavy neutrino masses. 
}
\end{figure}

As discussed in the introduction we wish to impose $\mu$-$\tau$ 
symmetry on our model in order to comply with the current 
neutrino data, which shows a preference for $\theta_{13}=0$ and 
$\theta_{23}=\pi/4$. Henceforth, we impose the \mt 
exchange symmetry on both the 
neutrino Yukawa matrix $Y_\Sigma$ and the Majorana 
mass matrix for the heavy fermions $M$. Therefore, 
the neutrino Yukawa matrix takes the form
\be
Y_\Sigma &=& \pmatrix{
a_4 & a_{11} & a_{11} \cr
a_{11}' & a_6 & a_8 \cr
a_{11}' & a_8 & a_6 },
\label{eq:mdirac}
\ee
In what follows we will assume (for simplicity) that 
$a_{11}' = a_{11}$. 
Note that the $\mu$-$\tau$ symmetry does not impose this 
condition. 
It only imposes that $Y_{\Sigma_{12}} = Y_{\Sigma_{13}}$ 
and $Y_{\Sigma_{21}} = Y_{\Sigma_{31}}$.  
We have put $a_{11}' = a_{11}$ 
in order to reduce the number of parameters in the theory. 
For the same reason, we assume all entries of $Y_\Sigma$ to be real. 
The heavy Majorana mass matrix is given by 
\be
M &=& \pmatrix{
M_{1} & 0 & 0 \cr
0 & M_{2} & 0 \cr
0 & 0 & M_{2}},
\label{eq:msigma}
\ee
where without loosing generality we have chosen to work 
in a basis where $M$ is real and diagonal. 
Here the condition $M_3 = M_2$ is imposed  
due to the $\mu-\tau$ symmetry.

The above choice of Yukawa and heavy fermion mass matrix 
lead to the following form of the light neutrino mass matrix 
\be
\tilde{m} \simeq \frac{v'^2}{2}\pmatrix{
\frac{a_4^2}{M_1}  + \frac{2a_{11}^2}{M_2} & 
a_{11}\bigg(\frac{a_4}{M_1}  + \frac{a_6+a_8}{M_2}\bigg) &
a_{11}\bigg(\frac{a_4}{M_1}  + \frac{a_6+a_8}{M_2}\bigg) \cr
a_{11}\bigg(\frac{a_4}{M_1}  + \frac{a_6+a_8}{M_2}\bigg) &
\frac{a_{11}^2}{M_1}  + \frac{a_6^2+a_8^2}{M_2} &
\frac{a_{11}^2}{M_1}  + \frac{2a_6a_8}{M_2} \cr
a_{11}\bigg(\frac{a_4}{M_1}  + \frac{a_6+a_8}{M_2}\bigg) &
\frac{a_{11}^2}{M_1}  + \frac{2a_6a_8}{M_2} &
\frac{a_{11}^2}{M_1}  + \frac{a_6^2+a_8^2}{M_2}}
~,
\label{eq:mnu}
\ee
where we have used the seesaw formula given by Eq. (\ref{eq:lownu}), 
which is valid up to second order in $1/m_\Sigma$. 
One can straightaway see from the above mass matrix that 
the scale of the neutrino masses emerges as 
$\sim v'^2 a^2/(2\,m_\Sigma)$, where $a$ is a typical 
value of the Yukawa coupling in Eq. (\ref {eq:mdirac}) and 
$m_\Sigma$ the scale of heavy fermion masses. 
As discussed in the introduction, we restrict the heavy 
fermion masses to be less than 1 TeV in order that they 
can be produced at the LHC. Therefore  
in principle, neutrino masses 
of $\sim 0.1$ eV could have been 
obtained with just the standard model Higgs doublet by reducing 
the Yukawa couplings to values $\sim 10^{-6}$.   
However, this is usually considered as extreme fine tuning as 
there is no reason why the Yukawa couplings of the neutrinos 
should be so much suppressed, and the motivation for the seesaw 
mechanism is lost. In order to circumvent this, 
we introduced a different Higgs 
doublet $\Phi_2$, 
which couples only to the exotic fermions. 
On the other hand, Yukawa coupling of the 
standard Higgs $\Phi_1$ with the exotic fermions
was forbidden in our model by the $Z_2$ symmetry. 
Hence, only the 
VEV of this new Higgs doublet appears in Eq. (\ref{eq:mnu}).  
Since this Higgs $\Phi_2$ is 
not coupled to any standard model particle, it  
could have a VEV which could be different. Therefore, we 
demand that $v'\sim 10^{5}$ eV in order to generate 
neutrino masses of $\sim 0.1$ eV keeping the Yukawa couplings 
$\sim 1$. 
We have checked that such low value of Higgs VEV is not in 
conflict with any experimental data. We will discuss in detail 
the scalar potential and the Higgs mass spectrum 
in Appendix A. 

We next turn to predictions of this model for the 
mass squared differences and the mixing 
angles. Since the neutrino mass matrix 
we obtained in Eq. (\ref{eq:mnu})
has $\mu$-$\tau$ symmetry it follows that 
\be
\theta_{13}=0 ~~{\rm and}~~\theta_{23}=\pi/4
.
\ee
To find the mixing angle $\theta_{12}$ and the mass 
squared differences $\ms$ and $\ma$ \footnote{We define 
$\Delta m_{ij}^2 = m_i^2 - m^2_j$.}, 
one needs to 
diagonalize the mass matrix $\tilde{m}$ given in Eq. (\ref{eq:mnu}).
In fact, the form of $\tilde{m}$ in Eq. (\ref{eq:mnu}) is the 
standard form of the neutrino mass matrix 
with  $\mu$-$\tau$ symmetry, and hence 
the expression of mixing angle $\theta_{12}$ as 
well $\ms$ and $\ma$ can be found 
in the literature (see for {\it e.g.}
\cite{s3us}). We show in Fig. \ref{fig:ssqth12} 
the variation of $\sss$ (upper panels) and $R=\ms/|\ma|$ 
(lower panels) 
with the Yukawa couplings $a_4$, 
$a_{11}$ and $a_6$. We do not show the 
corresponding dependence on $a_8$ since 
it looks almost identical to the panel corresponding 
to $a_6$. The figure is produced assuming 
inverted mass hierarchy for the neutrino, {\it i.e.}, 
$\ma < 0$. The neutrino mass matrix given by 
Eq. (\ref{eq:mnu}) could very easily yield $\ma > 0$ and 
hence the normal 
mass hierarchy (see for {\it e.g.} \cite{s3us}). 
However, for the sake of illustration, we will show 
results for  
only the inverted hierarchy in this paper. 
In every panel of Fig. \ref{fig:ssqth12}, all Yukawa couplings 
apart from the one plotted on the x-axis, are allowed to vary 
freely. The points show the predicted 
values of $\sss$ (upper panels) and $R$ (lower panels)  
as a function of the Yukawa couplings for 
which all oscillation parameters are within their current 
$3\sigma$ values \cite{limits},
 \be
7.1\times 10^{-5} eV^2 < \ms < 8.3\times 10^{-5} eV^2~,~~
2.0\times 10^{-3} eV^2 < |\ma| < 2.8\times 10^{-3} eV^2~,~~
\label{eq:dmconstraint}
\ee
\be
0.26<\sss<0.42~,~~\sin^22\theta_{23}>0.9~,~~\sch<0.05~.~~
\label{eq:thconstraint}
\ee
For this figure we take $M_1 = M_2$ 
for simplicity 
and define $m_0 = v'^2/(2M_1)$. The blue points are 
for $m=0.95$ eV while the green points are for $m_0=0.006$ eV.
 
Fig. \ref{fig:param} is a scatter plot showing the 
values of the Yukawa couplings which give all oscillation 
parameters within their current $3\sigma$ allowed ranges 
given in Eqs. (\ref{eq:dmconstraint}) and 
(\ref{eq:thconstraint}). 
Again as in the previous plot we assume $M_1 = M_2$, 
define $m_0 = v'^2/(2\,M_1)$ and show the allowed points 
for $m_0=0.96$ eV (blue), 0.006 eV (green) and 0.0021 eV (red). 
All Yukawa couplings apart from the ones shown in the x-axis 
and y-axis are allowed to vary freely, in each panel. 

Since the $\mu$ and $\tau$ charged lepton masses are different, 
we phenomenologically choose to not impose the $\mu$-$\tau$ 
symmetry on 
the charged lepton mass matrix\footnote{We reiterate that 
our choice of the lepton masses and mixing are dictated solely by 
observations.}. Hence, without loosing generality, the 
charged lepton Yukawa matrix can be taken as 
\be
Y_l = \pmatrix{
Y_e & 0 & 0 \cr
0 & Y_\mu & 0 \cr
0 & 0 & Y_\tau},
\label{eq:mcharged}
~.
\ee
The masses of the light charged leptons can then be 
obtained from Eqs. (\ref{eq:leptonmassesr}) and/or 
(\ref{eq:leptonmassesl}). For our choice 
of $Y_\Sigma$ and $M$, it turns out that 
$m_e \approx  v Y_e$  , $m_\mu \approx v Y_\mu$, 
and $m_\tau \approx v Y_\tau$,  
if we neglect terms proportional to ${v'}^2$. 
The mixing matrices $U_l$ and $U_r$ which diagonalize 
$\tilde{m}^\dagger_l\tilde{m}_l  $ 
(cf. Eq. (\ref{eq:leptonmassesl})) and 
$\tilde{m}_l \tilde{m}^\dagger_l $ (cf. Eq. (\ref{eq:leptonmassesr}))
respectively, turn out to be unit matrices at leading order. 
\be
U_l \simeq \pmatrix{
1 & 0 & 0 \cr
0 & 1 & 0 \cr
0 & 0 & 1}, 
~~~~
U_r \simeq \pmatrix{
1 & 0 & 0 \cr
0 & 1 & 0 \cr
0 & 0 & 1}, 
\label{eq:ul}
\ee

Finally, we show in Fig. \ref{fig:mutaubreak} the impact of 
$\mu$-$\tau$ symmetry breaking on the low energy 
neutrino parameters. For the sake of illustration we choose 
a particular form for this breaking, by taking 
$M_3\neq M_2$. Departure from $\mu$-$\tau$ symmetry 
results in departure of $U_{e3}$ from zero and $\sin^2\theta_{23}$ 
from 0.5. We show in Fig. \ref{fig:mutaubreak} the 
$U_{e3}$ (left hand panel) and $|0.5 - \sa|$ generated 
as a function of the symmetry breaking parameter 
$\epsilon = M_3 - M_2$. 
The figure is generated for 
$a_4=-0.066$, $a_{11}=0.171$, $a_6=0.064$, $a_8=0.0037$ 
and $m_0=0.745$ eV. We have fixed $M_1 = M_2=299$ GeV 
in this plot. 
For $\epsilon=0$, 
$\mu$-$\tau$ symmetry is restored and both 
$U_{e3}$ and $0.5 - \sa$ go to zero. 
We show only points in this figure for which the current data 
can be explained within $3\sigma$. We note that 
for $\epsilon > 0$
the curve extends to about $M_3 = M_2 + 2.6$ GeV, 
for this set of model parameters. For $\epsilon < 0$, 
the allowed range for $\epsilon$ is far more restricted.

We next turn our attention 
to the predictions of this model for the 
heavy fermion sector. The masses of the 
heavy fermions can be obtained using $Y_\Sigma$ and $M$, 
as discussed in the previous section. The $6\times 6$ 
mixing matrices, which govern the mixing of the 
heavy leptons with light ones, can also be obtained as discussed 
before. 
We will see in the next section that 
all the four $3\times 3$ blocks of the matrices $U$, $S$ and $T$ 
are extremely important for phenomenology at the LHC. 
We denote these $3\times 3$ blocks as 
\be
U = \pmatrix{U_{11} & U_{12} \cr
U_{21} & U_{22}} = 
\pmatrix{(W_\nu)_{11}U_0 & (W_\nu)_{12} U_\Sigma \cr
(W_\nu)_{21} U_0 & (W_\nu)_{22} U_\Sigma},
\ee
\be
S = \pmatrix{S_{11} & S_{12} \cr
S_{21} & S_{22}} = 
\pmatrix{(W_L)_{11}U_l & (W_L)_{12} U_h^L \cr
(W_L)_{21} U_l & (W_L)_{22} U_h^L},
\label{eq:smatrix}
\ee
\be
T = \pmatrix{T_{11} & T_{12} \cr
T_{21} & T_{22}} = 
\pmatrix{(W_R)_{11}U_r & (W_R)_{12} U_h^R \cr
(W_R)_{21} U_r & (W_R)_{22}U_h^R},
\ee
The matrices $W_\nu$, $W_L$ and 
$W_R$ have been given in Eqs. (\ref{eq:wnu}), (\ref{eq:wl})
and (\ref{eq:wr}) respectively. 
These can be 
estimated for our choice of $m_D$, $M$ and $m_l$. 
In particular, we note that $S_{11}$ and $T_{11}$ are 
close to 1, while $U_{11}$ is given almost by $U_{PMNS}$. 
The off-diagonal blocks $U_{12}$, $U_{21}$, $S_{12}$ and $S_{21}$, 
are suppressed by $\sim m_D/M$, while 
$T_{12}$ and $T_{21}$ are suppressed by 
$\sim m_Dm_l/M^2$. 
Finally, the matrices $U_{22} = (W_\nu)_{22} U_\Sigma$, 
$S_{22} = (W_L)_{22} U_h^L$, and 
while $T_{22} \simeq U_h^R $. To estimate these we need to 
evaluate first the matrices which diagonalize the heavy fermion 
mass matrices $\tilde{M}$, $\tilde{M}_H^\dagger \tilde{M}_H$, 
and $\tilde{M}_H \tilde{M}_H^\dagger$ respectively. For 
$M$ with \mt symmetry, it turns out that
\be
U_\Sigma \simeq U_h^L \simeq U_h^R \simeq 
\pmatrix{1 & 0 & 0\cr
0 & \frac{1}{\sqrt{2}} & -\frac{1}{\sqrt{2}} \cr
0 & \frac{1}{\sqrt{2}} & \frac{1}{\sqrt{2}}},
\label{eq:uh}
\ee
thereby yielding
\be
U_{22} &\simeq& \pmatrix{1 & 0 & 0\cr
0 & \frac{1}{\sqrt{2}} & -\frac{1}{\sqrt{2}} \cr
0 & \frac{1}{\sqrt{2}} & \frac{1}{\sqrt{2}}}.
\label{eq:u22}
\ee
\be
S_{22} &\simeq& \pmatrix{1 & 0 & 0\cr
0 & \frac{1}{\sqrt{2}} & -\frac{1}{\sqrt{2}} \cr
0 & \frac{1}{\sqrt{2}} & \frac{1}{\sqrt{2}}}.
\label{eq:s22}
\ee
\be
T_{22} &\simeq& 
\pmatrix{1 & 0 & 0\cr
0 & \frac{1}{\sqrt{2}} & -\frac{1}{\sqrt{2}} \cr
0 & \frac{1}{\sqrt{2}} & \frac{1}{\sqrt{2}}}.
\label{eq:t22}
\ee
This structure of $U_\Sigma$, $U_h^L$ and $U_h^R$ 
(and hence of $U_{22}$, 
$S_{22}$ and $T_{22}$),  
is an immediate consequence of the $\mu$-$\tau$ symmetry 
in $M$ and $m_D$. 
This is an extremely new and crucial 
feature.  
To the best of our knowledge, 
this has not been pointed out in any of the 
previous Type III seesaw models studies. The main reason is that 
no study so-far has considered the flavor aspect of Type III seesaw. 
As a result none of them considered imposing an underlying 
flavor symmetry group on the fermions such that the 
triplet fermion Majorana mass matrix and the Yukawa matrix 
would be \mt symmetric. 
They assume that 
$U_\Sigma$,  $U_h^L$ and $U_h^R $ are almost unit matrices since 
$M$ is real and diagonal. 
However, this is true only if $M_1 \ll M_2 \ll M_3$. 
Breaking of the 
$\mu$-$\tau$ symmetry either in $m_D$ or in $M$, 
will destroy this 
non-trivial form for $U_\Sigma$,  $U_{h}^R$ and $U_h^L$.
But having \mt symmetry in {\it both} $Y_\Sigma$ and $M$ is 
both theoretically as well as phenomenologically well motivated. 
We will see later that this non-trivial form of 
the matrices $U_\Sigma$,  $U_h^L$ and $U_h^R $ will lead to certain 
distinctive signatures at LHC.

One should note that while $U_\Sigma$, $U_{h}^L$ and $U_{h}^R$ have 
the form given by 
Eq. (\ref{eq:uh}), $U_l$ and $U_r \simeq I$, 
though both $M$ 
and $Y_l$ were taken as real and diagonal. The main reason for this 
drastic difference is the following. 
While we take exact $\mu$-$\tau$ 
symmetry for $M$, for $Y_l$ we take a 
large difference between $Y_\mu$ and $Y_\tau$ values. This 
choice was dictated 
by the observed charged lepton masses.

Finally, a comment regarding the extent of deviation from 
unitarity in our model is in order. 
It is clear from the discussion of the previous section and 
Eq. (\ref{eq:wnu}) that the deviation from 
unitarity of the light neutrino mixing matrix 
is $\propto m_D^2/m_\Sigma^2 \simeq m_\nu/m_\Sigma$, where 
$m_\nu$ and  $m_\Sigma$ are the light neutrino and 
heavy lepton mass scales respectively. Therefore, an 
important difference between our model and the usual 
GUT Type III seesaw models is that the extent of  
non-unitarity for our model is much larger. 
This will result in larger lepton flavor violating 
decays in our case. 
Detailed calculations for lepton flavor 
violating radiative as well as tree level decays of a generic 
Type III seesaw model have been published in 
\cite{type3lowE,type3muegamma} and we do not repeat them here. 
The authors of these papers have also worked out the 
current constraints on the deviation from unitarity. 
One can check that 
even for 100-1000 GeV mass range heavy leptons, the 
predicted non-unitarity and lepton flavor violating 
decay rates in our model 
are far below the current experimental bounds. 
In fact, the predicted decay rates can be seen to be far 
below the sensitivity reach of all forthcoming experiments.

\section{Heavy Fermion Production at LHC}

The triplet fermions couple to the standard model particles 
through the Yukawa couplings as well as gauge couplings. 
We give in Appendix B, the detailed Yukawa and gauge 
couplings of the neutral and charged heavy fermions with the 
standard model leptons, vector bosons, and Higgs particles. 
We have kept the masses of the heavy fermions in the 100 GeV to 
1 TeV range. Therefore, it should be possible to 
produce these fermions at LHC. In this section, we will 
study in depth the production and detection possibilities 
of the heavy leptons in our Type III seesaw model. 
Compared to the earlier papers, there are two distinctly 
new aspects in our analysis -- (i) presence of two Higgs 
doublet instead of one, leading to a far more rich 
collider phenomenology, (ii) presence of $\mu$-$\tau$ 
symmetry in our model. 

The heavy triplet fermion production at LHC has been discussed 
by many earlier papers \cite{type3lhc1,type3lhc2,type3lhchambyestrumia,
type3lhcaguila,type3lhc3}. At 
LHC we will be looking at the following production channels
$$pp\rightarrow \Sigma^{\pm}\Sigma^{\mp},
\Sigma^{\pm}\Sigma^{0}, \Sigma^{0}\Sigma^{0}.$$
The exotic fermions have both Yukawa couplings to Higgs as well as 
gauge couplings to vector bosons. Therefore, they could be 
in principle produced through either gauge mediated 
partonic processes (left diagram) 
or through Higgs mediated partonic processes (right diagram)
\vglue 0.5cm \hglue 2.0cm
\begin{picture}(800,100)
\ArrowLine(30,60)(10,90)
\ArrowLine(10,30)(30,60)
\Photon(30,60)(70,60){4}{4}
\ArrowLine(70,60)(90,90)
\ArrowLine(90,30)(70,60)
\Text(0,20)[]{$q$}
\Text(0,100)[]{$\bar{q}/q'$}
\Text(50,70)[]{\tiny{$Z$/$\gamma$/$W^{\pm}$}}
\Text(90,100)[]{$\Sigma^{\pm}$}
\Text(90,20)[]{$\Sigma^{0}$/ $\Sigma^{\mp}$}
\end{picture}
\vglue -3.5cm \hglue 8.5cm
\begin{picture}(800,100)
\ArrowLine(30,60)(10,90)
\ArrowLine(10,30)(30,60)
\DashLine(30,60)(70,60){4}
\ArrowLine(70,60)(90,90)
\ArrowLine(90,30)(70,60)
\Text(0,20)[]{$q$}
\Text(0,100)[]{$\bar{q}/q'$}
\Text(50,70)[]{\tiny{$h^0$/$A^0$}}
\Text(50,50)[]{\tiny{$H^0$/$H^{\pm}$}}
\Text(90,100)[]{$\Sigma^{\pm}$}
\Text(90,20)[]{$\Sigma^{0}$/ $\Sigma^{\mp}$}
\end{picture}
However, it turns out that 
the vertex factors for the couplings of heavy fermions to 
gauge bosons which are relevant 
for the formers production, {\it viz.},
$\Sigma^{+} \Sigma^{-} Z/ \gamma$ and 
$\Sigma^0 \Sigma^\pm W^{\mp}$,  
are much larger than those involving the 
Higgs, {\it viz.}, $\Sigma^{+}\Sigma^{-} H^0/h^0/A^0$  and 
$\Sigma^0 \Sigma^\pm H^{\mp}$. 
To illustrate this with a specific example, we compare the 
$\Sigma^{+} \Sigma^{-} Z$ coupling 
given in Eqs. (\ref{eq:S-S-ZR}) and (\ref{eq:S-S-ZL})
with the  
$\Sigma^{+} \Sigma^{-} h^0$ 
coupling given in Table 
\ref{tab:chvertex}. 
It is easy to see that the $\Sigma^{+} \Sigma^{-} Z/ \gamma$ 
coupling has terms proportional 
to $T_{22}^\dagger T_{22}$ and $S_{22}^\dagger S_{22}$, 
while the $\Sigma^{+} \Sigma^{-} h^0$ 
coupling depends on terms which have an 
off-diagonal mixing matrix block. Since the off-diagonal 
mixing matrix blocks are much 
smaller than the diagonal ones (as discussed in section 3), 
it is not surprising that 
the couplings of two exotic fermions to the Higgs particles 
are much smaller than to the gauge bosons. 
Hence the heavy exotic fermions will be produced predominantly 
via the gauge boson mediated processes. 
For the heavy fermion 
production cross-section at the LHC, we chose CTEQ6L \cite{CTEQ} 
as the parton distribution function (PDF) and partonic center of mass 
energy as the renormalization and 
factorization scale. We have explicitly checked 
that the production cross-sections do not change much with the 
PDF and scale. All cross-sections in this paper are 
calculated using the Calchep package \cite{calchep}.

\begin{figure}
\begin{center}
\includegraphics[width=0.54\textwidth,angle=270]{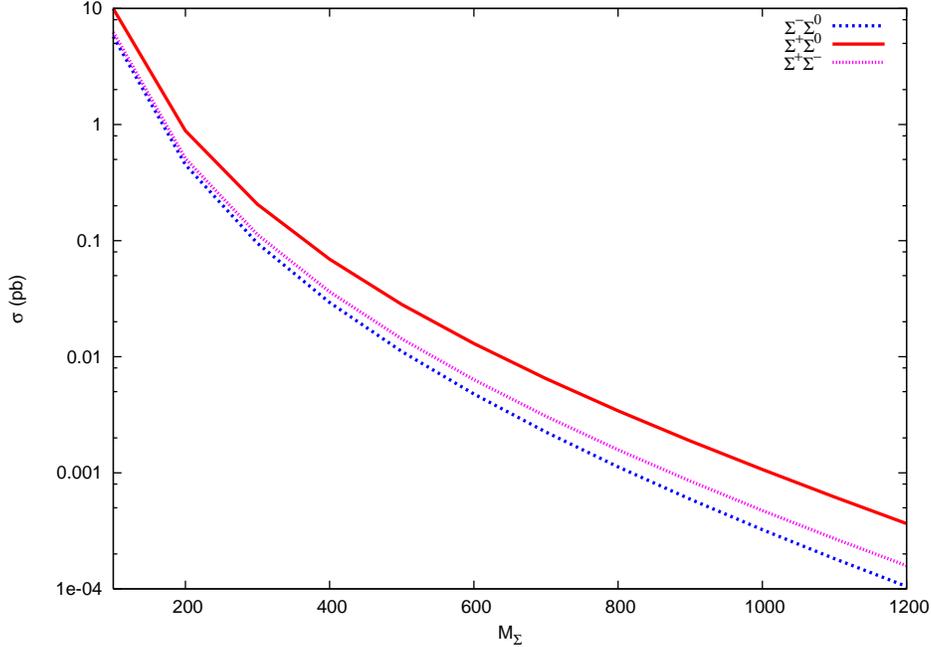}
\caption{\label{fig:crosssection}
Variation of production cross section of $\Sigma^{\pm}$, $\Sigma^{0}$ 
with the mass of exotic leptons.
}
\end{center}
\end{figure}

In Fig. \ref{fig:crosssection} we show the 
production cross-section 
for $\Sigma^-\Sigma^0$ (bold dotted line), 
$\Sigma^+\Sigma^0$ (solid line), and 
$\Sigma^+\Sigma^-$ (fine dotted line),  
at LHC as a 
function of the corresponding heavy fermion mass. 
It is straightforward to see 
that the ${\Sigma'}^0{\Sigma'}^0Z$ (and 
${\Sigma'}^0{\Sigma'}^0W^\pm$) couplings are absent. 
A very small ${\Sigma}^0{\Sigma}^0Z$ is generated through 
mixing from the ${\nu'}^0{\nu'}^0Z$ coupling. However, 
this is extremely small. 
Hence, $\Sigma^0\Sigma^0$ production through gauge 
interactions is heavily suppressed and is not shown in 
Fig. \ref{fig:crosssection}. 
One can see that the production 
cross-sections of the heavy fermions fall 
sharply with their mass. More precisely, 
$\sigma({\Sigma^{\pm}\Sigma^{\mp}})=112$
fb,  $\sigma({\Sigma^{+}\Sigma^{0}})=206$ fb and
$\sigma({\Sigma^{-}\Sigma^{0}})=95$ fb, for 
$M_{\Sigma_i}\simeq 300$ GeV. However, for 
$M_{\Sigma_i}\simeq 600$ GeV this quickly falls to 
$\sigma(\Sigma^\pm\Sigma^\mp)= 6$ fb,
$\sigma(\Sigma^+\Sigma^0)= 13$ fb, and 
$\sigma(\Sigma^-\Sigma^0)= 4$ fb. Therefore, 
it is obvious that the lightest amongst the 
three generation of heavy fermions will be 
predominantly produced at the collider, and will 
hence dominate the phenomenology. 
One can check that the production cross-sections that we 
get is almost identical to that obtained in earlier papers
\cite{type3lhchambyestrumia,type3lhcaguila}. This is not 
unexpected since 
our model is different from all the 
earlier models in the Higgs sector. However as discussed 
above, it is the gauge interactions which predominantly 
produce the exotic leptons. 
The gauge-lepton couplings in our model is same as  
in the earlier works. 
Since the heavy fermion production comes mostly from the gauge 
mediated sub-processes, we get the same production cross-sections 
as other papers.
 
\section{Heavy Fermion Decays}

Once produced at LHC, 
the heavy fermions will decay to different lighter states 
due to its interaction with 
different standard model particles. In particular, 
they could decay into light leptons and Higgs due to their
Yukawa couplings, or to light leptons and vector bosons 
due to their gauge interactions. The light leptons could be 
either the charged leptons or the neutrinos. The Higgs 
could be either the neutral Higgs $h^0$, $H^0$, $A^0$, 
or the charged Higgs $H^\pm$. The gauge bosons could be 
either $W^\pm$ or $Z$. The Higgs and gauge bosons would 
eventually give the final state particles in the detector, 
which will be tagged at the experiment. This will be 
studied in detail in the following sections. Here we concentrate 
on only the two body decay rates and branching ratios of the 
exotic leptons $\Sigma^\pm$ and $\Sigma^0$. 
All possible vertices and 
the corresponding vertex factors for the Yukawa 
interactions of $\Sigma^\pm$ and $\Sigma^0$ are given in 
Tables \ref{tab:chvertex}, \ref{tab:nuvertex}, \ref{tab:hcvertex}. 
The vertices and 
vertex factors for the charged and neutral current 
gauge interactions can be found in Appendix B.2. 
Presence of two Higgs doublets and $\mu$-$\tau$ symmetry in 
$Y_\Sigma$ and $M$ will have 
direct implications in the partial decay widths
for different decay processes. 

\subsection{Decay to Light Leptons and Higgs}

In this subsection, we perform a detailed 
study of all 
two-body decays of these fermions into a lepton 
and a Higgs. Since we have two Higgs doublets in our 
model, we have a pair of charged Higgs $H^\pm$, and 
three neutral Higgs -- $h^0$ and $H^0$ are CP even, 
while $A^0$ is CP odd. The Higgs mass spectrum and mixing 
is given in Appendix A. 
The decay width $\Gamma$ for 
$\Sigma_i \rightarrow l_j X $ is 
given by
\be
\Gamma = \frac{{M}_{\Sigma_i}}{32\pi} 
\bigg[1-\frac{{(M_{X}-m_{l_j})}^2}
{{{M}_{\Sigma_i}}^2}\bigg]^{\frac{1}{2}}\times\bigg[1-\frac{{(M_X+
m_{l_j})}^2}
{{{M}_{\Sigma_i}}^2}\bigg]^{\frac{1}{2}}\times A_{ji},
\label{eq:decayrate1}
\ee 
where ${M}_{\Sigma_i}$, $M_X$ and $m_{l_j}$ are the masses of 
$\Sigma^0_i/\Sigma^\pm_i$, $X$ 
and $l_j$, respectively, where $X$ is the 
relevant Higgs involved. The $l_j$ could be either a 
charged lepton or a neutrino. 
For the charged Higgs $H^\pm$ mode, and the neutral 
Higgs $h^0$ and $H^0$ mode, 
the factor $A_{ji}$ is given as  
\be
A_{ji}&=&\bigg(|{(C_{l\Sigma}^{X,L})}_{ji}|^2+
|{(C_{l\Sigma}^{X,R})}_{ji}|^2\bigg)
\bigg(1- \frac{(M_{X}^2-M_l^2) }{{M}_{\Sigma_i}^2}\bigg)
\nn\\
&&+\bigg({(C_{l\Sigma}^{X,L})}_{ji}^*{(C_{l\Sigma}^{X,R})}_{ji}
+{(C_{l\Sigma}^{X,R})}_{ji}^*{(C_{l\Sigma}^{X,L})}_{ji}\bigg)\frac{m_{l_j}}
{{M}_{\Sigma_i}}
\label{eq:expAji}
,
\ee
while for the CP-odd neutral Higgs $A^0$ the factor is
\be
A_{ji}&=&\bigg(|{(C_{l\Sigma}^{X,L})}_{ji}|^2+
|{(C_{l\Sigma}^{X,R})}_{ji}|^2\bigg)
\bigg(1- \frac{(M_{X}^2-M_l^2) }{{M}_{\Sigma_i}^2}\bigg)
\nn\\
&&-\bigg({(C_{l\Sigma}^{X,L})}_{ji}^*{(C_{l\Sigma}^{X,R})}_{ji}
+{(C_{l\Sigma}^{X,R})}_{ji}^*{(C_{l\Sigma}^{X,L})}_{ji}\bigg)\frac{m_{l_j}}
{{M}_{\Sigma_i}}
\label{eq:expAjiodd}
.
\ee
In the above equations ${(C_{l\Sigma}^{X,L})}/{(C_{l\Sigma}^{X,R})}$ 
are the relevant vertex factors given in Table \ref{tab:chvertex},
\ref{tab:nuvertex} and \ref{tab:hcvertex}, 
and $i,j$ represents the generation. In all numerical results that 
follow we will fix the model parameters 
(Yukawa couplings and entries of $M$ mass matrix) to their  
values given in Table \ref{tab:parambestfit}. This set of model 
parameters yield $\ms=7.67\times 10^{-5}$ eV$^2$, 
$\ma = -2.435\times 10^{-3}$ eV$^2$ and $\sss=0.33$. 
Of course $\theta_{13}=0$ and $\theta_{23}=\pi/4$.  
Throughout the rest of the paper, we also take 
the value of $M_{h^0}=40$ GeV, $M_{H^0}=150$ GeV, $M_{A^0}=140$ GeV and 
$M_{H^\pm}=170$ GeV. Also, for all cases where we present 
results for fixed values of the heavy fermion masses, we 
take $M_{\Sigma_1}=300$ GeV and $M_{\Sigma_2}=M_{\Sigma_3}=600$ GeV.
%
\begin{table}[h]
\begin{center}
\begin{tabular}{|c|c|c|c|c|c|c|}
\hline
$a_4$ & $a_6$ & $a_8$ & $a_{11}$ & $m_o$/eV & $\frac{M_{2}}{M_{_1}}$ 
& $\frac{M_{3}}{M_{1}}$ \cr
\hline
0.145 & $0.097$ & $0.109$ & $4 \times 10^{-4}$ & 2.356 & 2.0 & 2.0 \cr
\hline
\end{tabular}
\caption{\label{tab:parambestfit}
Model parameters used for all numerical results in section 
5 and 6.  
This set of model 
parameters yield $\ms=7.67\times 10^{-5}$ eV$^2$, 
$\ma = -2.435\times 10^{-3}$ eV$^2$ and $\sss=0.33$. 
Of course $\theta_{13}=0$ and $\theta_{23}=\pi/4$. Parameter 
$m_0=v'^2/(2M_1)$.  
}
\end{center}
\end{table}

\subsubsection{$\Sigma^\pm \rightarrow l^\pm \, h^0/H^0/A^0$}

\begin{figure}[t]
\begin{center}
\includegraphics[height=8.0cm,width=16.2cm,angle=0]{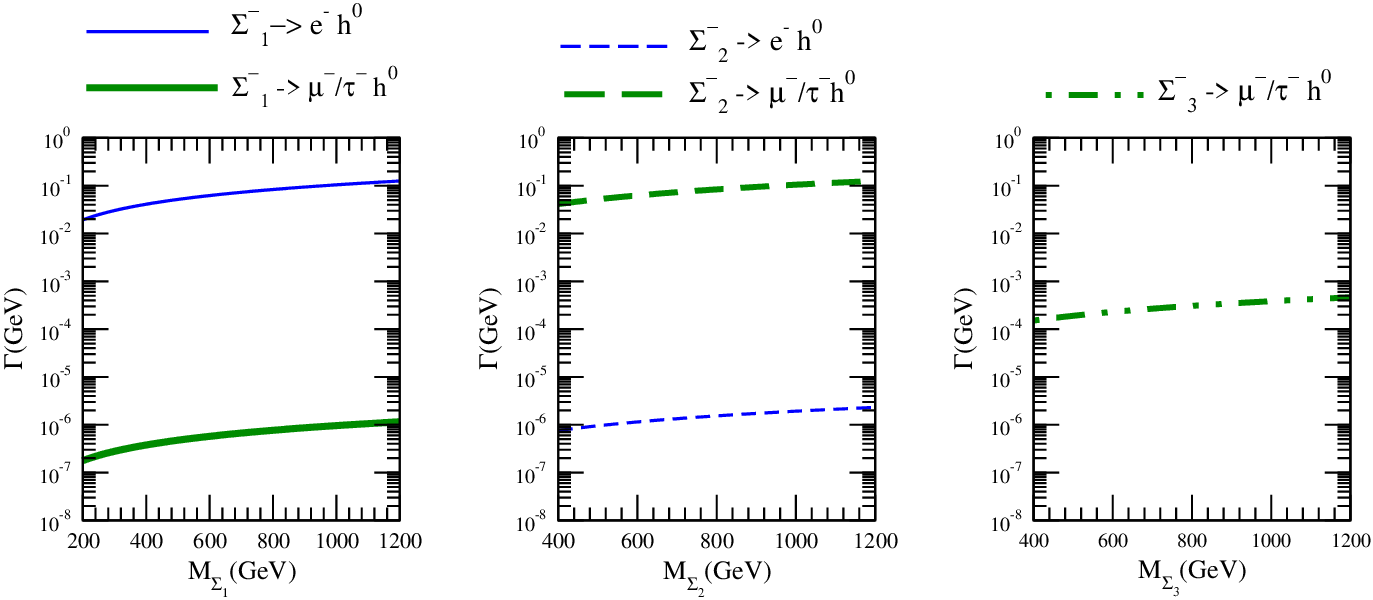}
\caption{\label{fig:dch}
Variation of $\Gamma(\Sigma^-_i \rightarrow l^{-}_j h^0)$ with ${M_{\Sigma_i}}$
}
\end{center}
\end{figure}

\begin{figure}[t]
\begin{center}
\includegraphics[height=8.0cm,width=16.2cm,angle=0]{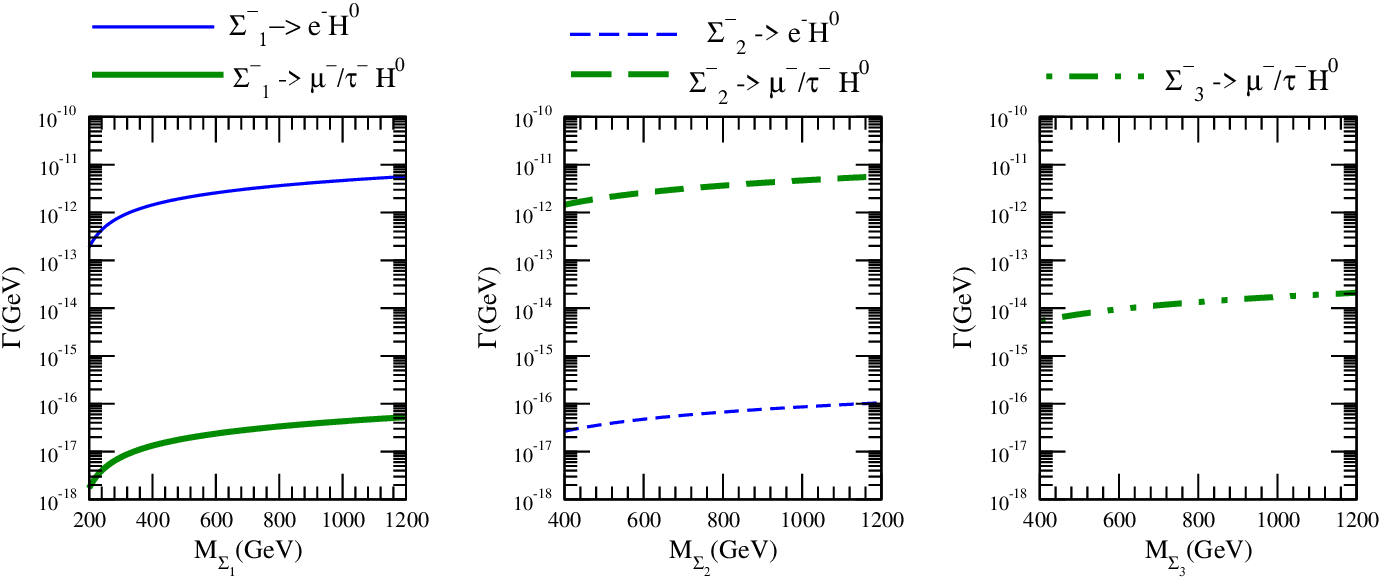}
\caption{\label{fig:dcH}
Variation of $\Gamma(\Sigma^-_i \rightarrow l_jH)$ with ${M_{\Sigma_i}}$}
\end{center}
\end{figure}

Let us begin with the decay of heavy 
charged leptons into light charged leptons and neutral Higgs. 
The Higgs concerned in this case could be $h^0$, $H^0$, or $A^0$.   
We start by probing the decay rate 
$\Sigma^{\pm}_i \rightarrow l_j^{\pm}h^0 $. 
From Eq. (\ref{eq:decayrate1}) we see that the 
decay rate is governed by the factor $A_{ji}$, which 
in turn depends on the vertex factors given in Table 
\ref{tab:chvertex}. The vertex factors are given in terms
of the $3\times 3$ block matrices $S_{ab}$ and $T_{ab}$, 
where $a,b=1,2$. We have seen in the 
earlier sections that $S_{12}$,  
$T_{12}$ and $T_{21}$ 
are heavily suppressed -- the first one by ${\cal O}(m_D/M)$ 
and $T_{12}$ and $T_{21}$ by ${\cal O}((m_l m_D)/M^2)$. The vertex 
factors also depend on the Higgs mixing angle $\alpha$. 
In Appendix A, we have shown how the neutrino mass constrains 
the neutral Higgs mixing such that 
$\sin\alpha \sim 10^{-6}$ and 
$\cos\alpha \sim 1$. 
Therefore, for the 
$\Sigma^{\pm}_i \rightarrow l^{\pm}h^0 $ decay the 
dominating vertex factor is
\be
C_{l^\pm\Sigma^\pm}^{h^0,R} \simeq 
\frac{1}{\sqrt{2}}S_{11}^\dagger Y_\Sigma^\dagger T_{22} \cos\alpha
.
\label{eq:vfactdch}
\ee 
We have seen in Eq. (\ref{eq:smatrix}) that 
$S_{11} \simeq 1$ if we neglect terms of the order of 
${\cal O}(m^2_D/M^2)$. Therefore,
\be
C_{l^\pm\Sigma^\pm}^{h^0,R}  \simeq \pmatrix{
a_4 & \sqrt{2} a_{11} & 0 \cr
a_{11} & \frac{1}{\sqrt{2}}(a_6 + a_8) &  \frac{1}{\sqrt{2}}(a_8 - a_6)\cr
a_{11} & \frac{1}{\sqrt{2}}(a_6 + a_8) &  \frac{1}{\sqrt{2}}(a_6 - a_8)}
.
\label{eq:cRdch}
\ee
From Eq. (\ref{eq:cRdch}) we can see that 
$(C_{l^\pm\Sigma^\pm}^{h^0,R})_{13} \simeq 0$. 
In fact one can check that this happens because $T_{22}$  
 given by Eq. (\ref{eq:t22}) has a specific form, which
is due to $\mu$-$\tau$ symmetry. 
The consequence of this is that decay of 
$\Sigma^-_3 \rightarrow e^-  h^0$ will be forbidden to leading 
order. Also note from Eq. (\ref{eq:cRdch}) that 
the decay rate of all heavy charged fermions into $\mu^\pm$
is predicted to be exactly equal to their decay rate into $\tau^\pm$. 
This is also an obvious consequence of the 
$\mu$-$\tau$ symmetry. 

The partial decay widths for 
$\Sigma^{-}_i \rightarrow l_j^{-}h^0 $
from an exact numerical calculation in shown in Fig. 
\ref{fig:dch}, as a function of the heavy charged fermion mass. 
The thin blue lines are decay into $e^-$, 
while the thick green lines are for decay into $\mu^-/\tau^-$. 
As expected, we notice the following two consequences of 
$\mu$-$\tau$ symmetry 
\begin{itemize}
\item Decay rate of $\Sigma^-_3 \rightarrow e^-h^0$ 
is almost zero. 
\item The decay rate of the heavy fermions into 
$\mu^-$ is exactly equal to that into $\tau^-$. 
\end{itemize}
We can also see that for $\Sigma^-_1$ decay, the decay 
rate into $e^-$ is about 5 orders of magnitude larger than 
into $\mu^-/\tau^-$. The trend is opposite for 
$\Sigma^-_2$ decay, where the decay is predominantly into 
$\mu^-/\tau^-$. Both of these features can be understood 
from Eq. (\ref{eq:cRdch}) and the values of the Yukawa 
couplings taken (cf. Table \ref{tab:parambestfit}). 
$\Sigma^-_1$ decay into $e^-$ and $\mu^-/\tau^-$ 
is proportional to $a_4^2$ and $a_{11}^2$, respectively. 
The ratio of the decay widths seen in the figure matches 
the ratio $a_4^2/(a_{11}^2) \sim 10^5$. Similarly, 
one can check that for $\Sigma^-_2$ decay, the corresponding ratio 
is $4 a_{11}^2/(a_6+a_8)^2$, which agrees with the middle panel of 
Fig. \ref{fig:dch}. Finally, the fact that 
the decay rate of $\Sigma^-_3 \rightarrow \mu^-h^0$ 
is less than that of $\Sigma^-_2 \rightarrow \mu^-h^0$ 
can also be understood in terms of Eq. (\ref{eq:cRdch}) and 
the Yukawa coupling values taken for the calculation. 

We next turn to the decay width for 
$\Sigma^{\pm}_i \rightarrow l_j^{\pm}H^0$. Expression for the 
decay rate is same as that given by Eq. (\ref{eq:decayrate1}) 
except that now $M_{h^0}$ is replaced by the $H^0$ mass $M_{H^0}$. 
For this decay channel the $A_{ji}$ factor is dominantly 
given by 
\be
C_{l^\pm\Sigma^\pm}^{H^0,R} 
\simeq \frac{1}{\sqrt{2}}S_{11}^\dagger Y_\Sigma^\dagger T_{22} \sin\alpha
.
\label{estim}
\ee 
Note that compared to the effective vertex factor 
for $\Sigma^{\pm}_i \rightarrow l_j^{\pm}h^0$ given in 
Eq. (\ref{eq:vfactdch}), the effective vertex factor 
given above for $\Sigma^{\pm}_i \rightarrow l_j^{\pm}H^0$ 
is suppressed by $\sin\alpha$. Since $\sin\alpha \sim 10^{-6}$, 
the decay rate of $\Sigma^{\pm}_i$ into $H^0$ are heavily suppressed. 
We show in Fig. \ref{fig:dcH} this decay rate 
calculated from exact numerical results. 
Comparing Fig. \ref{fig:dch} with Fig. \ref{fig:dcH}, we see 
that decays into $H^0$ are suppressed by a factor of about 
$\sim 10^{10}$, as expected from the order of magnitude estimate. 
Therefore, we can neglect 
$\Sigma^{\pm}_i \rightarrow l_j^{\pm}H^0$ for all 
practical purposes. 

From Table \ref{tab:chvertex} it is easy to 
see that the decay rate $\Sigma^{\pm}_i \rightarrow l_j^{\pm}A^0$
will be almost identical to that that predicted for 
$\Sigma^{\pm}_i \rightarrow l_j^{\pm}h^0$. The vertex factors for the 
two process are the same and hence the only difference 
could come from the difference between the Higgs masses. 
However, it is easy to see from Eq. (\ref{eq:decayrate1}) that 
the effect of the Higgs mass on the decay rate is not 
very significant, especially for relatively heavy fermions. 
 
\subsubsection{$\Sigma^0 \rightarrow l^\mp \, H^\pm$}

\begin{figure}[t]
\begin{center}
\includegraphics[height=8.0cm,width=16.2cm,angle=0]{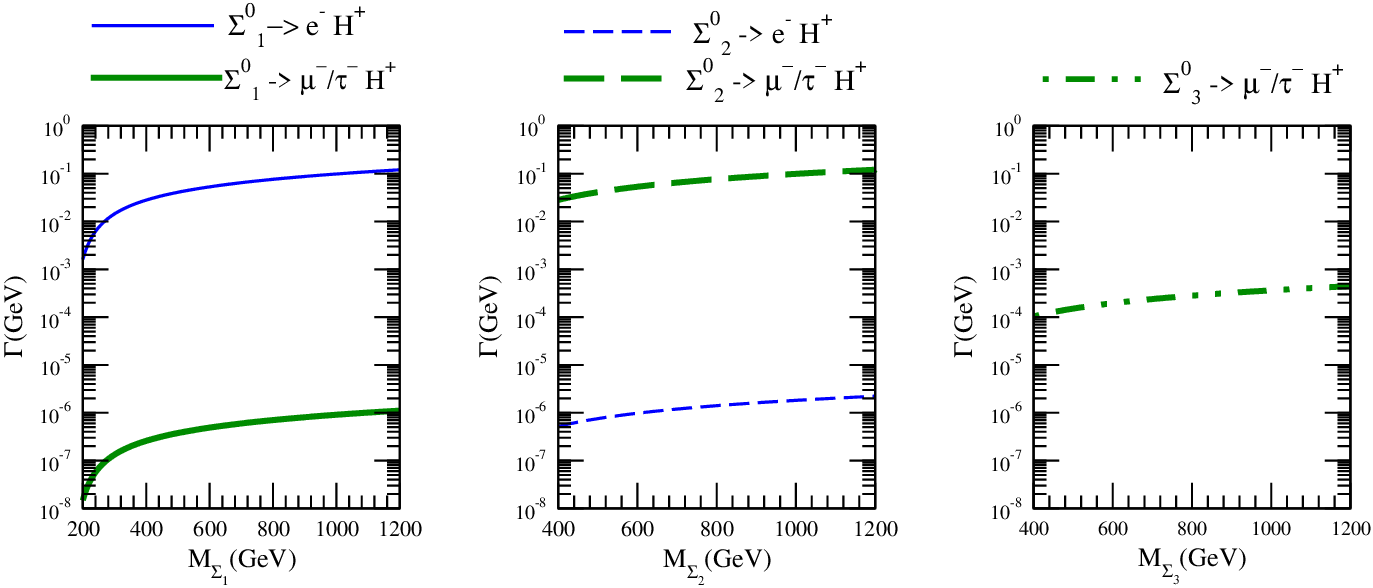}
\caption{\label{fig:dnH}
Variation of $\Gamma(\Sigma^0_i \rightarrow l_jH^+)$ with ${M_{\Sigma_i}}$}
\end{center}
\end{figure}

The decay rate for this channel is also given by 
Eq. (\ref{eq:decayrate1}), and 
is governed primarily 
by the vertex factor
\be
C_{l^\pm\Sigma^0}^{H^\pm,R}  \simeq 
\frac{1}{\sqrt{2}} S_{11}^\dagger Y_\Sigma^\dagger U_{22}^*
\cos\beta
.
\label{eq:vfactdch1}
\ee 
Recall that $\cos\beta \sim 1$. 
As discussed before, the matrix  $U_{22}$ displays 
features similar to the matrix $T_{22}$. Therefore, the form of 
dominant vertex factor 
for this case is similar to that given in 
Eq. (\ref{eq:cRdch}). The corresponding decay 
rates are shown in Fig. \ref{fig:dnH}. All features 
seen for $\Sigma^-_i \rightarrow l_jh^0$ is also seen here. 
Decay channel $\Sigma^{0}_3 \rightarrow e_j^{\mp}H^\pm $
is forbidden. Decay rates to $\mu^\mp$ is equal to 
decay rate to $\tau^\mp$. The huge hierarchy in the 
decay rates of $\Sigma^0_1$ and $\Sigma^0_2$ into 
$e$ and $\mu/\tau$ are also present due to same reason as 
given for $\Sigma^-\rightarrow l^{-}h^0 $ decays. The decay 
rate and flavor structure for the final state charged leptons 
is therefore seen to be same here as for the decay of 
charged heavy fermions into charged light leptons and $h^0$. 
However, in this case we have a charged Higgs in the final state 
and it should be easy to tag this and differentiate the two 
processes in the detector at LHC. We will also 
discuss in the following sections that the $h^0$ also 
has a much longer lifetime than $H^\pm$, which can be 
observed in the detector.   
In addition, the heavy 
lepton itself is charged in one case and uncharged in the 
other. The two processes should hence be separable at the 
collider experiment.


\subsubsection{$\Sigma^0 \rightarrow \nu \,h^0/H^0/A^0$}

We next turn to the decay channels with a light neutrino in 
the final state. This will give missing energy in the 
final state. Decay of the neutral $\Sigma^0$   
will create a neutrino and a neutral 
Higgs. As in the case of decay of $\Sigma^\pm$ to charged leptons and 
neutral Higgs, one can check from Table \ref{tab:nuvertex} 
that the decay to the Higgs $H^0$ is heavily suppressed due to 
smallness of $U_{21}$ as well as the $\sin\alpha$ term. 
However, decay to $h^0$ is driven by the vertex factor
\be
C_{\nu\Sigma^0}^{h^0,R} 
= \frac{1}{{2}}U_{11}^\dagger Y_\Sigma^\dagger U_{22}^* \cos\alpha
.
\label{eq:vfactdch2}
\ee   

\noindent
For the decay $\Sigma^0_i \rightarrow \nu_j A^0$ we find 
from Table \ref{tab:nuvertex} that the dominant vertex factor is 
\be
C_{\nu\Sigma^0}^{A^0,R} 
= -\frac{i}{{2}} U_{11}^\dagger Y_\Sigma^\dagger U_{22}^*
\cos\beta 
.
\label{eq:vfactdch3}
\ee 
Since $\cos\beta \simeq \cos\alpha$, the decay rate 
and flavor structure for 
this channel will be similar to what we found for 
the $\Sigma^0_i \rightarrow \nu_j h^0$ channel. The main  
difference comes in the difference between the 
masses of the $h^0$ and $A^0$ Higgs.

\noindent
For the $\Sigma^0_i \rightarrow \nu_j H^0$ 
decay, one can see from Table \ref{tab:nuvertex} 
that the vertex factors for both $P_L$ as well as 
$P_R$ vertices, are suppressed by $\sin\alpha$. 
Therefore, this decay rate can be neglected.

\subsubsection{$\Sigma^\pm \rightarrow \nu\, H^\pm$}

From Table \ref{tab:hcvertex} the vertex factor 
for this decay will be 
\be
C_{\nu\Sigma^\pm}^{H^\pm,L} 
\simeq  U_{11}^T Y_\Sigma^T S_{22}
\cos\beta 
.
\label{eq:vfactdch4}
\ee 
As we have seen in section 3, the structure of 
$S_{22}$ is very similar to that of $U_{22}$. Hence, a comparison 
of the vertex factor for this process with the one from 
$\Sigma^0_i \rightarrow \nu_j h^0$ and 
$\Sigma^0_i \rightarrow \nu_j A^0$ shows that all three will have 
decay rates of comparable magnitude, modulo the difference in the 
masses of the scalars $h^0$, $A^0$ and $H^\pm$. Since we assume 
masses of $h^0$, $A^0$ and $H^\pm$ as 40 GeV, 140 GeV and 170 GeV 
respective, the decay rate for $\Sigma^\pm \rightarrow \nu H^\pm$
is predicted to be the lowest. 

\subsection{ Decay to Light Leptons and Vector Bosons}

The exotic heavy leptons have gauge interactions. Therefore, 
it is expected that they will also decay into final state 
particles with vector bosons, $W^\pm$ and $Z$. 
The decay width $\Gamma$ for 
$\Sigma^{\pm}_i \rightarrow l_j^{\pm}/ \nu V$ and  
$\Sigma^{0}_i \rightarrow l_j^{\pm}/ \nu V $ in the $m_l=0$ 
limit is given by
\be
\Gamma = \frac{M_{\Sigma_i}}{32\pi} 
\bigg[1-\frac{{M_{V}}^2}{{M_{\Sigma}}^2}\bigg]^2 \,
\bigg[2+\frac{{M_{\Sigma}}^2}{{M_{V}}^2}\bigg]
\bigg(|{(C_{l^\pm \Sigma}^{V,L})}_{ji}|^2+
|{(C_{l^\pm \Sigma}^{V,R})}_{ji}|^2\bigg)
,
\label{eq:decayrate2}
\ee
where ${C_{l^\pm\Sigma}^{V,L}}$ and ${C_{l^\pm\Sigma}^{V,R}}$ 
are the relevant vertex factors given in Appendix B.2, 
and $M_V$ is the mass of the vector boson involved. 
The dominant vertex factor relevant for 
$\Sigma^{\pm} \rightarrow l^{\pm}Z$ and 
$\Sigma^0 \rightarrow l^{\pm}W^{\mp}$ in terms of $Y_\Sigma$, $M$, 
$v'$ and the mixing matrices are given respectively by 
\be
C_{l^\pm \Sigma^\pm}^{Z,L} 
\simeq \frac{v^{\prime}}{2}\,\frac{g}{c_w}\, 
{U_l}^{\dagger}Y_{\Sigma}^{\dagger}M^{-1}U_{h}^L
,~~{\rm and}~~
C_{l^\pm \Sigma^0}^{W^\mp,L}
\simeq -\frac{v^{\prime}}{2}\,g\, {U_l}^{\dagger}Y_{\Sigma}^{\dagger}M^{-1}
U_{\Sigma}
.
\label{eq:vert1}
\ee
For the other two channels  $\Sigma^{0} \rightarrow \nu Z$ 
and  $\Sigma^{\pm} \rightarrow \nu W^{\pm}$, they are 
given respectively by
\be
C_{\nu \Sigma^0}^{Z,L} 
\simeq \frac{v^{\prime}}{2\sqrt{2}}(gc_w+g^{\prime}s_w)
{U_{PMNS}^{\dagger}}Y_{\Sigma}^{\dagger}M^{-1} U_{\Sigma}
,~~{\rm and}~~
C_{\nu \Sigma^\pm}^{W^\mp,R}
\simeq -\frac{v^{\prime}}{\sqrt{2}}\,g\,
{U_{PMNS}^T }Y_{\Sigma}^{T}M_{}^{-1}U_{h}^R
.
\label{eq:vert2}
\ee
As mentioned before, the gauge interaction part of our model is 
identical to that for the one Higgs doublet Type III seesaw 
considered earlier. Some of 
these vertex factors\footnote{Vertex factor for 
$\Sigma^{0} \rightarrow \nu Z$ is not given in \cite{type3muegamma}.}
can therefore can be 
seen to agree with that given in \cite{type3muegamma}. The only 
difference is that we include the matrices $U_l$, 
$U_h^R$ and $U_\Sigma$ in our general 
expressions, while these were taken 
as unit matrices in \cite{type3muegamma}. 

\subsection{ Comparing $\Sigma^{\pm/0}$ Decays to Higgs and Gauge Bosons}

\begin{figure}[t]
\begin{center}
\includegraphics[height=10.0cm,angle=0]{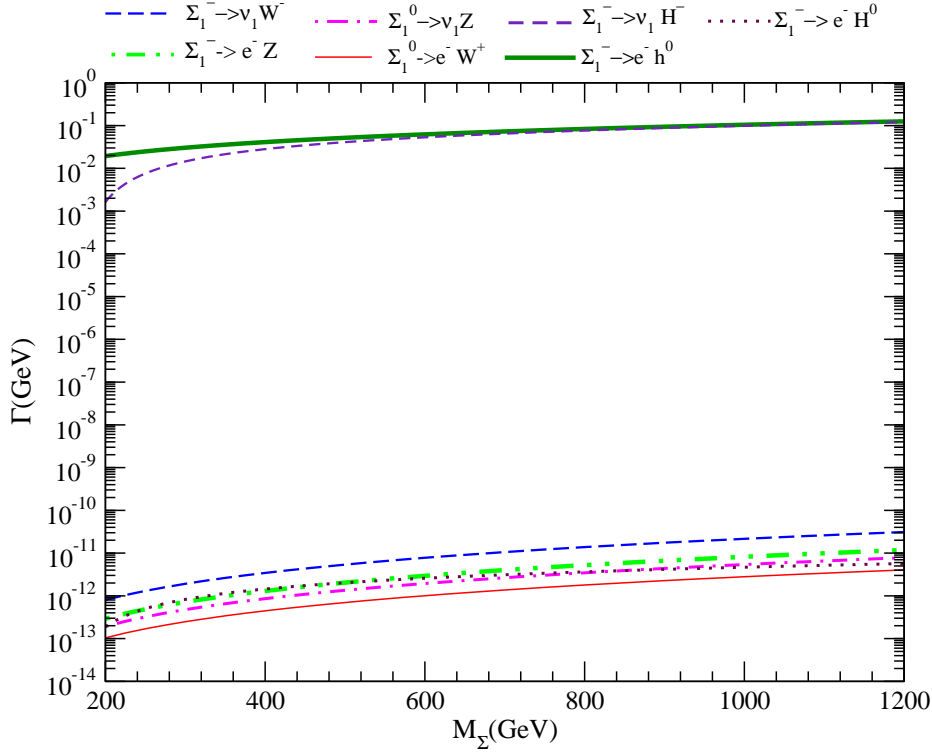}
\caption{\label{fig:compareHZ}
Comparison of the decay rate of the heavy fermion into (i) Higgs and 
(ii) vector bosons, in our two Higgs doublet model. 
}
\end{center}
\end{figure}

In Fig. \ref{fig:compareHZ} we show the decay rates 
$\Sigma^-_1 \rightarrow \nu_1 W^-$ (long-dashed blue line), 
$\Sigma^-_1 \rightarrow e^- Z$ (dot-dashed green line), 
$\Sigma^0_1 \rightarrow \nu_1 Z$ (dot-dashed magenta line), 
$\Sigma^0_1 \rightarrow e^- W^+$ (thin solid red line), 
$\Sigma^-_1 \rightarrow e^- H^0$ (dotted maroon line), 
$\Sigma^-_1 \rightarrow \nu_1 H^-$ (dashed violet line), and 
$\Sigma^-_1 \rightarrow e^- h^0$ (thick solid dark green line). 
One can clearly see that all decays to gauge bosons are 
suppressed with respect to decays to $h^0$ (and $A^0$) and 
$H^\pm$ by a factor of more than $10^{10}$. The reason for 
this can be seen by comparing the vertex factors 
involved in decays to Higgs $h^0$, $A^0$ and $H^\pm$
(cf. Eqs. (\ref{eq:vfactdch}), (\ref{eq:vfactdch1}), 
(\ref{eq:vfactdch2}), (\ref{eq:vfactdch3}), (\ref{eq:vfactdch4})), 
with decays to gauge bosons 
(cf. Eqs. (\ref{eq:vert1}) and (\ref{eq:vert2})). 
It is clear that while the former vertex factors 
do not have any suppression factor, 
the latter are all suppressed by $v'/M_\Sigma$. Another important 
difference between the decay rates to Higgs given in Eq. 
(\ref{eq:decayrate1}), and 
gauge bosons given in Eq. (\ref{eq:decayrate2}), is in the 
kinematic factors. Comparison of the two equations reveals 
that (for $m_l=0$), there is an additional factor of 
$(2+M_\Sigma^2/M_V^2)$ for the gauge boson decays. 
This extra $M_\Sigma^2$ in the numerator cancels out the 
$1/M_\Sigma^2$ in the denominator coming from the square of 
the vertex factors. However, the suppression of the 
gauge boson decay rates due to $g^2{v'}^2/M_V^2 \propto v'^2/V^2$ 
remains, where $V^2=v'^2 + v^2$. Since 
we have taken $v' \sim 10^{-3}$-$10^{-4}$ GeV, the decays to gauge 
bosons are suppressed by a factor of $\sim 10^{10}$-$10^{12}$ 
compared to 
the decays to Higgs. Therefore, branching ratios of the 
heavy fermion decay to $W^\pm$ and $Z$ can be neglected 
in our model and we concentrate on only decays to $h^0$, 
$A^0$ and $H^\pm$ in our next section. Note that the decay to 
$H^0$ is also suppressed by a factor of $10^{10}$-$10^{12}$, as 
was also pointed out earlier. We had seen that this suppression 
is due to $\sin^2\alpha$ coming from the vertex factor 
for this process. Since $\sin^2\alpha \sim 10^{-12}$, we find 
that the decay rate for this case is of the same order of magnitude 
as the decays to the gauge bosons. Hence, this is also 
neglected henceforth.

\subsection{ Comparison Between One and Two Higgs Doublet Models}

It is pertinent to 
compare the two-body decays of the heavy fermions 
in our two Higgs doublet model with the usual 
Type III seesaw models considered 
earlier which have one Higgs doublet. The expressions for heavy 
fermion decays to Higgs and gauge bosons 
in the one Higgs doublet models have been given 
before in the literature 
\cite{type3lhc2,type3lhchambyestrumia,type3lhcaguila,type3lhc3}, 
and we give them here for the sake of comparison. 
The decay rates to gauge bosons 
in the one Higgs doublet model is given as (for $m_l=0$)
\be
\Gamma^{1HDM}(\Sigma^0 \to \nu Z ) &\simeq& 
\frac{\lambda^2\,M_\Sigma}{64\pi}
(1- \frac{M_Z^2}{{M_{\Sigma}}^2})^{2}(1+2\frac{M_Z^2}{{M_{\Sigma}}^2}),\\
\Gamma^{1HDM}(\Sigma^0 \rightarrow l^{\mp} W^{\pm} )&\simeq& 
\frac{\lambda^2\,M_\Sigma}{32\pi}
(1- \frac{M_W^2}{{M_{\Sigma}}^2})^{2}(1+2\frac{M_W^2}{{M_{\Sigma}}^2}),\\ 
\Gamma^{1HDM}(\Sigma^{\pm} \rightarrow l^{\pm}Z ) &\simeq&
 \frac{\lambda^2\,M_\Sigma}{32\pi}
(1- \frac{M_Z^2}{{M_{\Sigma}}^2})^{2}(1+2\frac{M_Z^2}{{M_{\Sigma}}^2}),\\ 
\Gamma^{1HDM}(\Sigma^{\pm} \rightarrow \nu W^{\pm} ) &\simeq&
  \frac{\lambda^2\,M_\Sigma}{16\pi}
(1- \frac{M_Z^2}{{M_{\Sigma}}^2})^{2}(1+2\frac{M_Z^2}{{M_{\Sigma}}^2}).
\ee
where $\lambda$ is the triplet fermion -- lepton doublet -- 
Higgs doublet  
Yukawa coupling in the one Higgs doublet model, and all 
mixing terms are neglected. This should be compared with the 
corresponding expression given in Eq. (\ref{eq:decayrate2}), 
which on neglecting all mixing and hence flavor effects reduces 
to (for $m_l=0$)
\be
\Gamma^{2HDM}(\Sigma^0 \to \nu Z ) &\simeq& 
\frac{y_\Sigma^2\,M_\Sigma}{64\pi}\frac{{v'}^2}{V^2}
(1- \frac{M_Z^2}{{M_{\Sigma}}^2})^{2}(1+2\frac{M_Z^2}{{M_{\Sigma}}^2}),\\
\Gamma^{2HDM}(\Sigma^0 \rightarrow l^{\mp} W^{\pm} )&\simeq& 
\frac{y_\Sigma^2\,M_\Sigma}{32\pi}\frac{{v'}^2}{V^2}
(1- \frac{M_W^2}{{M_{\Sigma}}^2})^{2}(1+2\frac{M_W^2}{{M_{\Sigma}}^2}),\\ 
\Gamma^{2HDM}(\Sigma^{\pm} \rightarrow l^{\pm}Z ) &\simeq&
 \frac{y_\Sigma^2\,M_\Sigma}{32\pi}\frac{{v'}^2}{V^2}
(1- \frac{M_Z^2}{{M_{\Sigma}}^2})^{2}(1+2\frac{M_Z^2}{{M_{\Sigma}}^2}),\\ 
\Gamma^{2HDM}(\Sigma^{\pm} \rightarrow \nu W^{\pm} ) &\simeq&
  \frac{y_\Sigma^2\,M_\Sigma}{16\pi}\frac{{v'}^2}{V^2}
(1- \frac{M_Z^2}{{M_{\Sigma}}^2})^{2}(1+2\frac{M_Z^2}{{M_{\Sigma}}^2}).
\ee
where $V^2=v^2 + {v'}^2$ is the electroweak breaking scale. 
The scale of the Yukawa coupling constants and VEVs are fixed by 
the neutrino mass $m_\nu \sim \lambda^2 V^2/M_\Sigma$ for the 
one Higgs doublet model and $m_\nu \sim y_\Sigma^2 {v'}^2/M_\Sigma$. 
If one replaces $\lambda^2$ and $y_\Sigma^2 {v'}^2/V^2$ with 
$m_\nu M_\Sigma/V^2$ in both set of expressions, one can see that the 
the decay rates of heavy fermions into gauge bosons are 
identical for both models.

\vglue 0.4cm
\noindent
The rates for decay into Higgs for the one Higgs doublet
model neglecting flavor effects, is given by (for $m_l=0$)
\be
\Gamma^{1HDM}(\Sigma^0 \to \nu H^0) &\simeq& 
\frac{\lambda^2\,M_\Sigma}{64\pi}
(1- \frac{M_H^2}{{M_{\Sigma}}^2})^{2},\\
\Gamma^{1HDM}(\Sigma^{\pm} \to l^\pm H^0 )
&\simeq& 
\frac{\lambda^2\,M_\Sigma}{32\pi}
(1- \frac{M_H^2}{{M_{\Sigma}}^2})^{2}. 
\ee
For the two Higgs doublet model, the corresponding decay rates are 
given by Eq. (\ref{eq:decayrate1}), 
which on neglecting all flavor effects reduces 
to (for $m_l=0$)
\be
\Gamma^{2HDM}(\Sigma^0 \to \nu h^0/A^0) &\simeq& 
\frac{y_\Sigma^2\,\cos^2\alpha\,M_\Sigma}{64\pi}
(1- \frac{M_{h/A}^2}{{M_{\Sigma}}^2})^{2},\\
\Gamma^{2HDM}(\Sigma^{\pm} \to l^\pm h^0 /A^0)
&\simeq& 
\frac{y_\Sigma^2\,\cos^2\alpha\,M_\Sigma}{32\pi}
(1- \frac{M_{h/A}^2}{{M_{\Sigma}}^2})^{2},
\\
\Gamma^{2HDM}(\Sigma^0 \to \nu H^0) &\simeq& 
\frac{y_\Sigma^2\,\sin^2\alpha\,M_\Sigma}{64\pi}
(1- \frac{M_H^2}{{M_{\Sigma}}^2})^{2},\\
\Gamma^{2HDM}(\Sigma^{\pm} \to l^\pm H^0 )
&\simeq& 
\frac{y_\Sigma^2\,\sin^2\alpha\,M_\Sigma}{32\pi}
(1- \frac{M_H^2}{{M_{\Sigma}}^2})^{2},
\ee
where the first two expressions are for decays to $h^0$ or $A^0$ 
and the last two for decays to $H^0$. Again, for the same value of 
$M_\Sigma \sim 100$ GeV in both models, one requires 
$\lambda \sim 10^{-5}$-$10^{-6}$ 
for the one Higgs doublet model in order to produce $m_\nu\sim 0.1$ eV, 
while $y_\Sigma \sim 1$ for our  
two Higgs doublet model. Therefore, clearly 
$$\Gamma^{2HDM}(\Sigma^0 \to \nu h^0/A^0) \sim 10^{11}\,\times
\Gamma^{1HDM}(\Sigma^0 \to \nu H^0),$$
$$\Gamma^{2HDM}(\Sigma^\pm \to l^\pm h^0/A^0) \sim 10^{11}\,\times
\Gamma^{1HDM}(\Sigma^\pm \to l^\pm H^0).$$
Hence, the the exotic fermions decay about $ 10^{11}$ times 
faster in our model compared to the one Higgs doublet 
model\footnote{We reiterate that the decays $\Sigma^0 \to \nu H^0$ and 
$\Sigma^{\pm} \to l^\pm H^0$ are suppressed by the 
$\sin^2\alpha \sim 10^{-12}$ factor and hence turn out to be 
comparable to the decay rates in the one Higgs doublet model. 
However, the branching ratio to this mode is negligible 
and can be neglected.}.
This could lead to observational consequences at 
LHC. In particular, authors of \cite{type3lhchambyestrumia} 
talk about using ``displaced vertices'' as a signature of the 
Type III seesaw mechanism. In our model the lifetime of the exotic 
fermions is a factor of $10^{11}$ shorter and so will be the 
gap between their primary production vertex and the decay vertex. 
Our model therefore predicts no displaced vertex for the 
heavy fermion decays. In addition, decay to $h^0$ are 
predominant. The $h^0$ decay predominantly into $b\bar b$ pairs, 
but with a very long lifetime, as 
we will discuss in section 6. This will give a distinctive 
signature of our model at LHC. We will discuss 
displaced vertices from $h^0$ decay in section 7.

\subsection{Flavor Structure and the Decay Branching Ratios}

\begin{table}[h]
\begin{center}
\begin{tabular}{||c|c|c|c||} \hline\hline
\
Decay modes & $\Sigma^\pm_1$ & $\Sigma^\pm_2$& $\Sigma^\pm_3$\\ \hline
\hline
$\nu\,H^\pm$ & 0.363 & 0.473 &0.473\\
\hline
$ e^\pm\,A^0$ & 0.247 & $2.28 \times 10^{-6}$&0.0\\
\hline
$\mu^\pm\,A^0 $ & $2.3 \times 10^{-6}$ & $0.125$ &$0.125$\\
\hline
$\tau^\pm\,A^0 $ & $2.3 \times 10^{-6}$ & $0.125$ &$ 0.125$\\
\hline
$e^\pm\,h^0 $ & 0.389 & $2.5 \times 10^{-6}$&0.0\\
\hline
$\mu^\pm\,h^0 $ & $3.6 \times 10^{-6}$ & 0.139& 0.139\\
\hline
$\tau^\pm\,h^0 $ &$3.6 \times 10^{-6}$ & 0.139&0.139\\

\hline
\hline
\end{tabular}
\caption{\label{tab:sigcbr}Decay branching fractions of $\Sigma^\pm_1$, 
$\Sigma^\pm_2$ and $\Sigma^\pm_3$ for $M_{h^0}$=40, $M_{H^0}$=150, 
$M_{H^{\pm}}=170$ GeV  and $M_{A^0}=140$ GeV. We have taken 
model parameters $M_1=300$ GeV and $M_2=M_3=600$ GeV.}
\end{center}
\end{table}
\begin{table}[]
\begin{center}
\begin{tabular}{||c|c|c|c||} \hline\hline
\
Decay modes & $\Sigma^0_1$ & $\Sigma^0_2$& $\Sigma^0_3$\\ \hline
\hline
$e^{\mp}\,H^{\pm}$ & 0.368 & $4.3\times 10^{-6}$ &0.0\\
\hline
$\mu^{\mp}\,H^{\pm}$ & $3.4 \times 10^{-8}$ & 0.236 &0.236\\
\hline
$\tau^{\mp}\,H^{\pm}$ & $3.4 \times 10^{-8}$ & 0.236&0.236\\
\hline
$ \nu\,A^0$ & 0.243 & 0.250&0.250\\
\hline
$ \nu\,h^0$ & 0.386 & 0.277&0.277\\
\hline
\hline
\end{tabular}
\caption{\label{tab:sigmazbr}Decay branching fractions of $\Sigma^0_1$,
  $\Sigma^0_2$ and $\Sigma^0_3$ for $M_{h^0}$=40, $M_{H^0}$=150,
  $M_{H^{\pm}}=170$ GeV   and $M_{A}=140$ GeV. We have taken 
model parameters $M_1=300$ GeV and $M_2=M_3=600$ GeV.}
\end{center}
\end{table}

Finally, we present the branching fractions of the heavy fermion 
decays. Table \ref{tab:sigcbr} shows the branching fractions for 
the $\Sigma^\pm$, while Table \ref{tab:sigmazbr} gives the 
branching fraction for $\Sigma^0$ decays. For the channels with 
neutrino in the final state, we give the sum of the branching 
fraction into all the three generations, as observationally 
it will be impossible to see the neutrino generations at LHC. 
We do not show decays to gauge bosons and $H^0$ as they are 
suppressed by a factor of $10^{11}$ with respect to the 
decays into $h^0$, $A^0$ and $H^\pm$. 
As a result of the inherent $\mu$-$\tau$ symmetry in the model, 
$\Sigma_3^{\pm/0}$ decays to electrons is strictly forbidden and 
branching ratios of their decay into $\mu$ and $\tau$ leptons are equal. 
We find that due to the form of $U_{22}$, $S_{22}$ and $T_{22}$ 
given in Eqs. (\ref{eq:u22}), (\ref{eq:s22}) and (\ref{eq:t22}),
$\Sigma_2^{\pm/0}$ 
decays to electrons is also negligible and 
their probability to decay into $\mu$ and $\tau$ leptons is equal. 
We also find that the branching 
fractions of $\Sigma_2^\pm$ is equal to the branching 
fractions of $\Sigma_3^\pm$, and similarly for the 
neutral heavy fermions. We also notice that 
$\Sigma_1^{\pm/0}$ decays only to electrons and 
their decay to $\mu$ and $\tau$ lepton is almost zero. This 
as pointed out before, comes due to the constraint on the Yukawa 
couplings from the low energy neutrino oscillation data. 
The difference between the branching 
fraction to $h^0$, $A^0$ and $H^\pm$ is mainly driven 
by the difference in the masses which we have chosen for these 
Higgses.

\section{Higgs Decay}

In the previous section we concluded that the heavy fermions will 
all decay into $h^0$, $A^0$ or $H^\pm$.  
We next turn to the subsequent decay of these 
Higgs particles. 
We concentrate on 
the possible decay modes of $h^0$, $A^0$ 
and $H^{\pm}$ and tabulate only
those few which have significant branching ratios. 
The branching ratios obviously depend on our choice for the 
Higgs masses as well as our choice of the mixing angles 
$\alpha$ and $\beta$, which appear in the coupling. 
The part of the Lagrangian containing the 
interaction terms of the Higgs with the leptons and quarks are 
given in Appendix B. The interaction of Higgs fields with the 
gauge fields comes from the Higgs kinetic terms and is the 
same as the general two Higgs doublet model. Possible 
decay channels for the charged Higgs involve the $W^\pm$ and 
the neutral CP even Higgs. 
It is well known that in the two Higgs doublet model, 
the $W^{\pm}-H^{\mp}-H^0$ coupling is 
proportional to $\sin(\beta-\alpha)$, whereas 
$W^{\pm}-H^{\mp}-h^0$ coupling is proportional to $\cos(\beta-\alpha)$
\cite{hunter}. In Appendix A, we have 
shown how constraint from neutrino mass drives 
$\sin \alpha \sim \sin \beta \sim
10^{-6}$. Therefore, in our model $H^{\pm}\to W^{\pm}H^0$ is 
always suppressed, irrespective of the Higgs mass\footnote{One 
can see that this channel is also kinematically forbidden for our 
choice of the Higgs masses, whereby with $M_{H^{\pm}}=170$, it is 
impossible to create an on-shell pair of $W^{\pm}$ and $H^0$.}. 
In fact, the only decay channel possible for the charged 
Higgs in our model is $H^{\pm}\to W^{\pm}h^0$, for which the 
decay branching fraction
\be
BR(H^{\pm}\to W^{\pm}h^0) = 1.0
\label{eq:brchargedhiggs}
.
\ee
The $W^\pm$ next decay into either $qq'$ pairs or 
$l^\pm \nu_l/\bar\nu_l$ pairs with the following decay branching fractions
\be
BR(W^\pm \to qq') = 0.67,
\nn\\
BR(W^\pm \to e^\pm \nue/\anue) = 0.11,
\nn\\
BR(W^\pm \to \mu^\pm \numu/\anumu) = 0.11,
\nn\\
BR(W^\pm \to \tau^\pm \nutau/\nutau) = 0.11
.
\ee
The branching fractions of the neutral Higgs $h^0$, $H^0$ and $A^0$ 
are given in Table \ref{tab:tab1}. Though $H^0$ is almost never produced 
through heavy fermion decays in our model, we have included them 
in the table for completeness. We find that the neutral Higgs 
decay to $b \bar{b}$ pairs almost 90\% of the times. The second 
largest decay fraction is to $\tau\bar\tau$ pairs, while 
decays to $c\bar c$ happens less than few percent of the times. 
In our following sections where we look for collider signatures, we 
will consider $h^0$ (and $A^0$) decays to only $b \bar{b}$ 
and $\tau\bar\tau$ pairs. 

Finally, a short discussion on direct production of  
$h^0$, without involving the heavy fermion decays, is in order. 
In our model, the lightest Higgs has a mass as low as 40 GeV. 
This might appear to be a cause of concern, given that 
such a Higgs was not observed at LEP. 
However, it is easy to see that this Higgs mass is  
not excluded by the direct Higgs searches at 
LEP-2. This is because the 
coupling corresponding to $Z-Z-h^0$ vertex is given by 
$(gM_Z/\cos\theta_w)\sin(\beta-\alpha)$. Since 
in our model $\sin(\beta-\alpha)$ is almost zero, the LEP-2 
bound on Higgs mass does not pose any serious threat to our model, 
irrespective of the mass of $h^0$.



\begin{table}
\begin{center}
\begin{tabular}{||c|c|c|c||} \hline\hline

Decay modes & $h^0$ & $H^0$& $A^0$\\ \hline
\hline
$b\bar{b}$ & 0.89 & 0.87 &0.87\\
\hline
$\tau\bar{\tau}$ & 0.07 & 0.09 &0.09\\
\hline
$c\bar{c}$ & 0.04 & 0.04&0.04\\
\hline\hline
\end{tabular}
\caption{\label{tab:tab1}Decay branching fractions of 
$h^0$, $H^0$ and $A^0$ for $M_{h^0}=40$ GeV, $M_{H^0}=150$ GeV,   
and $M_{A^0}=140$ GeV.}
\end{center}
\end{table}

\section{Displaced $h^0$ Decay Vertex}

Amongst the most significant difference of our model with the 
usual Type III seesaw model are the decay lifetimes of 
the heavy fermions and $h^0$ (as well as $A^0$). 
We had seen in section 5 that the total decay rate for 
300 GeV $\Sigma^{0}$ is about $4\times 10^{-2}$ GeV. This 
gives the corresponding rest frame lifetime as 
$4.97\times 10^{-13}$ cm. The lifetime for $\Sigma^\pm$ is 
similar. 
One can check that 
for the usual one Higgs doublet models, the rest frame lifetime
for the heavy fermions 
is $\simeq 0.5$ cm for $m_\nu=0.1$ eV and $M_\Sigma\sim 100$ GeV, 
which is rather large.  
The authors of \cite{type3lhchambyestrumia} therefore proposed 
that the displaced decay vertex of heavy fermion could be a 
typical signature of the one Higgs Type III seesaw model. 
Clearly, for our model with two Higgs doublets, the decay 
lifetime is $10^{11}$ times smaller and hence we predict no 
displaced vertex for the heavy fermion decay. This can be 
used as a distinguishing signature between the two models. 

Another very important and unique feature of our model is the 
very long lifetime of our neutral Higgs $h^0$, which comes 
due to the smallness of $\sin\alpha$. In fact, since 
$\sin\alpha\sim 10^{-6}$, the lifetime for $h^0$ in our model 
is $10^{12}$ times larger compared to the standard model Higgs. 
In particular, the $h^0$ total decay rate is $4\times 10^{-15}$ GeV. 
This gives $h^0$ a rest frame lifetime of $4.97$ cm. For a
$h^0$ with 200 GeV of energy, the lifetime in the lab frame 
is seen to be close to 25 cm. 
Therefore, we expect a big gap between the decay vertices of 
the heavy fermion and the $h^0$. This displaced $h^0$ 
decay vertex should be detectable at the LHC detectors 
ATLAS and CMS. 

We would like 
to make just a few qualitative remarks about the prospects 
of detecting the displaced $h^0$ vertex. 
Like stressed many times before, 
the $h^0$ decay predominantly into $b\bar b$ pairs. 
While $b$-tagging is a very important and 
standard tool for collider experiments, and while 
both ATLAS \cite{atlas}  and CMS  \cite{cms} have been developing 
algorithms for tagging the $b$, there is an additional complication 
with $b$-tagging in our model which should be pointed out here. 
Since the $h^0$ lifetime is a few 10s of cm in the lab frame, 
it is expected to decay inside the silicon tracker of ATLAS and 
CMS. In particular, the pixel tracker of CMS and ATLAS which are only 
few cm from the center of the beam pipe, will miss the $h^0$ decay 
vertex. However, the silicon strip trackers would be useful in 
observing the $b$-jets. The tracks from 
the primary and secondary vertices of the $b$-hadron should be seen. 
In addition, one could use the two other 
standard tools for tagging the 
$b$-jets. Firstly, one could the tag the lepton in the jet coming from 
the semi-leptonic decays of the $b$-hadron. These leptons are expected 
to have smaller $p_T$ compared to the ones coming from 
$W^\pm$ and $Z$ decays, and hence this is called soft-lepton 
tagging \cite{atlas,cms}. More importantly, one could construct the 
invariant mass distribution of the 2 $b$-jets. This should give us a 
sharp peak corresponding to the $h^0$. We therefore expect
that ATLAS and CMS should be able to detect the displaced $h^0$ 
decay vertex. This would give a characteristic and unambiguous 
signal of our model. 

Note that while the lifetime of $h^0$ is constrained to be 
large due to the smallness of $\sin\alpha$ alone 
in our model, things are 
slightly more complicated for the lifetime of $A^0$. This is because 
in principle $A^0$ could decay through the mode $A^0 \to Z h^0$. 
While this is forbidden kinematically for the $A^0$ mass we assume, 
one could argue that for a large 
enough mass for $A^0$, the lifetime of $A^0$ could be shorter. 
However, we stress that even if $A^0$ decays fast into $h^0$, 
that would still produce a displaced vertex, since the $h^0$ 
would still have a very long lifetime.

\section{Model Signatures at the LHC}

Having discussed in details the production and subsequent decays 
of the exotic fermions, as well as the decay branching fractions of 
the intermediate Higgs into final state particles, we next 
describe the signatures of the two Higgs doublet 
Type III seesaw model at the LHC. We will present a comprehensive 
list of final state particles and their corresponding collider 
signatures. 

The most important characteristics of our model are the following:
\begin{enumerate}
\item Presence of $\mu$-$\tau$ symmetry in $Y_\Sigma$ and 
$M$. This is expected to show-up in the flavor of the 
final state lepton coming directly from the $\Sigma^{\pm/0}$ decay
vertex. 
\item Presence of two CP even neutral Higgses ($h^0$ and $H^0$), 
one CP odd neutral
Higgs ($A^0$), and a pair of charged Higgs ($H^\pm$).
\item Predominant decay of the heavy fermions into  
light leptons, and $h^0$, $A^0$ or $H^\pm$. 
Decays into $H^0$ and gauge bosons almost never happen. 
\item Very short lifetime for the heavy fermion due to the 
very large Yukawa couplings. 
\item Predominant decay of $h^0$ and $A^0$ into 
$b\bar b$ pairs 89\% and 87\% of the time, respectively. 
They decay also into $\tau\bar\tau$ 7-9\% of the time. 
\item Very large lifetime of $h^0$ and $A^0$.  
\item The Higgs $H^\pm$ decays into $W^\pm h^0$ and almost 
never into  $W^\pm H^0$. 
\item Short predicted lifetime for $H^\pm$. 
\end{enumerate}
In what follows, we will use these model characteristics to 
identify distinctive final state channels at the collider. 
We identify {\it all} possible channels in the collider 
for our model and calculate the respective effective cross-sections.  
The results are given in Tables \ref{tab:sigpm}, \ref{tab:sig0ch}
and \ref{tab:sig0ch2}. We will also discuss some of the 
most important channels and the characteristic backgrounds, if 
any, associated with them.  
In this section we have only 
given results for effective cross-sections 
for the decay of $\Sigma^{\pm/0}_1$ with 
$M_{\Sigma_1}=300$ GeV.  
Results for the other heavy fermion generations 
can be similarly obtained. 



\subsection{Signatures from $\Sigma^{+}\Sigma^{-}$ decays}
\label{sec:colldk1}

\begin{table}[p]
\begin{center}
\renewcommand{\arraystretch}{1.4}
\begin{tabular}{||c|c|c||}
\hline\hline
Sl no &Channels &Effective cross-section (in fb) \\
\hline\hline
1&$\Sigma^+ \Sigma^- \to l^+l^-h^0h^0 \to 4b+OSD$& 35.84  \\\hline
2&$\Sigma^+ \Sigma^- \to l^+l^-h^0h^0 \to 2b+OSD+2\tau$&3.67 \\\hline
3&$\Sigma^+ \Sigma^- \to l^+l^-h^0h^0 \to OSD+4\tau$ & 0.37\\ \hline
\hline
4&$\Sigma^+ \Sigma^- \to l^+h^0 H^{-} \nu \to  4b+l+2j+\not{p_T}$& 26.88
\\\hline
5&$\Sigma^+ \Sigma^- \to l^+h^0 H^{-} \nu \to  4b+OSD(l+l')+\not{p_T}$& 8.92
\\\hline
6&$\Sigma^+ \Sigma^- \to l^+h^0 H^{-} \nu \to  4b+l+\tau+\not{p_T}$& 4.48
\\\hline
7&$\Sigma^+ \Sigma^- \to l^+h^0 H^{-} \nu \to  2b+l+2\tau+2j+\not{p_T}$& 2.69
\\\hline
8&$\Sigma^+ \Sigma^- \to l^+h^0 H^{-} \nu \to  2b+l+3\tau+\not{p_T}$& 0.45
\\\hline
9&$\Sigma^+ \Sigma^- \to l^+h^0 H^{-} \nu \to  2b+OSD(l+l')+2\tau+\not{p_T}$&0.9
\\\hline
10&$\Sigma^+ \Sigma^- \to l^+h^0 H^{-} \nu \to  l+4\tau+2j+\not{p_T}$& 0.28
\\\hline
11&$\Sigma^+ \Sigma^- \to l^+h^0 H^{-} \nu \to  OSD(l+l')+4\tau+\not{p_T}$&0.04
\\\hline
12&$\Sigma^+ \Sigma^- \to l^+h^0 H^{-} \nu \to  l+5\tau+\not{p_T}$&0.02
\\ \hline\hline
13&$\Sigma^+ \Sigma^- \to H^{+}\nu H^{-} \nu \to  4b+4j+\not{p_T}$& 15.68
\\\hline
14&$\Sigma^+ \Sigma^- \to H^{+}\nu H^{-} \nu \to  4b+2j+l'+\not{p_T}$& 10.52
\\\hline
15&$\Sigma^+ \Sigma^- \to H^{+}\nu H^{-} \nu \to  4b+2j+\tau+\not{p_T}$& 5.26
\\\hline
16&$\Sigma^+ \Sigma^- \to H^{+}\nu H^{-} \nu \to  4b+OSD'+\not{p_T}$& 0.86
\\\hline
17&$\Sigma^+ \Sigma^- \to H^{+}\nu H^{-} \nu \to  4b+2\tau+\not{p_T}$& 0.43
\\\hline
18&$\Sigma^+ \Sigma^- \to H^{+}\nu H^{-} \nu \to  4b+1\tau+1l'+\not{p_T}$& 0.53
\\\hline
19&$\Sigma^+ \Sigma^- \to H^{+}\nu H^{-} \nu \to  2b+2\tau+4j+\not{p_T}$& 3.25
\\\hline
20&$\Sigma^+ \Sigma^- \to H^{+}\nu H^{-} \nu \to  2b+2\tau+2j+l'+\not{p_T}$& 2.12
\\\hline
21&$\Sigma^+ \Sigma^- \to H^{+}\nu H^{-} \nu \to  2b+3\tau+2j+\not{p_T}$& 1.06
\\\hline
22&$\Sigma^+ \Sigma^- \to H^{+}\nu H^{-} \nu \to  2b+2\tau+OSD'+\not{p_T}$& 0.32
\\\hline
23&$\Sigma^+ \Sigma^- \to H^{+}\nu H^{-} \nu \to  2b+4\tau+\not{p_T}$& 0.08
\\\hline
24&$\Sigma^+ \Sigma^- \to H^{+}\nu H^{-} \nu \to  2b+3\tau+l'+\not{p_T}$& 0.02
\\\hline
25&$\Sigma^+ \Sigma^- \to H^{+}\nu H^{-} \nu \to  4\tau+4j+\not{p_T}$& 0.15
\\\hline
26&$\Sigma^+ \Sigma^- \to H^{+}\nu H^{-} \nu \to  4\tau+2j+l'+\not{p_T}$& 0.10
\\\hline
27&$\Sigma^+ \Sigma^- \to H^{+}\nu H^{-} \nu \to  5\tau+2j+\not{p_T}$& 0.05
\\\hline
28&$\Sigma^+ \Sigma^- \to H^{+}\nu H^{-} \nu \to  5\tau+l'+\not{p_T}$& 0.006
\\\hline
29&$\Sigma^+ \Sigma^- \to H^{+}\nu H^{-} \nu \to  4\tau+OSD'+\not{p_T}$& 0.02
\\\hline
30&$\Sigma^+ \Sigma^- \to H^{+}\nu H^{-} \nu \to  6\tau+\not{p_T}$& 0.005
\\\hline\hline
\end{tabular}
\caption{\label{tab:sigpm}Effective cross-sections (in fb) 
for different $\Sigma^+\Sigma^-$ 
decay channels for $M_{\Sigma_1}=300$ GeV.}
\end{center}
\end{table}

We give in Table \ref{tab:sigpm} all possible collider 
signatures coming from the decay of $\Sigma^+\Sigma^-$ pairs, 
for our two Higgs doublet Type III seesaw model. In the last 
column we also give the corresponding effective 
cross-sections for these channels in units of fb. 
Of course the final cross-sections can be 
obtained only after putting in the various cuts and 
efficiency factors. These efficiency factors will have
to be folded with the cross-sections given in Table \ref{tab:sigpm} 
to get the final  effective 
cross-sections for the various channels. Few clarifications 
on our notation is in order. Light charged leptons could be  
released in the final state through two ways: (i) from the 
decay of the heavy fermions $\Sigma^{\pm} \to l^\pm h^0$ and 
$\Sigma^{0} \to l^\pm H^\mp$, (ii) from 
the decays of $W \to l\bar\nu_l$. The charged leptons  
released from the $\Sigma^{\pm/0}$ decays are different  from those 
from $W^\pm$ in two respects. Firstly, the former 
carry the information on the flavor structure of the model as 
discussed in the previous sections, while the latter do not. 
Secondly, since they come from decays of the heavier $\Sigma ^{\pm/0}$, 
they are expected to be harder than the ones from $W^\pm$ decays. 
We refer to the charged leptons from the $\Sigma^{\pm/0}$ decays as 
$l$ and the ones from $W^\pm$ decays as $l'$. The notation $OSD$ stands 
for opposite sign dileptons from $\Sigma^{\pm/0}$ decays, while 
$OSD'$ stands for opposite sign dileptons from $W^\pm$ decays. 
When we have one charged lepton from $\Sigma^{\pm/0}$ decay 
and an opposite sign charged lepton from $W^\pm$ decay, then 
it is denoted as $OSD(l+l')$ and so on.  
\footnote{
We should also mention at this point 
that for some cases  
whether the charged lepton in the final state is a 
$l$ or $l'$ can be said from the detector topology 
only after proper cuts have been imposed on the 
lepton transverse momentum. This will however require 
detailed simulation, which is outside the scope of this 
paper and will be done in an independent work.}

While we provide an 
exhaustive list of channels for the $\Sigma^+\Sigma^-$ decay 
mode in Table \ref{tab:sigpm}, obviously not all of them can 
be effectively used at the LHC to provide smoking gun evidence
for our two-Higgs doublet Type III seesaw model. We will highlight 
below a few of these channels which appear to be 
particularly promising. 

\begin{itemize}
\item 
As discussed in details before, one of the main 
decay channels of $\Sigma^{\pm}$ is ${\Sigma^{\pm} \rightarrow l^{\pm}\,h^0}$. 
The $h^0$ with mass of 40 GeV, then decays subsequently to 
$b\bar{b}$ pairs giving rise 
to a final state signal of a pair of opposite sign dileptons (OSD) 
+ 4 $b$-jets.
$$ \Sigma^{+} \Sigma^{-} \to l^+l^- h^0h^0 \to l^+l^-b\bar bb\bar b 
\to 4b +OSD.$$
We have seen from Table \ref{tab:sigcbr} that the branching 
ratio for $\Sigma^{\pm} \to l^{\pm}\,A^0$ is also comparable. 
This will also produce the same collider signature of 
$4b+OSD$. 
The only observable difference will be that the $b$-jets produced 
from the $A^0$ decay will be harder as $A^0$ is much more massive 
than $h^0$. Here and everywhere else in this section, we will 
ignore the information on the hardness of the $b$-jets and 
present the sum of the cross-sections with $h^0$ and $A^0$ 
in the intermediate state. We should also stress that while we 
write only $h^0$ explicitly in the intermediate channels in the 
Tables, the cross-sections given in the final column always 
also include $A^0$ as well as $h^0$. One finds that the effective 
cross-section for this channel is 35.84 fb, which is rather high. 
The OSD released are expected to 
be hard, as they come from the decay of the massive fermions. 

\vglue 0.4cm
\noindent
Instead of decaying into $b\bar b$ pair, the $h^0$s could decay 
into $\tau\bar \tau$. If one of the $h^0$ decays into $b\bar b$ 
and the other into $\tau\bar \tau$, we will get 
$$ \Sigma^{+} \Sigma^{-} \to l^+l^- h^0h^0 \to l^+l^-b\bar b \tau\bar \tau 
\to 2b +OSD+2\tau.$$
This has an effective cross-section of 3.67 fb, which will reduce 
further due to the lower $\tau$ detection efficiency. A third 
possibility exists where both the $h^0$ decay into 
$\tau\bar \tau$ pairs. The effective cross-section for this channel 
is small as can be seen from the Table \ref{tab:sigpm}, and will 
get smaller once the $\tau$ detection efficiencies are folded.

\item 
The other dominant decay channel for $\Sigma^{\pm}$ decay is 
$\Sigma^{\pm} \rightarrow \nu H^{\pm}$. The neutrino will give missing 
energy while $H^{\pm}$ will decay into $H^{\pm} \to W^\pm h^0$. 
The $W^\pm$ could decay hadronically giving 2 jets or leptonically 
giving either a $\tau$-jet + missing energy or $e$/$\mu$ lepton 
+ missing energy. Since the lepton released in the 
${\Sigma^{\pm} \rightarrow l^{\pm}\,h^0}$ is important both 
for understanding the flavor structure of the mixing matrix 
as well as for tagging the channel in order to reduce the 
background, we consider first the case where one of heavy 
charged fermion decays into a hard charged lepton and $h^0$ and 
the other into a neutrino and $H^{\pm}$. The most 
interesting channels 
in this case turn out to be:
 $$ \Sigma^{+}\Sigma^{-} \to l^+h^0H^-\nu \to l^+h^0 h^0 W^-\nu 
\to 4b + l + 2j + \not{p_T},$$
$$\Sigma^{+}\Sigma^{-} \to l^+h^0H^-\nu \to l^+h^0 h^0 W^-\nu \to  
4 b + l+ \tau + \not{p_T},$$
where for the former, 
the two $h^0$ (one from the $\Sigma^+$ decay and another 
from $H^-$ decay) produce 4 $b$-jets, and the $W^-$ 
decays produce two hadronic jets. In the latter channel, 
the $W^-$ decays into $\tau\nutau$, producing a $\tau$-jet. 
The effective cross-section for the former channel is 
26.88 fb, while that for the latter is 4.48 fb. The 
effective cross-sections for the other 
channels with $l^+h^0 h^0 W^-\nu $ in the intermediate states 
are given in Table \ref{tab:sigpm}. However, their 
cross-sections are smaller.

\item
Finally, both the charged heavy fermions could decay through the  
$H^\pm \nu$ mode. In this case 
we have the following 
leading order possibilities:
$$\Sigma^{+}\Sigma^{-} \to H^+\nu H^-\nu \to
h^0h^0W^+W^-\nu\nu \to 4b + 4j  + \not{p_T},$$
$$\Sigma^{+}\Sigma^{-} \to H^+\nu H^-\nu \to
h^0h^0W^+W^-\nu\nu \to 4b + 2j  + l'+ \not{p_T},$$
$$\Sigma^{+}\Sigma^{-} \to H^+\nu H^-\nu \to
h^0h^0W^+W^-\nu\nu \to 4b + 2j  + \tau+ \not{p_T}.$$
The mode $\Sigma^{+}\Sigma^{-} \to 4b+OSD'+\not{p_T}$,  
appearing at serial number 16 in  Table \ref{tab:sigpm} 
could have been easy to tag as it contains 4$b$-jets and 
pair of opposite sign dileptons coming from $W^\pm$ 
decay, and missing energy. However, the effective 
cross-section for this channel is relatively low. Note that 
none of the channels with $H^+\nu H^-\nu $ in their 
intermediate state have $l$ in their final state. For these 
channels therefore, it is impossible to say anything about the 
flavor structure of the model. 

\end {itemize}

\subsection{$\Sigma^{\pm}\Sigma^{0}$ decay}
\label{sec:colldk2}

\begin{table}
\begin{center}
\renewcommand{\arraystretch}{1.4}
\begin{tabular}{||c|c|c||}
\hline\hline
Sl no &Channels &Effective cross-section (in fb) \\
\hline\hline
1&$\Sigma^{\pm}\Sigma^0\to l^\pm h^0h^0\nu\to 4b+l+\not{p_T}$&96.3
\\\hline
2&$\Sigma^{\pm}\Sigma^0\to l^\pm h^0h^0\nu\to 2b+l+2\tau+\not{p_T}$&19.7
\\\hline
3&$\Sigma^{\pm}\Sigma^0\to l^\pm h^0h^0\nu\to l+2\tau+\not{p_T}$&0.99
\\\hline\hline
4&$\Sigma^{\pm}\Sigma^0\to H^{\pm}\nu h^0\nu\to 4b+1l^{\prime}+\not{p_T}$&107.4
\\\hline
5&$\Sigma^{\pm}\Sigma^0\to H^{\pm}\nu h^0\nu \to 4b+\tau+\not{p_T}$&53.7
\\\hline
6&$\Sigma^{\pm}\Sigma^0\to H^{\pm}\nu h^0\nu \to 4b+2j+\not{p_T}$&35.98
\\\hline
7&$\Sigma^{\pm}\Sigma^0\to H^{\pm}\nu h^0\nu \to 2b+2\tau+2j+\not{p_T}$&7.36
\\\hline
8&$\Sigma^{\pm}\Sigma^0\to H^{\pm}\nu h^0\nu \to 2b+2\tau+l'+\not{p_T}$&2.42
\\\hline
9&$\Sigma^{\pm}\Sigma^0\to H^{\pm}\nu h^0\nu \to 2b+3\tau+\not{p_T}$&1.21
\\\hline
10&$\Sigma^{\pm}\Sigma^0\to H^{\pm}\nu h^0\nu \to 4\tau+2j+\not{p_T}$&0.38
\\\hline
11&$\Sigma^{\pm}\Sigma^0\to H^{\pm}\nu h^0\nu \to l'+4\tau+\not{p_T}$&0.12
\\\hline
12&$\Sigma^{\pm}\Sigma^0\to H^{\pm}\nu h^0\nu \to 5\tau+\not{p_T}$&0.06
\\\hline\hline
13&$\Sigma^{\pm}\Sigma^0\to l^\pm H^\mp l^\pm h^0\to4b+2l+2j$&36.12
\\\hline
14&$\Sigma^{\pm}\Sigma^0\to l^\pm H^\mp l^\pm h^0\to4b+3l(2l+l')+\not{p_T}$&12.04
\\\hline
15&$\Sigma^{\pm}\Sigma^0\to l^\pm H^\mp l^\pm h^0\to4b+2l+1\tau+\not{p_T}$&6.02
\\\hline
16&$\Sigma^{\pm}\Sigma^0\to l^\pm H^\mp l^\pm h^0\to2b+2l+2\tau+2j$&7.4
\\\hline
17&$\Sigma^{\pm}\Sigma^0\to l^\pm H^\mp l^\pm
h^0\to2b+3l(2l+l')+2\tau+\not{p_T}$&2.4
\\\hline
18&$\Sigma^{\pm}\Sigma^0\to l^\pm H^\mp l^\pm
h^0\to2b+2l+3\tau+\not{p_T}$&1.20
\\\hline
19&$\Sigma^{\pm}\Sigma^0\to l^\pm H^\mp l^\pm
h^0\to2l+4\tau+2j$&0.36
\\\hline
20&$\Sigma^{\pm}\Sigma^0\to l^\pm H^\mp l^\pm
h^0\to3l(2l+l')+4\tau+\not{p_T}$&0.12
\\\hline
21&$\Sigma^{\pm}\Sigma^0\to l^\pm H^\mp l^\pm
h^0\to 2l+5\tau+\not{p_T}$&0.06
\\\hline\hline
\end{tabular}
\caption{\label{tab:sig0ch}Effective cross-sections (in fb) 
of different $\Sigma^{\pm} \Sigma^0$ decay channels 
for $M_{\Sigma_1}=300$ GeV.}
\end{center}
\end{table}

\begin{table}
\begin{center}
\renewcommand{\arraystretch}{1.4}
\begin{tabular}{||c|c|c||}
\hline\hline
Sl no &Channels &Effective cross-section (in fb) \\
\hline\hline
1&$\Sigma^{\pm}\Sigma^0\to H^\pm\nu H^\pm l^\mp \to 
4b+l+4j+\not{p_T}$&13.36
\\\hline
2&$\Sigma^{\pm}\Sigma^0\to H^\pm\nu H^\pm l^\mp \to 
4b+l+\tau+2j+\not{p_T}$&4.38
\\\hline
3&$\Sigma^{\pm}\Sigma^0\to H^\pm\nu H^\pm l^\mp \to 
4b+OSD(l+l')+2j+\not{p_T}$&6.57
\\\hline
4&$\Sigma^{\pm}\Sigma^0\to H^\pm\nu H^\pm l^\mp \to 
4b+LSD(l+l')+2j+\not{p_T}$&2.19
\\\hline
5&$\Sigma^{\pm}\Sigma^0\to H^\pm\nu H^\pm l^\mp \to 
4b+OSD(l+l')+\tau+\not{p_T}$&1.09
\\\hline
6&$\Sigma^{\pm}\Sigma^0\to H^\pm\nu H^\pm l^\mp \to 
4b+LSD(l+l')+\tau+\not{p_T}$&0.37
\\\hline
7&$\Sigma^{\pm}\Sigma^0\to H^\pm\nu H^\pm l^\mp \to 
2b+OSD(l+l')+2\tau+2j+\not{p_T}$&1.35
\\\hline
8&$\Sigma^{\pm}\Sigma^0\to H^\pm\nu H^\pm l^\mp \to 
2b+LSD(l+l')+2\tau+2j+\not{p_T}$&0.45
\\\hline
9&$\Sigma^{\pm}\Sigma^0\to H^\pm\nu H^\pm l^\mp \to 
2b+OSD(l+l')+3\tau+\not{p_T}$&0.23
\\\hline
10&$\Sigma^{\pm}\Sigma^0\to H^\pm\nu H^\pm l^\mp \to 
2b+LSD(l+l')+3\tau+\not{p_T}$&0.08
\\\hline
11&$\Sigma^{\pm}\Sigma^0\to H^\pm\nu H^\pm l^\mp \to 
OSD(l+l')+4\tau+2j+\not{p_T}$&0.06
\\\hline
12&$\Sigma^{\pm}\Sigma^0\to H^\pm\nu H^\pm l^\mp \to 
LSD(l+l')+4\tau+2j+\not{p_T}$&0.02
\\\hline
13&$\Sigma^{\pm}\Sigma^0\to H^\pm\nu H^\pm l^\mp \to 
2b+l+2\tau+4j+\not{p_T}$&2.78
\\\hline
14&$\Sigma^{\pm}\Sigma^0\to H^\pm\nu H^\pm l^\mp \to 
l+4\tau+4j+\not{p_T}$&0.14
\\\hline
15&$\Sigma^{\pm}\Sigma^0\to H^\pm\nu H^\pm l^\mp \to 
4b+3l(l+2l')+\not{p_T}$&1.68
\\\hline
16&$\Sigma^{\pm}\Sigma^0\to H^\pm\nu H^\pm l^\mp \to 
2b+3l(l+2l')+2\tau+\not{p_T}$&0.32
\\\hline
15&$\Sigma^{\pm}\Sigma^0\to H^\pm\nu H^\pm l^\mp \to 
3l(l+2l')+4\tau+\not{p_T}$&0.02
\\\hline
16&$\Sigma^{\pm}\Sigma^0\to H^\pm\nu H^\pm l^\mp \to 
4b+l+2\tau+\not{p_T}$&0.42
\\\hline
17&$\Sigma^{\pm}\Sigma^0\to H^\pm\nu H^\pm l^\mp \to 
2b+l+4\tau+\not{p_T}$&0.08
\\\hline
18&$\Sigma^{\pm}\Sigma^0\to H^\pm\nu H^\pm l^\mp \to 
l+5\tau+2j+\not{p_T}$&0.04
\\\hline
19&$\Sigma^{\pm}\Sigma^0\to H^\pm\nu H^\pm l^\mp \to 
l+6\tau+\not{p_T}$&0.004
\\\hline
20&$\Sigma^{\pm}\Sigma^0\to H^\pm\nu H^\pm l^\mp \to 
l+l'+5\tau+\not{p_T}$&0.008
\\\hline\hline
\end{tabular}
\caption{\label{tab:sig0ch2}
Effective cross-sections (in fb) 
of different $\Sigma^{\pm} \Sigma^0$ decay channels 
for $M_{\Sigma_1}=300$ GeV.}
\end{center}
\end{table}

We give in Tables 
\ref{tab:sig0ch} and \ref{tab:sig0ch2}, 
all possible 
decay channels, final state 
configurations of particles, 
and their corresponding effective cross-sections for the 
$\Sigma^\pm\Sigma^0$ production and decays. We reiterate that the final 
effective cross-section after cuts 
will be obtained once these cross-sections 
are folded with the efficiency functions. 
For the leptons we follow the same convention for 
our notation as done for the previous section.

\begin{itemize}

\item We begin by looking at the $\Sigma^\pm\Sigma^0$ decays 
where $\Sigma^\pm \to l^\pm h^0$ and $\Sigma^0 \to \nu h^0$. 
This would lead to the following final state configuration
$$\Sigma^{\pm}\Sigma^{0} \to l^\pm h^0 h^0 \nu \to 
4b + l + \not{p_T},$$
with a very large effective cross-section of 96.3 fb. This 
channel should be easy to tag. The 4 hard 
$b$-jets come from the displaced 
$h^0$ vertices, and the lepton released is hard. This lepton 
will also carry information on the $\mu$-$\tau$ 
symmetric flavor structure of the model. Another 
unambiguous channel with significant effective cross-section 
coming from the $l^\pm h^0 h^0 \nu$ intermediate state is
$$\Sigma^{\pm}\Sigma^{0} \to l^\pm h^0 h^0 \nu \to 
2b + l + 2\tau+ \not{p_T},$$
where one of the $h^0$ decays into $\tau\bar\tau$. 

\item The other intermediate state which has very large 
effective cross-sections is 
$\Sigma^\pm\Sigma^0 \to \nu H^\pm \,\nu h^0$. 
The $H^\pm$ would decay into $W^\pm h^0$, and $W^\pm$ into 
a lepton $l'$ finally giving
$$\Sigma^{\pm}\Sigma^{0} \to H^\pm \nu h^0 \nu \to 
4b + l' + \not{p_T},$$
with an effective cross-section of 107.4 fb. 
Alternatively, the $W^-$ could instead decay into $\tau\anutau$ giving 
$$\Sigma^{\pm}\Sigma^{0} \to H^\pm \nu h^0 \nu \to 
4b + \tau + \not{p_T},$$
with effective cross-section of 53.7 fb, or decay into $qq'$ 
giving 
$$\Sigma^{\pm}\Sigma^{0} \to H^\pm \nu h^0 \nu \to 
4b + 2j + \not{p_T},$$
with an effective cross-section of 35.98 fb. 

\item Large effective cross-section in the $\Sigma^{\pm}\Sigma^{0}$ 
channel is also expected from the following decay chain 
$$\Sigma^{\pm}\Sigma^{0} \to l^\pm h^0 l^\pm H^\mp \to 
4b + 2l + 2j,$$
with effective cross-section of 36.12 fb. Both the leptons 
in this channel come from the heavy fermion decay vertices and 
carry the flavor information of the model. 

\item  $\Sigma^{\pm}\Sigma^{0}$ could also decay through the 
intermediate states $H^\pm \nu H^\pm l^\mp$. This leads to 20 
possible final state particles and collider signatures. These 
are listed in Table \ref{tab:sig0ch2}.  
However, the only one which has sizable effective cross-section is
$$\Sigma^{\pm}\Sigma^{0} \to l^\pm h^0 l^\pm H^\mp \to 
4b + l + 4j + \not{p_T}.$$
However, this channel has 4 light quark jets, which is always 
prone to problems with backgrounds.

\end{itemize}

\subsection{Backgrounds}

\begin{table}[h]
\begin{center}
\renewcommand{\arraystretch}{1.4}
\begin{tabular}{||c|c|c||}
\hline\hline
Sl no&Channels & Effective cross-section  \\
& &in fb \\
\hline\hline
1&4b +OSD & 35.84 \\\hline
2&$4 b + l + \not{p_T}$ &96.3\\\hline
3&$4 b + l' + \not{p_T}$ &107.4\\\hline
4&$4 b + \tau + \not{p_T}$ &53.7\\\hline
5&$4 b + l + 2j + \not{p_T}$ &26.88\\\hline
6&$4 b + 2l + 2j$ &36.12\\\hline
7&$4 b +3l(2l+l^{\prime}) +\not{p_T}$&12.04\\
\hline\hline
\end{tabular}
\caption{\label{tab:sigfinal}
Effective cross-sections in fb for $M_{\Sigma_1}=300$ GeV,
for the most important channels for our model.}
\end{center}
\end{table}

In Tables \ref{tab:sigpm}, \ref{tab:sig0ch} and 
\ref{tab:sig0ch2} we provided a comprehensive list of 
collider signature channels for 
the heavy fermions, and their corresponding 
effective cross-sections. In the previous subsection
we had also discussed some of the 
most important channels with large effective cross-sections. 
In Table \ref{tab:sigfinal} we give a subset of those highlighted 
in sections \ref{sec:colldk1} and \ref{sec:colldk2}. These 
are expected to be the most unambiguous channels, with smallest 
backgrounds and the largest signal cross-sections. In almost 
all channels listed in Table \ref{tab:sigfinal}, the final 
collider signature contains 
4 $b$-jets and a hard lepton coming from the primary 
heavy fermion decay vertex. In addition, the 4 $b$-jets 
come from the $h^0$ decay vertex which is significantly 
displaced with respect to the heavy fermion decay vertex. 
The main source of standard model background for the channels with 
4 $b$-jets and a lepton are the $t\bar{t}b\bar{b}$ modes, 
which can give multiple $b$-jets, leptons and missing energy. 
However, as mentioned many times before, 
the $b$-jets come from $h^0$ displaced vertex and should 
not have any standard model background. Having the hard lepton 
in the final state further cuts down the background. Therefore, 
each of these collider channels are expected to have very 
little to no backgrounds. For a detailed 
signal to background analysis one requires a detailed 
simulation for the final state topology, which is outside 
the scope of this work. 
Nevertheless we add a few lines discussing qualitatively the  
possibility of backgrounds for some of the listed channels in 
Table \ref{tab:sigfinal}.

\begin{itemize}
\item 4b +OSD: Here the two opposite sign 
dileptons come from the $\Sigma^+\Sigma^-$
decays. Since the $\Sigma^\pm$ are heavy with $M_{\Sigma^{\pm}}=300$ GeV, 
the leptons will 
be very hard and we can put a cut of $p_T \gtap 100$ GeV. 
The displaced $h^0$ vertex should remove all backgrounds. 

\item $4 b + l + \not{p_T}$: Here $t\bar{t}b\bar{b}$ does not directly
give any background, unless one of the leptons from the final state  
is missed. However, the $p_T$ cut on the hard lepton and the displaced 
$h^0$ vertices should effectively remove any residual background.

\item $4 b + l' + \not{p_T}$: Here the $p_T$ cut on the lepton cannot be 
imposed as the lepton here comes from $W^\pm$ decay. However, the 
4 $b$-jets still come from the displaced $h^0$ vertices and that 
should anyway take care of killing all backgrounds to a large extent.

\item $4 b + l +2 jet + \not{p_T}$: The main background could
again come from standard model 
$t\bar{t}b\bar{b}$ channels. This can also be removed 
by the displaced $h^0$ vertex and 
a cut of $p_T\gtap 100$ GeV for the lepton. 

\item $4 b + 2l + 2j$: Similar to the first case, but with 2 extra jets.

\item $4 b +3l(2l+l') +\not{p_T}$: Out of the 3 leptons in this channel, 
two are hard and one is relatively soft. In addition we have the 
$h^0$ displaced vertex. Therefore, 
this channel is expected to be absolutely background free. 

\end{itemize}

\section{Conclusions}

The seesaw mechanism has remained 
the most elegant scheme to explain the smallness 
of the neutrino masses without having to 
unnaturally fine tune the Yukawa couplings to 
very small values. In the so-called Type III seesaw, three 
self-conjugate ($Y=0$) SU(2) triplet fermions are added to the 
standard model particle content.
These exotic fermions are color singlets and  belong to the 
adjoint representation of SU(2). 
These exotic fermions have Yukawa couplings with the 
standard model lepton doublet and the Higgs doublet. 
They also have a Majorana mass term. Once these 
heavy leptons are integrated out from the theory, 
we are left with a Majorana mass term for the neutrino 
given by the famous seesaw formula, where the smallness 
of the neutrino mass is explained by the largeness of the 
heavy fermion mass, without having to fine tune the 
Yukawa couplings to very small values. To generate 
neutrino masses $m_\nu\sim 0.1$ eV, one requires 
that the heavy fermion mass should be $\sim 10^{14}$ GeV. 
Being in the adjoint representation of SU(2), 
one of the most interesting feature of these 
exotic fermions is that they have gauge couplings, and 
therefore can be produced at collider experiments. 
The only constraint for the production of these particles 
at LHC is that their mass should be in a few 100 GeV range. 
However, in order to produce neutrino masses 
$m_\nu \sim 0.1$ eV, one would then have to tune the 
Yukawa couplings to be $\sim 10^{-6}$, which ruins 
completely the very spirit 
and motivation for seesaw. 

In order to circumvent this problem and preserve the 
motivation of the seesaw mechanism, we propose an 
extended Type III seesaw model with two SU(2) Higgs doublets 
along with the three self-conjugate SU(2) fermion triplets. 
We impose an additional $Z_2$ symmetry such that 
one of the Higgs doublets, called 
$\Phi_1$, has positive charge while 
the other, called $\Phi_2$, has negative charge under this 
symmetry. 
In addition, we demand that 
all standard model particles have positive charge 
with respect to $Z_2$ while the three new exotic fermion 
triplets are negatively charged. Therefore, 
$\Phi_1$ behaves like the standard model 
Higgs, while $\Phi_2$ is coupled {\it only} 
to the exotic fermion triplets. As a result, the neutrino mass 
term coming from the seesaw formula 
depends on the VEV of $\Phi_2$ ($v'$), while all 
other fermion masses are dependent on the VEV of $\Phi_1$ ($v$). 
We can therefore choose a value for $v'$ such that 
$m_\nu \sim 0.1$ eV for exotic fermion masses $\sim 100$ GeV, 
without having to fine tune the Yukawa couplings to very small 
values. 

Another typical feature about neutrinos concern their 
peculiar mixing pattern which should be explained by 
the underlying theory. The current neutrino oscillation 
data suggests an inherent $\mu$-$\tau$ symmetry in the 
low energy neutrino mass matrix. It is therefore expected 
that this $\mu$-$\tau$ symmetry should also exist at the 
high scale, either on its own or as a sub-group of a bigger 
flavor group. 
We imposed an exact $\mu$-$\tau$ symmetry on 
both the Yukawa coupling of the triplet fermions $Y_\Sigma$ 
as well as on their Majorana mass matrix $M$. Therefore the  
low energy neutrino matrix 
$\tilde m$ obtained after the seesaw 
had an in-built $\mu$-$\tau$ symmetry. As a result our 
model predicts $\theta_{23}=\pi/4$ and $\theta_{13}=0$. 
The mixing angle $\theta_{12}$ as well as the mass squared 
differences $\ms$ and $\ma$ are given in terms of the entries of 
$Y_\Sigma$ and $M$. We showed how the oscillation parameters 
depend on the model parameters in $Y_\Sigma$ and $M$. 

A very important and new aspect which emerged from our study 
concerns the mixing in the heavy fermion sector. It had 
been assumed in all past studies that the mixing matrices 
$U_\Sigma$, $U_h^L$ and $U_h^R$ which 
diagonalize the heavy fermion mass 
matrices $\tilde M_\Sigma$, $M_H^\dagger M_H$ and 
$M_H M_H^\dagger$ respectively, are almost unit matrices. 
However, we showed that for the case where 
$M$ was $\mu$-$\tau$ symmetric, these matrices 
were highly non-trivial, and in particular had the 
last column as $(0,1/\sqrt{2},1/\sqrt{2})$. 
We showed that this has observable 
consequences for the heavy fermion decays at the 
collider. We showed that flavor structure of our 
model was reflected in the pattern of heavy fermion 
decays into light charged leptons. These showed a 
$\mu$-$\tau$ symmetry. The state $\Sigma_3^{\pm/0}$ 
decayed equally into muons and taus and 
almost never decayed into electrons. 
We have checked that this feature 
exists not only for our model, but for any model with an 
underlying flavor symmetry group that imposes 
$\mu$-$\tau$ symmetry on the heavy Majorana mass matrix $M$. 
In fact, we have made explicit checks on the seesaw model 
proposed by Altarelli and Feruglio \cite{AFseesaw}, where $A_4$ was 
imposed as the flavor symmetry group. Though the 
proposed $A_4$ model by Altarelli and Feruglio was a 
Type I seesaw model, it can be easily adapted to the 
Type III seesaw case. We found that the mixing matrices 
$U_\Sigma$, $U_h^L$ and $U_h^R$ even for that case 
had  $(0,1/\sqrt{2},1/\sqrt{2})$ as their last column.

Having established the flavor structure of our model, we 
next turned to the production and detection of heavy fermions 
at LHC. We discussed quantitatively and in details
the cross-section for 
the heavy fermion production at LHC and their decay rates. 
While the production cross-sections for our 
model turned out to be same as that in all earlier 
calculations done in the context of the one Higgs doublet 
model, the decay pattern for the heavy fermions in our case 
was found to be extremely different and unique. 
The \mt permutation symmetry showed up in the flavor 
pattern of the heavy fermion decays due to the typical 
last column $(0,1/\sqrt{2},1/\sqrt{2})$ of the matrices 
$U_\Sigma$, $U_h^L$ and $U_h^R$. 
We also showed that in our case the decay rate of the 
heavy fermions was about $10^{11}$ times larger than 
that found for the one Higgs doublet 
model. 
In fact, the heavy fermion decay rate for our 
model is $5.8\times 10^{-2}$ GeV and $4\times 10^{-2}$ GeV 
for 300 GeV charged and neutral heavy fermions, respectively. 
Therefore, while for the one Higgs doublet 
case one could attempt to look for displaced 
heavy fermion decay vertices, in our case they will 
decay almost instantaneously. We found that this 
tremendous decay rate came from the very fast decays of 
$\Sigma^{\pm/0}$ into light leptons and Higgs $h^0$, 
$A^0$ or $H^\pm$. 
The very large decay rate was shown to stem from the 
very large Yukawa couplings in our model. As the 
Yukawa couplings are a factor of $10^{5}-10^{6}$ 
larger in our model, 
the decay rates which depend on the square of 
the Yukawa couplings are a factor $10^{10}$-$10^{12}$ higher. 
Decays into $H^0$ and gauge bosons in our model was shown 
to be same as in the one Higgs doublet 
case, and the reason explained. 

Another distinctive feature of our model appeared in the 
pattern of the Higgs decays. We showed that the smallness 
of the neutrino masses constrained 
the neutral Higgs mixing angle $\alpha$ to be very small. 
This resulted in a very small decay rate for the 
$h^0$ Higgs. For a mass of $M_{h^0}=40$ GeV, 
the $h^0$ lifetime in the Higgs rest frame comes out to be 
about 5 cm. This will give a displaced decay vertex 
in the LHC detectors, ATLAS and CMS. The lifetime 
for $H^\pm$ turned out to be small. 

Finally, we discussed in detail the expected collider 
signatures for our two Higgs doublet Type III seesaw model with 
$\mu$-$\tau$ symmetry. We tabulated a comprehensive and 
exhaustive list of all possible collider signature 
channels for the heavy fermions at LHC. We gave the 
effective cross-sections for each of these channels. 
The effective cross-section for some of these channels 
were seen to be very high. We made a short-list of 
channels with very high effective cross-section and 
low background at LHC. This was presented in 
Table \ref{tab:sigfinal}. In all the listed channels 
we have 4 $b$-jets coming from the decay of two $h^0$ 
which have displaced vertices. In addition, all (except one)
of them have a hard lepton in the final state coming from 
the $\Sigma^{\pm/0}$ decay. This lepton in addition to being 
hard, also carries information on the $\mu$-$\tau$ symmetry 
of our model. These collider signature channels are 
hence very distinctive of our model and suffer from almost 
no standard model background. 

In conclusion, we proposed a Type III seesaw model with 
large Yukawa couplings and 
triplet fermion masses light enough to be produced at the 
LHC. This could be achieved through a unique two Higgs 
doublet model. The very large Yukawa couplings resulted in 
very fast decays of the heavy fermions, with a decay rate 
about $10^{11}$ times faster than 
obtained in the earlier Type III seesaw models. 
We imposed a $\mu$-$\tau$ symmetry on 
our model in order to comply with the low energy neutrino 
oscillation data. This flavor pattern is reflected also 
in the mixing matrices of the heavy fermions, which are no
longer unity, and which have observable consequences for the 
heavy fermion decays at the LHC. 
The neutrino mass constrains also the mixing angle of the 
neutral Higgs to be very small. This nearly forbids the decay 
of the heavy fermions into the heavier CP even neutral Higgs 
$H^0$. Thus they decay almost always into $h^0$ (and $A^0$) 
and $H^\pm$. More importantly, the small neutral Higgs mixing 
angle increases the lifetime of the $h^0$. These are 
expected to live for more than 10 cm at LHC before decaying  
predominantly into $b\bar b$, producing $b$-jets. This would  
be seen as a  displaced $h^0$ decay vertex in the detector. 
We identified collider signature channels at LHC which 
have very large effective cross-section and 
almost no standard model background. These could be used to 
provide smoking gun evidence for our model.

\vglue 0.8cm
\noindent
{\Large{\bf Acknowledgments}}\vglue 0.3cm
\noindent
The authors wish to thank 
A. Abada, B. Bajc, S. Bhattacharyya,  
F. Bonnet, A. Datta, R. Foot, 
S. Goswami, T. Hambye, A. Ibarra, K. Matchev, B. Mukhopadhyaya, 
A. Raychaudhuri and A. Sen 
for discussions and valuable comments. 
The authors acknowledge the HRI cluster facilities for computation.
This work has been supported by the Neutrino Project and 
the RECAPP Project 
under the XI Plan of Harish-Chandra Research Institute.

\vglue 1.0cm
\begin{center}
{\LARGE{\bf Appendix}}
\end{center}

\begin{appendix}

\section{The Scalar Potential and Higgs Spectrum}
\renewcommand{\theequation}{A\arabic{equation}}
\setcounter{equation}{0}

Our model has two SU(2) complex  
Higgs doublets $\Phi_1$ and $\Phi_2$, with hypercharge $Y=1$. The scalar 
potential can then be written as 
\be
V &=& \lambda_1\left({\Phi_1^\dagger}{\Phi_1} - v^2\right)^2 
   + \lambda_2\left({\Phi_2^\dagger}{\Phi_2} - v'^2\right)^2 
   +\lambda_3\left(({\Phi_1^\dagger}{\Phi_1} - v^2)
+({\Phi_2^\dagger}{\Phi_2} - v'^2)\right)^2 
\nonumber \\
&&
+ \lambda_4\left(({\Phi_1^\dagger}\Phi_1)({\Phi_2^\dagger}\Phi_2 ) - 
({\Phi_1^\dagger}\Phi_2)(\Phi_2^\dagger\Phi_1)\right) +
\lambda_5\left({\rm Re}(\Phi_1^\dagger \Phi_2) - vv'\cos\xi\right)^2 
\nonumber \\
&&
+ \lambda_6\left({\rm Im}(\Phi_1^\dagger\Phi_2) - vv'\sin\xi\right)^2
,
\label{eq:scalar}
\ee
where
\be
\langle\Phi_1 \rangle =
\pmatrix
{0\cr v},~~
 \langle \Phi_2 \rangle =
\pmatrix
{0\cr v'e^{i \xi}},
~~{\rm and}~~ 
\tan\beta = \frac{v'}{v}
\label{eq:tanbeta}
.
\ee
Recall that under the 
imposed $Z_2$ symmetry, $\Phi_1$ 
carries charge $+1$, while $\Phi_2$ has $-1$ charge. 
Therefore, the $\lambda_5$ term is zero when the symmetry is exact. 
We will discuss shortly the phenomenological consequences of this 
and argue in favor of a mild breaking of this $Z_2$ symmetry. 
With the scalar potential Eq. (\ref{eq:scalar}) it is straightforward 
to obtain the Higgs mass matrix and obtain the corresponding 
mass spectrum. The  physical degrees of freedoms contain the 
charged Higgs $H^\pm$ and the neutral Higgs 
$H^0$, $h^0$, and $A^0$.  While 
$H^0$ and $h^0$ are CP even, $A^0$ is CP odd. 
If we work in a simplified scenario where $\xi$ is taken as zero, 
then it is is quite straightforward to derive the mass of the 
charged Higgs $H^\pm$ and the CP-odd Higgs $A^0$. The masses are 
given as 
\be
M_{H^{\pm}}^2= \lambda_4 (v^2+{v^{\prime}}^2)
,~~
{\rm and}~~
M_{A^0}^2= \lambda_6 (v^2+{v^{\prime}}^2),
\label{eq:hmass}
\ee
respectively. The mass matrix for the neutral CP-even Higgs is
\be
M' &=& \pmatrix{
4 v^2(\lambda_1+ \lambda_3)+{v^{\prime}}^2\lambda_5 &  (4 \lambda_3 + \lambda_5)v v^{\prime}\cr
(4 \lambda_3+\lambda_5)v v^{\prime} & 4 {v^{\prime}}^2(\lambda_2 + \lambda_3)+v^2\lambda_5}
.
\ee
The mixing angle, obtained from diagonalizing the above matrix is given by
\be
\tan 2 \alpha=  \frac{2\,M_{12}}{M_{11}-M_{22}}
,
\label{eq:tan2alpha}
\ee
and the corresponding masses are 
\be
M_{H^0,h^0}^2= \frac{1}{2} \{M_{11}+M_{22}\pm 
\sqrt{(M_{11}-M_{22})^2+4M_{12}^2}.
\label{eq:higgsmass}
\ee
The physical Higgs are given in terms of components of 
$\Phi_1$ and $\Phi_2$ as follows. 
The neutral Higgs are given as  
\be
{H}^0 &=& \sqrt{2}\left(({\rm Re}\Phi_1^0 - v)\cos\alpha 
+ ({\rm Re}\Phi_2^0 - v') \sin\alpha \right),
\\
h^0 &=& \sqrt{2} \left( -({\rm Re}  \Phi_1^0 - v)\sin\alpha 
+ ({\rm Re}\Phi_2^0 - v') \cos\alpha \right),
\\
A^0 &=& \sqrt{2}(-{\rm Im}\Phi_1^0 \sin\beta + {\rm Im}\Phi_2^0\cos\beta),
\ee
while the charged Higgs are 
\be
H^\pm = -\Phi_1^\pm \sin\beta + \Phi_2^\pm\cos\beta
.
\ee
The Goldstones turn out to be 
\be
G^\pm &=&\Phi_1^\pm \cos\beta +\Phi_2^\pm\sin\beta
\\
G^0 &=& \sqrt{2}({\rm Im}\Phi_1^0 \cos\beta + 
{\rm Im}\Phi_2^0\sin\beta).
\ee
Recall that the 
requirement from small neutrino masses $m_\nu \sim 0.1$ eV
constrains 
$v'\sim 10^{-4}$ GeV. Therefore, for our model we get from 
Eqs. (\ref{eq:tanbeta}) and (\ref{eq:tan2alpha})
\be
\tan\beta \sim 10^{-6},
~~{\rm and}~~
\tan2\alpha \sim \tan\beta \sim 10^{-6}
.
\label{eq:betaalpha}
\ee
One can estimate from Eq. (\ref{eq:higgsmass}), 
that in the limit $v^{\prime} \ll v$, 
\be
M_{H^0}^2 \simeq (\lambda_1+\lambda_3) v^2,
~~{\rm and}~~
M_{h^0}^2 \simeq \lambda_5 v^2
.
\label{eq:higgsm}
\ee
We should point out here that in the limit 
of exact $Z_2$ symmetry, $\lambda_5=0$ exactly, and 
in that case $M_{h^0}^2 \propto \frac{{\lambda_3}^2}
{(\lambda_1+\lambda_3)^2}{v^{\prime}}^2$. 
Since $v'\sim 10^{-4}$ GeV, this would give a
very tiny mass for the neutral Higgs $h^0$. 
To prevent that, we introduce a mild explicit breaking of 
the $Z_2$ symmetry, by taking 
$\lambda_5 \neq 0$ in the scalar potential. This not only alleviates the 
problem of an extremely light Higgs boson, it also 
circumvents spontaneous breaking of $Z_2$, when 
the Higgs develop vacuum expectation value. This 
saves the model from complications such as 
creation of domain walls, due to the spontaneous 
breaking of a discrete symmetry. The extent of 
breaking of $Z_2$ is determined by the strength of 
$\lambda_5$. Since we wish to impose only a mild breaking, 
we take $\lambda_5 \sim 0.05$. This gives us 
a light neutral Higgs mass of $M_h^0 \simeq 40$ GeV from 
Eq. (\ref{eq:higgsm}). Since all other $\lambda_i \sim 1$, 
the mass of the other CP even neutral Higgs, the CP odd neutral 
Higgs and the charged Higgs are all seen to be $\sim v$ GeV 
from Eqs. (\ref{eq:hmass}) and (\ref{eq:higgsm}). 
We will work with $M_H^0=150$ GeV, $M_A^0 = 140$ GeV and 
$M_H^\pm = 170$ GeV. 
 
Also required are the couplings  
of our Higgs with the gauge bosons. This is needed in order to 
understand the Higgs decay and the subsequent collider signatures 
of our model. These are standard expressions and are well 
documented
(see for instance \cite{hunter}). One 
can check that certain couplings depend on 
$\sin\alpha$ and $\sin(\beta-\alpha)$. From Eq. (\ref{eq:betaalpha}) 
we can see that these couplings are almost zero. Others depend on 
$\cos\alpha$ and $\cos(\beta-\alpha)$ and therefore large. We 
refer the reader to \cite{hunter} for a detailed discussion on 
the general form for the coulings.


\section{Appendix B: The Interaction Lagrangian}
\renewcommand{\theequation}{B\arabic{equation}}
\setcounter{equation}{0}

\subsection{ Lepton-Higgs Coupling}

The lepton Yukawa part of the Lagrangian for our two Higgs doublet model 
was given in Eq. (\ref{eq:yukawa}) as, 
\be
-{\cal L}_Y = \left[Y_{l_{ij}}\overline l'_{R_i} \Phi_1^\dagger L'_j 
+ Y_{\Sigma_{ij}}
{\tilde{\Phi}_2}^\dagger \overline \Sigma'_{R_i} L'_j +h.c. \right]
+ \frac{1}{2} M_{ij} \,{\rm Tr}\left[\overline {\Sigma'}_{R_i}  
{\tilde{\Sigma}_{R_j}^C} + h.c. \right]
.
\label{eq:yukawaA}
\ee
From this one can extract the 
individual Yukawa coupling vertex factors between two fermions and 
a Higgs. We have three generations of heavy and light neutral 
leptons and three generations of heavy and light charged leptons. 
In addition, we have three neutral and a pair of charged Higgs. 
The Yukawa interaction between any pair of fermions and a corresponding 
physical Higgs field can 
be extracted from Eq. (\ref{eq:yukawaA}). We list 
below all Yukawa possible interactions {\it in the mass basis} of the 
particles. The vertex factors are denoted as $C_{FI}^{X,L/R}$, where 
$F$ and $I$ are the initial and final state fermions respectively, 
$X$ is the physical Higgs involved and $L/R$ are for either the 
vertex with $P_L$ or $P_R$ respectively, where $P_L$ and $P_R$ are 
the left and right chiral projection operators respectively. Note 
that we have suppressed the generation indices for clarity of the 
expressions. But the generation indices are implicitly there and the 
vertex factors are all $3\times 3$ matrices. 

\be
-{\cal L}^{H^0}_{l,\Sigma^{-}} &=& H^0\{ \overline{l}({C}^{{H}^0,L}_{ll}P_L+C^{H^0,R}_{ll}P_R)l+ \{\overline{l}({C}^{H^{0},L}_{l\Sigma^{-}}P_L+{C}^{{H}^0,R}_{l\Sigma^{-}}P_R){\Sigma}^{-}+ h.c \} \\ \nn  && + \overline{\Sigma^{-}}({C}^{H^0,L}_{{\Sigma}^{-} {\Sigma}^{-}}P_L+{C}^{H^0,R}_{{\Sigma}^{-} {\Sigma}^{-}}P_R){\Sigma}^{-}\}
\label{eq:1yukawa}
\ee

\be
-{\cal L}^{h^0}_{l,\Sigma^{-}} &=& h^0 \{ \overline{l}(C^{h^0,L}_{ll}P_L+C^{h^0,R}_{ll}P_R)l +\{\overline{l}(C^{h^0,L}_{l\Sigma^{-}}P_L+C^{h^0,R}_{l\Sigma^{-}}P_R)\Sigma^{-}+ h.c\} \\ \nn && + \overline{\Sigma^{-}}({C}^{h^0,L}_{{\Sigma}^{-} {\Sigma}^{-}}P_L+ C^{h^0,R}_{{\Sigma}^{-}{\Sigma}^{-}}P_R)\Sigma^{-}\}
\label{eq:2yukawa}
\ee

\be
-{\cal L}^{A^0}_{l,\Sigma^{-}} &=& A^0\{\overline{l}(C^{A^0,L}_{ll}P_L+C^{A^0,R}_{ll}P_R)l+\{\overline{l}(C^{A^0,L}_{l\Sigma^{-}}P_L+C^{A^0,R}_{l\Sigma^{-}}P_R)\Sigma^{-}+ h.c \}\\ \nn && + \overline{{\Sigma}^{-}}(C^{A^0,L}_{\Sigma^{-}\Sigma^{-}}P_L+C^{A^0,R}_{\Sigma^{-}\Sigma^{-}}P_R)\Sigma^{-}\}
\label{eq:3yukawa}
\ee

\be
-{\cal L}^{G^0}_{l,\Sigma^{-}} &=& G^0\{\overline{l}(C^{G^0,L}_{ll}P_L+C^{G^0,R}_{ll}P_R)l+\{\overline{l}(C^{G^0,L}_{l\Sigma^{-}}P_L+C^{G^0,R}_{l\Sigma^{-}}P_R)\Sigma^{-}+ h.c \}\\ \nn && + \overline{{\Sigma}^{-}}(C^{G^0,L}_{\Sigma^{-}\Sigma^{-}}P_L+C^{G^0,R}_{\Sigma^{-}\Sigma^{-}}P_R)\Sigma^{-}\}
\label{eq:4yukawa}
\ee

\be
-{\cal L}^{H^0}_{\nu,\Sigma^{0}} &=& H^0\{ \overline{\nu^{\prime}}(C^{H^0,L}_{\nu\nu}P_L+C^{H^0,R}_{\nu\nu}P_R)\nu +\{\overline{\nu^{\prime}}(C^{H^0,L}_{\nu\Sigma^{0}}P_L+C^{H^0,R}_{\nu\Sigma^{0}}P_R)\Sigma^{0}+ h.c \}\\ \nn  && + \overline{{\Sigma}^{0}}(C^{H^0,L}_{\Sigma^{0}\Sigma^{0}}P_L+C^{H^0,R}_{\Sigma^{0}\Sigma^{0}}P_R)\Sigma^{0}\}
\label{eq:5yukawa}
\ee

\be
-{\cal L}^{h^0}_{\nu,\Sigma^{0}} &=& h^0\{ \overline{\nu^{\prime}}(C^{h^0,L}_{\nu\nu}P_L+C^{h^0,R}_{\nu\nu}P_R)\nu+\{\overline{\nu}(C^{h^0,L}_{\nu\Sigma^{0}}P_L+C^{h^0,R}_{\nu\Sigma^{0}}P_R)\Sigma^{0}+ h.c \}\\ \nn  && + \overline{{\Sigma}^{0}}(C^{h^0,L}_{\Sigma^{0}\Sigma^{0}}P_L+C^{h^0,R}_{\Sigma^{0}\Sigma^{0}}P_R)\Sigma^{0}\}
\label{eq:6yukawa}
\ee

\be
-{\cal L}^{A^0}_{\nu,\Sigma^{0}} &=& A^0\{ \overline{\nu}(C^{A^0,L}_{\nu\nu}P_L+C^{A^0,R}_{\nu\nu}P_R)\nu+\{\overline{\nu}(C^{A^0,L}_{\nu\Sigma^{0}}P_L+C^{A^0,R}_{\nu\Sigma^{0}}P_R)\Sigma^{0}+ h.c \}\\ \nn && + \overline{{\Sigma}^{0}}(C^{A^0,L}_{\Sigma^{0}\Sigma^{0}}P_L+C^{A^0,R}_{\Sigma^{0}\Sigma^{0}}P_R)\Sigma^{0}\}
\label{eq:7yukawa}
\ee
\be
-{\cal L}^{G^0}_{\nu,\Sigma^{0}} &=& G^0\{ \overline{\nu}(C^{G^0,L}_{\nu\nu}P_L+C^{A^0,R}_{\nu\nu}P_R)\nu+\{\overline{\nu}(C^{G^0,L}_{\nu\Sigma^{0}}P_L+C^{G^0,R}_{\nu\Sigma^{0}}P_R)\Sigma^{0}+ h.c \}\\ \nn && + \overline{{\Sigma}^{0}}(C^{G^0,L}_{\Sigma^{0}\Sigma^{0}}P_L+C^{G^0,R}_{\Sigma^{0}\Sigma^{0}}P_R)\Sigma^{0}\}
\label{eq:8yukawa}
\ee

\be
-{\cal L}^{H^{-}}_{l,\Sigma^{0},\nu,\Sigma^{-}} &=& H^{-}\{\overline{l}(C^{H^{-},L}_{l\nu}P_L+C^{H^{-},R}_{l\nu}P_R)\nu+ \overline{l}(C^{H^{-},L}_{l\Sigma^{0}}P_L+C^{H^{-},R}_{l\Sigma^{0}}P_R)\Sigma^{0}\\\nn  && + \overline{{\Sigma}^{-}}(C^{H^{-},L}_{\Sigma^{-}\nu}P_L+C^{H^{-},R}_{\Sigma^{-}\nu}P_R)\nu+  \overline{{\Sigma}^{-}}(C^{H^{-},L}_{\Sigma^{-}\Sigma^{0}}P_L+C^{H^{-},R}_{\Sigma^{-}\Sigma^{0}}P_R)\Sigma^{0}\}\\ \nn && +h.c
\label{eq:9yukawa}
\ee

\be
-{\cal L}^{G^{-}}_{l,\Sigma^{0},\nu,\Sigma^{-}} &=& H^{-}\{\overline{l}(C^{G^{-},L}_{l\nu}P_L+C^{G^{-},R}_{l\nu}P_R)\nu+ \overline{l}(C^{G^{-},L}_{l\Sigma^{0}}P_L+C^{G^{-},R}_{l\Sigma^{0}}P_R)\Sigma^{0}\\\nn  && + \overline{{\Sigma}^{-}}(C^{G^{-},L}_{\Sigma^{-}\nu}P_L+C^{G^{-},R}_{\Sigma^{-}\nu}P_R)\nu+  \overline{{\Sigma}^{-}}(C^{G^{-},L}_{\Sigma^{-}\Sigma^{0}}P_L+C^{G^{-},R}_{\Sigma^{-}\Sigma^{0}}P_R)\Sigma^{0}\}\\ \nn && +h.c
\label{eq:10yukawa}
\ee

The exact vertex factors $C_{FI}^{X,L/R}$ for our 
two Higgs doublet Type III seesaw  model are listed in 
Tables \ref{tab:chvertex}, \ref{tab:nuvertex}, \ref{tab:hcvertex}. 

\begin{table}
\begin{center}
\begin{tabular}{||c|c||c|c||} \hline
$ C^{H^0,L}_{ll}$ & $\!\!\frac{1}{\sqrt{2}}
(T^\dagger_{11}Y_l S_{11} \cos\alpha 
+ T^\dagger_{21}Y_\Sigma S_{11} \sin\alpha)\!\!$ &$C^{H^0,R}_{ll}$ & $
\!\!\frac{1}{\sqrt{2}}(S^\dagger_{11}Y^\dagger_l T_{11} \cos\alpha 
+ S^\dagger_{11}Y^\dagger_\Sigma T_{21} \sin\alpha)\!\!$ \cr

$C^{H^0,L}_{l{\Sigma^{-}}}$ & $\!\!\frac{1}{\sqrt{2}}
(T^\dagger_{11}Y_l S_{12} \cos\alpha 
+ T^\dagger_{21}Y_\Sigma S_{12} \sin\alpha)\!\!$ &$ C^{H^0,R}_{l{\Sigma^{-}}}$ & $  
\!\!\frac{1}{\sqrt{2}}(S^\dagger_{11}Y^\dagger_l T_{12} \cos\alpha 
+ S^\dagger_{11}Y^\dagger_\Sigma T_{22} \sin\alpha)\!\!$ \cr

$C^{H^0,L}_{ \Sigma^{-} \Sigma^{-}}$ & $\!\!\frac{1}{\sqrt{2}}
(T^\dagger_{12}Y_l S_{12} \cos\alpha 
+ T^\dagger_{22}Y_\Sigma S_{12} \sin\alpha)\!\!$ &$ C^{H^0,R}_{\Sigma^{-} \Sigma^{-} }$ & $ 
\!\!\frac{1}{\sqrt{2}}(S^\dagger_{12}Y^\dagger_l T_{12} \cos\alpha 
+ S^\dagger_{12}Y^\dagger_\Sigma T_{22} \sin\alpha)\!\!$ \cr

$C^{h^0,L}_{ll}$  & $\!\!\frac{-1}{\sqrt{2}}
(T^\dagger_{11}Y_l S_{11} \sin\alpha 
- T^\dagger_{21}Y_\Sigma S_{11} \cos\alpha)\!\!$ &$ C^{h^0,R}_{ll}$ & $
\!\!\frac{-1}{\sqrt{2}}(S^\dagger_{11}Y^\dagger_l T_{11} \sin\alpha 
- S^\dagger_{11}Y^\dagger_\Sigma T_{21} \cos\alpha)\!\!$ \cr

$C^{h^0,L}_{l{\Sigma^{-}}}$ & $\!\!\frac{-1}{\sqrt{2}}
(T^\dagger_{11}Y_l S_{12} \sin\alpha 
- T^\dagger_{21}Y_\Sigma S_{12} \cos\alpha)\!\!$ &$C^{h^0,R}_{l{\Sigma^{-}}} $ & $
\!\!\frac{-1}{\sqrt{2}}(S^\dagger_{11}Y^\dagger_l T_{12} \sin\alpha 
- S^\dagger_{11}Y^\dagger_\Sigma T_{22} \cos\alpha)\!\!$ \cr

$C^{h^0,L}_{ \Sigma^{-} \Sigma^{-}}$ & $\!\!\frac{-1}{\sqrt{2}}
(T^\dagger_{12}Y_l S_{12} \sin\alpha 
- T^\dagger_{22}Y_\Sigma S_{12} \cos\alpha)\!\!$&$C^{h^0,R}_{ \Sigma^{-} \Sigma^{-}}$  & $
\!\!\frac{-1}{\sqrt{2}}(S^\dagger_{12}Y^\dagger_l T_{12} \sin\alpha 
- S^\dagger_{12}Y^\dagger_\Sigma T_{22} \cos\alpha)\!\!$ \cr

$C^{A^0,L}_{ll}$  & $\!\!\frac{i}{\sqrt{2}}
(T^\dagger_{11}Y_l S_{11} \sin\beta 
+ T^\dagger_{21}Y_\Sigma S_{11} \cos\beta)\!\!$ &$ C^{A^0,R}_{ll}$ & $
\!\!\frac{-i}{\sqrt{2}}(S^\dagger_{11}Y^\dagger_l T_{11} \sin\beta 
+ S^\dagger_{11}Y^\dagger_\Sigma T_{21} \cos\beta)\!\!$ \cr

$C^{A^0,L}_{l{\Sigma^{-}}}$ & $\!\!\frac{i}{\sqrt{2}}
(T^\dagger_{11}Y_l S_{12} \sin\beta 
+ T^\dagger_{21}Y_\Sigma S_{12} \cos\beta)\!\!$ &$C^{A^0,R}_{l{\Sigma^{-}}} $ & $
\!\!\frac{-i}{\sqrt{2}}((S^\dagger_{11}Y^\dagger_l T_{12} \sin\beta 
+ S^\dagger_{11}Y^\dagger_\Sigma T_{22} \cos\beta)\!\!$ \cr

$C^{A^0,L}_{ \Sigma^{-} \Sigma^{-}}$ & $\!\!\frac{i}{\sqrt{2}}
(T^\dagger_{12}Y_l S_{12} \sin\beta 
+ T^\dagger_{22}Y_\Sigma S_{12} \cos\beta)\!\!$&$C^{A^0,R}_{ \Sigma^{-} \Sigma^{-}}$  & $
\!\!\frac{-i}{\sqrt{2}}(S^\dagger_{12}Y^\dagger_l T_{12} \sin\beta 
+ S^\dagger_{12}Y^\dagger_\Sigma T_{22} \cos\beta)\!\!$ \cr

$C^{G^0,L}_{ll}$  & $\!\!\frac{-i}{\sqrt{2}}
(T^\dagger_{11}Y_l S_{11} \cos\beta 
 -T^\dagger_{21}Y_\Sigma S_{11} \sin\beta)\!\!$ &$ C^{G^0,R}_{ll}$ & $
\!\!\frac{i}{\sqrt{2}}(S^\dagger_{11}Y^\dagger_l T_{11} \cos\beta 
- S^\dagger_{11}Y^\dagger_\Sigma T_{21} \sin\beta)\!\!$ \cr

$C^{G^0,L}_{l{\Sigma^{-}}}$ & $\!\!\frac{-i}{\sqrt{2}}
(T^\dagger_{11}Y_l S_{12} \cos \beta 
- T^\dagger_{21}Y_\Sigma S_{12} \sin\beta)\!\!$ &$C^{G^0,R}_{l{\Sigma^{-}}} $ & $
\!\!\frac{i}{\sqrt{2}}(S^\dagger_{11}Y^\dagger_l T_{12} \cos\beta 
- S^\dagger_{11}Y^\dagger_\Sigma T_{22} \sin\beta)\!\!$ \cr

$C^{G^0,L}_{ \Sigma^{-} \Sigma^{-}}$ & $\!\!\frac{-i}{\sqrt{2}}
(T^\dagger_{12}Y_l S_{12} \cos\beta 
- T^\dagger_{22}Y_\Sigma S_{12} \sin\beta)\!\!$&$C^{G^0,R}_{ \Sigma^{-} \Sigma^{-}}$  & $
\!\!\frac{i}{\sqrt{2}}(S^\dagger_{12}Y^\dagger_l T_{12} \cos\beta 
-S^\dagger_{12}Y^\dagger_\Sigma T_{22} \sin\beta)\!\!$ \cr
\hline
\end{tabular}
\caption{\label{tab:chvertex}
The vertex factors for $P_L$ ($P_R$) and their corresponding 
exact expression in terms of the Yukawa couplings and mixing matrices 
are given in the first (third) and second 
(forth) column respectively. The vertex factors listed here are 
for Yukawa interactions of the charged leptons with neutral Higgs.}
\end{center}
\end{table}

\begin{table}
\begin{center}
\begin{tabular}{|c|c||c|c|} \hline

$C^{H^0,L}_{\nu \nu}$ & $\frac{\sin\alpha}{{2}}
(U_{21}^T Y_\Sigma U_{11})$ &$ C^{H^0,R}_{\nu \nu}$ & $
\frac{\sin\alpha}{{2}} (U_{11}^\dagger Y^\dagger_\Sigma U^*_{21})$ \cr

$C^{H^0,L}_{\nu \Sigma^0}$ & $\frac{\sin\alpha}{{2}}
(U_{21}^T Y_\Sigma U_{12})$ &$ C^{H^0,R}_{\nu \Sigma^0}$ & $
\frac{\sin\alpha}{{2}} (U_{11}^\dagger Y^\dagger_\Sigma U^*_{22})$ \cr

$C^{H^0,L}_{\Sigma^0 \Sigma^0}$  & $\frac{\sin\alpha}{{2}}
(U_{22}^T Y_\Sigma U_{12})$ &$ C^{H^0,R}_{\nu \Sigma^0}$ & $
\frac{\sin\alpha}{{2}} (U_{12}^\dagger Y^\dagger_\Sigma U^*_{22})$ \cr

$C^{h^0,L}_{\nu \nu}$ & $\frac{\cos\alpha}{{2}}
(U_{21}^T Y_\Sigma U_{11})$ &$C^{h^0,R}_{\nu \nu} $ & $
\frac{\cos\alpha}{{2}} (U_{11}^\dagger Y^\dagger_\Sigma U^*_{21})$ \cr

$C^{h^0,L}_{\nu\Sigma^0}$ & $\frac{\cos\alpha}{{2}}
(U_{21}^T Y_\Sigma U_{12})$ &$C^{h^0,L}_{\nu \Sigma^0}$ & $
\frac{\cos\alpha}{{2}} (U_{11}^\dagger Y^\dagger_\Sigma U^*_{22})$ \cr

$C^{h^0,L}_{\Sigma^0 \Sigma^0}$ & $\frac{\cos\alpha}{{2}}
(U_{22}^T Y_\Sigma U_{12})$ & $C^{h^0,R}_{\Sigma^0 \Sigma^0}$ &$
 \frac{\cos\alpha}{{2}} (U_{12}^\dagger Y^\dagger_\Sigma U^*_{22})$ \cr

$C^{A^0,L}_{\nu \nu}$&  $\frac{i \cos\beta}{{2}}
(U_{21}^T Y_\Sigma U_{11})$ & $C^{A^0,R}_{\nu \nu}$ &$
- \frac{i \cos\beta}{{2}}(U_{11}^\dagger Y^\dagger_\Sigma U^*_{21})$ \cr

$C^{A^0,L}_{\nu\Sigma^0}$ & $\frac{i\cos\beta}{{2}}
(U_{21}^T Y_\Sigma U_{12})$ &$ C^{A^0,R}_{\nu\Sigma^0}$ & $
- \frac{i \cos\beta}{{2}}(U_{11}^\dagger Y^\dagger_\Sigma U^*_{22})$ \cr

$C^{A^0,L}_{\Sigma^0 \Sigma^0}$ & $\frac{i\cos\beta}{{2}}
(U_{22}^T Y_\Sigma U_{12})$ &$ C^{A^0,R}_{\Sigma^0 \Sigma^0}$ & $
-\frac{i \cos\beta}{{2}} (U_{12}^\dagger Y^\dagger_\Sigma U^*_{22})$ \cr

$C^{G^0,L}_{\nu \nu}$&  $\frac{i \sin\beta}{{2}}
(U_{21}^T Y_\Sigma U_{11})$ & $C^{G^0,R}_{\nu \nu}$ &$
- \frac{i \sin\beta}{{2}}(U_{11}^\dagger Y^\dagger_\Sigma U^*_{21})$ \cr

$C^{G^0,L}_{\nu\Sigma^0}$ & $\frac{i\sin\beta}{{2}}
(U_{21}^T Y_\Sigma U_{12})$ &$ C^{G^0,R}_{\nu\Sigma^0}$ & $
- \frac{i \sin\beta}{{2}}(U_{11}^\dagger Y^\dagger_\Sigma U^*_{22})$ \cr

$C^{G^0,L}_{\Sigma^0 \Sigma^0}$ & $\frac{i\sin\beta}{{2}}
(U_{22}^T Y_\Sigma U_{12})$ &$ C^{G^0,R}_{\Sigma^0 \Sigma^0}$ & $
-\frac{i \sin\beta}{{2}} (U_{12}^\dagger Y^\dagger_\Sigma U^*_{22})$ \cr
\hline
\end{tabular}
\caption{\label{tab:nuvertex}
The vertex factors for $P_L$ ($P_R$) and their corresponding 
exact expression in terms of the Yukawa couplings and mixing matrices 
are given in the first (third) and second 
(forth) column respectively. The vertex factors listed here are 
for Yukawa interactions of the neutral leptons with neutral Higgs.}
\end{center}
\end{table}

\begin{table}
\begin{center}
\begin{tabular}{||c|c||c|c||} \hline

$ C^{H^{-},L}_{l\nu}$ & $
-T^\dagger_{11}Y_l U_{11} \sin\beta $
 &$C^{H^{-},R}_{l\nu}$ & $
(\frac{1}{\sqrt{2}}S^\dagger_{11}Y^\dagger_{\Sigma} {U_{21}}^* - 
S^\dagger_{21}{Y_{\Sigma}}^*  {U_{11}}^* )\cos\beta$ \cr


$C^{H^{-},L}_{l{\Sigma^{0}}}$ & $
-T^\dagger_{11}Y_l U_{12} \sin\beta $
&$ C^{H^{-},R}_{l{\Sigma^{0}}}$ & $  
(\frac{1}{\sqrt{2}}S^\dagger_{11}Y^\dagger_{\Sigma} {U_{22}}^* - 
 S^\dagger_{21}{Y_{\Sigma}}^*  {U_{12}}^* )\cos\beta$ \cr

$C^{H^{-},L}_{\nu \Sigma^{-}}$ & 

$(\frac{1}{\sqrt{2}}U^T_{21}Y_{\Sigma} {S_{12}} - 
 U^T_{11}{Y_{\Sigma}}^T  {S_{22}}) \cos\beta$

&$ C^{H^{-},R}_{\Sigma^{-}{\nu}}$ & 

$-U^\dagger_{11}Y_l^{\dagger} T_{12} \sin\beta $ \cr

$C^{H^{-},L}_{\Sigma^{-}{\Sigma^{0}}}$ & $
-T^\dagger_{12}Y_l U_{12} \sin\beta $
&$ C^{H^{-},R}_{\Sigma^{-}{\Sigma^{0}}}$ & $  
(\frac{1}{\sqrt{2}}S^\dagger_{12}Y^\dagger_{\Sigma} {U_{22}}^* - 
 S^\dagger_{22}{Y_{\Sigma}}^*  {U_{12}}^*) \cos\beta$ \cr

$ C^{G^{-},L}_{l\nu}$ & $
T^\dagger_{11}Y_l U_{11} \cos\beta $
 &$C^{G^{-},R}_{l\nu}$ & $
(\frac{1}{\sqrt{2}}S^\dagger_{11}Y^\dagger_{\Sigma} {U_{21}}^* - 
 S^\dagger_{21}{Y_{\Sigma}}^*  {U_{11}}^* )\sin\beta$ \cr

$C^{G^-,L}_{l{\Sigma^{0}}}$ & $
T^\dagger_{11}Y_l U_{12}  \cos\beta$
&$ C^{G^{-},R}_{l{\Sigma^{0}}}$ & $  
(\frac{1}{\sqrt{2}}S^\dagger_{11}Y^\dagger_{\Sigma} {U_{22}}^* - 
 S^\dagger_{21}{Y_{\Sigma}}^*  {U_{12}}^*) \sin\beta$ \cr

$C^{G^{-},L}_{\nu\Sigma^{-}}$ & $
(\frac{1}{\sqrt{2}}U^T_{21}Y_{\Sigma} {S_{12}} - 
 U^T_{11}{Y_{\Sigma}}^T  {S_{22}}) \sin\beta$

&$ C^{G^{-},R}_{\Sigma^{-}{\nu}}$ & 
$
U_{11}^{\dagger}Y_l^{\dagger}T_{12}  \cos\beta$ \cr


$C^{G^{-},L}_{\Sigma^{-}{\Sigma^{0}}}$ & $
T^\dagger_{12}Y_l U_{12} \cos\beta$
&$ C^{G^{-},R}_{\Sigma^{-}{\Sigma^{0}}}$ & $  
(\frac{1}{\sqrt{2}}S^\dagger_{12}Y^\dagger_{\Sigma} {U_{22}}^* - 
 S^\dagger_{22}{Y_{\Sigma}}^*  {U_{12}}^* )\sin\beta$ \cr
\hline
\end{tabular}
\caption{\label{tab:hcvertex}
The vertex factors for $P_L$ ($P_R$) and their corresponding 
exact expression in terms of the Yukawa couplings and mixing matrices 
are given in the first (third) and second 
(forth) column respectively. The vertex factors listed here are 
for Yukawa interactions of the charged as well as neutral 
leptons with charged Higgs.}
\end{center}
\end{table}

\subsection{Lepton-Gauge coupling}

The lepton-gauge couplings come from the kinetic energy terms 
for the $\Sigma$ fields in the Lagrangian. 
The kinetic energy terms are given as 
\be
-{\cal L}_k &=& \overline{\Sigma'_R}i\gamma^{\mu}
D_{\mu} {\Sigma_R}'+ {L_k^{SM}},
\label{eq:qyukawa2}
\ee
where the first term is for heavy triplet fermion field and the 
second term contains the corresponding contributions from all 
standard model fields. The $\Sigma'_R$ field is  
defined in Eq. (\ref{eq:sigmadef}). 
The covariant derivative is defined as 
\be 
D_{\mu}= {\partial_{\mu}}- \sqrt{2} g \pmatrix{
W_{\mu}^3 & W_{\mu}^+  \cr
 W_{\mu}^- & -W_{\mu}^3 \cr
}.
\ee 
Inserting the covariant derivative in Eq. (\ref{eq:qyukawa2}) 
one obtains the following interaction terms between 
leptons and gauge fields
\be
{\cal L}_{int} = {\cal L}_{NC}^{l,\Sigma^{-}} + 
{\cal L}_{NC}^{\nu,\Sigma^{0}} + 
{\cal L}_{CC},
\ee
where the first two terms contain the neutral current interactions 
between $l^\pm$ and $\Sigma^\pm$ (first term) and 
between $\nu$ and $\Sigma^0$ (second term) respectively. The 
last term gives the charged current interaction between the 
leptons. 
The neutral current interaction Lagrangian involving $l$ and 
$\Sigma^-$ is
given by 
\be
{\cal L}_{NC}^{l,\Sigma^{-}} &=&\overline{l}\gamma^{\mu}\{c^{Z,R}_{ll}P_R+c^{Z,L}_{ll}P_L\}l\,Z_{\mu} 
+\{\overline{l}\gamma^{\mu}\{c^{Z,R}_{l\Sigma^{-}}P_R+c^{Z,L}_{l\Sigma^{-}}P_L\}{\Sigma}^{-}\,Z_{\mu}
+h.c\} 
\nn \\ 
&&
+\overline{{\Sigma}^{-}}\gamma^{\mu}\{c^{Z,R}_{\Sigma^{-}\Sigma^{-}}P_R+c^{Z,L}_{\Sigma^{-}\Sigma^{-}}P_L\}\Sigma^{-}\,Z_{\mu}
,
\label{eq:nccurrent}
\ee
where 
\be 
c^{Z,R}_{ll}&=&\frac{g}{c_w}s_w^2(T_{11}^{\dagger}T_{11})-c_wg(T_{21}^{\dagger}T_{21})\}, \nn\\
c^{Z,R}_{l\Sigma^{-}}&=&\frac{g}{c_w}s_w^2(T_{11}^{\dagger}T_{12})-c_wg(T_{21}^{\dagger}T_{22}), \nn\\
c^{Z,R}_{\Sigma^{-}\Sigma^{-}}&=&\frac{g}{c_w}s_w^2(T_{12}^{\dagger}T_{12})-c_wg(T_{22}^{\dagger}T_{22}),
\label{eq:S-S-ZR}
\ee 
\be
c^{Z,L}_{ll}&=&\frac{g}{c_w}(-\frac{1}{2}+s_w^2)(S_{11}^{\dagger}S_{11})-c_wg(S_{21}^{\dagger}S_{21}), \nn\\
c^{Z,L}_{l\Sigma^{-}}&=&\frac{g}{c_w}(-\frac{1}{2}+s_w^2)(S_{11}^{\dagger}S_{12})-c_wg(S_{21}^{\dagger}S_{22}),\nn\\ 
c^{Z,L}_{\Sigma^{-}\Sigma^{-}}&=&\frac{g}{c_w}(-\frac{1}{2}+s_w^2)(S_{12}^{\dagger}S_{12})-c_wg(S_{22}^{\dagger}S_{22}).
\label{eq:S-S-ZL}
\ee
The neutral current interaction Lagrangian 
involving the neutral leptons is given by
\be
{\cal L}_{NC}^{\nu,\Sigma^{0}}&=&(gc_w+g^{\prime}s_w)\frac{1}{2}\overline{\nu} \gamma^{\mu}\{(U_{11}^{\dagger}U_{11})P_L\} \nu Z_{\mu}\nn\\   && +(gc_w+g^{\prime}s_w)\frac{1}{2}\overline{{\Sigma}^0} \gamma^{\mu}\{(U_{12}^{\dagger}U_{12})P_L\} {\Sigma}^0 Z_{\mu} \nn \\ && +\{(gc_w+g^{\prime}s_w)\frac{1}{2}\overline{\nu} \gamma^{\mu}\{(U_{11}^{\dagger}U_{12})P_L\} \Sigma^0 Z_{\mu}+{\rm h.c} \}.
\label{eq:nncurrent}
\ee
The charged current interaction Lagrangian is given by
\begin{eqnarray}
\hspace*{-0.2cm}
{\cal L}_{CC} &=& g \overline{\nu} \gamma^{\mu} \{\{(U_{21}^{\dagger}S_{21})+\frac{1}{\sqrt{2}}(U_{11}^{\dagger}S_{11})\}P_L+(U_{21}^TT_{21})P_R\}lW_{\mu}^{+}\\ \nonumber
&& +g \overline{\nu} \gamma^{\mu} \{\{(U_{21}^{\dagger}S_{22})+\frac{1}{\sqrt{2}}(U_{11}^{\dagger}S_{12})\}P_L +(U_{21}^TT_{22})P_R\}{\Sigma}^{-}W_{\mu}^{+}\\ \nonumber
&& + g\overline{{\Sigma}^0} \gamma^{\mu}\{\{(U_{22}^{\dagger}S_{21})+\frac{1}{\sqrt{2}}(U_{12}^{\dagger}S_{11})\}P_L+(U_{22}^TT_{21})P_R\}lW_{\mu}^{+}\\ \nonumber
&& + g\overline{{\Sigma}^0} \gamma^{\mu}\{\{(U_{22}^{\dagger}S_{22})+\frac{1}{\sqrt{2}}(U_{12}^{\dagger}S_{12})\}P_L+(U_{22}^TT_{22})P_R\}{\Sigma}^- W_{\mu}^{+}+h.c
\label{eq:chcurrent}
\end{eqnarray}

\subsection{Quark-Higgs coupling}

Finally, we discuss the the Yukawa Lagrangian for quark sector, 
which is given by
\be
-{\cal L}_Q &=& Y_{U_{ij}}\overline{u'_{R_i}}\tilde{\Phi}_1^\dagger Q'_j 
+ Y_{D_{ij}}\overline{d'_{R_i}} \Phi_1^\dagger Q'_j + h.c
\label{eq:qyukawa1}
,
\ee
where $Q'$ is the left-handed quark doublet and $u'_R$ and $d'_R$ 
are the right-handed ``up'' and ``down'' types of quark fields. 
Again, primes denote the flavor bases. 
After the electroweak spontaneous symmetry breaking the 
up and down quark mass matrices are obtained as
\be
M_U=Y_Uv\\ \nn
M_D=Y_Dv
\ee
Note that only $\Phi_1$ couples to both the up and down 
quark fields due to the 
imposed $Z_2$ symmetry,  
while the Yukawa couplings of $\Phi_2$ to quarks is 
forbidden\footnote{This is a major difference 
between our model and 
other two Higgs doublet models where the Higgs which 
couples to the neutrinos also couples to the up type quarks, 
while the one which couples to the charged leptons couples to 
the down type quarks.}. 
However, due to the mixing between Higgs fields as discussed in 
Appendix A, 
all the physical Higgs particles would couple to the quark fields.
Here we list all the interaction vertices between quarks and Higgs fields,
which are specific to our model. The fields represents 
the fields in the mass basis.
\be
-{\cal L}^{H^0}_{u,d} &=& \frac{1}{\sqrt{2}}\frac{cos{\alpha}}{v }\overline{u}M_uuH^0 +\frac{1}{\sqrt{2}}\frac{cos{\alpha} }{v } \overline{d}M_ddH^0
\ee
\be
-{\cal L}^{h^0}_{u,d} &=& -\frac{1}{\sqrt{2}}\frac{sin{\alpha} }{v } \overline{u}M_uuh^0-\frac{1}{\sqrt{2}}\frac{sin{\alpha} }{v }\overline{d}M_ddh^0
\ee
\be
-{\cal L}^{A^0}_{u,d} &=& i \frac{1}{\sqrt{2}}\frac{sin{\beta} }{v }\overline{u}\gamma^5 M_uuA^0-i \frac{1}{\sqrt{2}} \frac{sin{\beta} }{v }\overline{d}\gamma^5 M_ddA^0
\ee
\be
-{\cal L}^{G^0}_{u,d} &=& -i \frac{1}{\sqrt{2}} \frac{cos{\beta} }{v }\overline{u}\gamma^5 M_uuG^0+i \frac{1}{\sqrt{2}} \frac{cos{\beta} }{v }\overline{d}\gamma^5 M_ddG^0
\ee
\be
-{\cal L}^{G^{\pm}}_{u,d} &=& \frac{cos{\beta}}{v}\overline{u}(V_{CKM}M_dP_R-M_uV_{CKM}P_L)d +h.c 
\ee
\be
-{\cal L}^{H^{\pm}}_{u,d} &=& -\frac{sin{\beta}}{v}\overline{u}(V_{CKM}M_dP_R-M_uV_{CKM}P_L)d + h.c 
\ee

\end{appendix}


\end{document}